\documentclass[12pt,a4paper,twoside]{article}
\usepackage{geometry}
\usepackage[round]{natbib}
\usepackage{cancel}
\usepackage{mathtools}
\allowdisplaybreaks
\usepackage{mathrsfs}
\usepackage{amsfonts}
\usepackage{amssymb}
\usepackage{booktabs}
\usepackage{graphicx}
\usepackage{placeins}
\usepackage{stmaryrd}
\usepackage{lipsum}  
\usepackage{epstopdf}
\usepackage{xcolor}
\usepackage{setspace} 
\usepackage{bm}
\usepackage[labelfont=bf]{caption}
\usepackage[skip=-0.25\textwidth]{subcaption}
\usepackage{enumitem}
\usepackage{titlesec}

\colorlet{DDcolor}{black}
\newcommand{\DD}[1]{\textcolor{DDcolor}{{#1}}}

\titleformat{\section}
{\normalfont\fontsize{12}{15}\bfseries}{\thesection .}{0.3em}{}
\titleformat{\subsection}
{\normalfont\fontsize{12}{15}\itshape\bfseries\centering}{\upshape\thesubsection .}{0.3em}{}
\titleformat{\subsubsection}
{\normalfont\fontsize{12}{15}\itshape\bfseries}{\upshape\thesubsubsection .}{0.3em}{}

\usepackage{color}
\usepackage[titletoc,title]{appendix}
\usepackage{xr-hyper}
\usepackage{hyperref}%
\hypersetup{
	colorlinks   = true,
	citecolor    = blue,
    linkcolor    = blue
}
\usepackage{xpatch}
\makeatletter
\patchcmd{\NAT@citex}
  {\@citea\NAT@hyper@{%
     \NAT@nmfmt{\NAT@nm}%
     \hyper@natlinkbreak{\NAT@aysep\NAT@spacechar}{\@citeb\@extra@b@citeb}%
     \NAT@date}}
  {\@citea\NAT@nmfmt{\NAT@nm}%
     \NAT@aysep\NAT@spacechar
     \NAT@hyper@{\NAT@date}}
  {}{}

\patchcmd{\NAT@citex}
  {\@citea\NAT@hyper@{%
     \NAT@nmfmt{\NAT@nm}%
     \hyper@natlinkbreak{\NAT@spacechar\NAT@@open\if*#1*\else#1\NAT@spacechar\fi}%
     {\@citeb\@extra@b@citeb}%
     \NAT@date}}
  {\@citea\NAT@nmfmt{\NAT@nm}%
     \NAT@spacechar\NAT@@open\if*#1*\else#1\NAT@spacechar\fi
     \NAT@hyper@{\NAT@date}}
  {}{}
\makeatother
\usepackage{placeins}
\renewenvironment{abstract}  
{\normalfont
	\begin{center}
		\bfseries \abstractname\vspace{-.5em}\vspace{0pt}
	\end{center}
	\list{}{%
		\setlength{\leftmargin}{0.5 mm}
		\setlength{\rightmargin}{\leftmargin}%
	}%
	\item\relax}
{\endlist}  
\def\equationautorefname~#1\null{%
	(#1)\null
}   
\xpretocmd{\eqref}{equation\,}{}{}
\usepackage{authblk}

\title{\textbf{\Large Correlative effects of induced magnetic field-buoyancy on reactive solute dispersion dynamics in couple-stress fluids}}
\author[1]{Aritra Roy}
\author[2]{Debabrata Das}
\author[2]{Pranab Kumar Mondal\thanks{Email: pranabm@iitg.ac.in, mail2pranab@gmail.com}$^{,}$}
\affil[1]{Department of Mechanical Engineering, Indian Institute of Technology Bombay, Powai - 400076 , India }
\affil[2]{Microfluidics and Microscale Transport Processes Laboratory, Department of Mechanical Engineering,
IIT Guwahati, Guwahati 781039, India}
\date{}                     
\usepackage[utf8]{inputenc}
\DeclareUnicodeCharacter{2212}{\textminus}
\begin{document}
	\maketitle
	\linespread{1.1}
    \begin{abstract}
		\noindent
				We investigate, in this study,  the dispersion of a reactive solute in a couple-stress fluid flowing between two parallel plates under the combined effects of pressure-driven flow, buoyancy, and an induced magnetic field. The model includes first-order heterogeneous reactions at both channel walls, and a bulk reaction is incorporated into the model. We use  Mei’s multiscale homogenization technique, accurate to $\mathcal{O}(\varepsilon^3)$, to develop a higher-order asymptotic formulation as well as to determine the effective longitudinal dispersion coefficient and concentration field. The analytical predictions are complemented by the results obtained from Brownian dynamics simulations and finite-difference solutions, while the Aris method of moments is employed to quantify the transient mean displacement, spatial variance, and effective dispersivity. The hydrodynamic analysis reveals a singular branch in the velocity solution at $Ha=\delta/2$ and identifies a characteristic scaling, $Ha\sim\delta^{1/4}$, separating couple-stress- and magnetically dominated regimes. The analytical dispersion model recovers the classical Taylor-dispersion behavior in the limiting case of a Newtonian fluid in the non-reactive regime and agrees well with experimental measurements. The results show that couple-stress rheology and magnetic damping suppress shear-induced dispersion, whereas buoyancy enhances dispersion through additional transverse velocity gradients. A distinct saturation regime of the dispersion coefficient is identified with increasing $\delta$, while unequal wall absorption produces persistent transverse asymmetry and stronger absorption enhances solute removal, particularly near the source. The numerical and stochastic results alongside statistical inferences are used to further validate the analytical framework developed in this endeavor and resolve higher-order concentration structures and particle-scale wall adsorption.

	\end{abstract}
	
\section{Introduction}
\label{sec:Intro}
The mass transport dynamics of fluids exhibiting complex rheology play an instrumental role in modern engineering, bridging the microscopic precision of analytical chemistry in lab-on-a-chip devices \citep{santiago_rama,santiago_jian} with the macroscopic realities of large-scale environmental flows \citep{mckinley2002filament,dittrich2006lab,ramanaiah1978squeeze,brenner2013macrotransport}. The industrial utility of such flows expands exponentially when coupled with electromagnetism; specifically, applied and induced magnetic fields are now actively leveraged to regulate several crucial biochemical processed in applications consisting of MHD based chromatography \citep{eijkel2003circular} and polymerase chain reaction (MHD-PCR devices) \citep{west2002application}. Consequently, establishing a rigorous mathematical framework for reactive solute dispersion, particularly under the simultaneous highly coupled influence of both applied and induced magnetic fields alongside first-order catalytic wall reactions, which has emerged as a vital imperative for advancing foundational fluid mechanics and optimizing next-generation technologies.

\DD{The classical theory of hydrodynamic dispersion was established by \cite{taylor1953dispersion}, who demonstrated that the combined action of molecular diffusion and shear rate enhances the longitudinal spreading of solutes in laminar flows. Building on this work, \cite{aris_statement} developed the method of moments, providing an effective one-dimensional description of solute transport and an analytical expression for the effective Taylor dispersion coefficient, comprising the molecular diffusivity, $D$, and the shear-induced contribution, $Ka^{2}U^{2}/D$, where $U$ is the mean flow velocity, $a$ is the characteristic cross-sectional length and $K$ is a geometry dependent constant. Since then, Taylor-Aris dispersion theory has become the cornerstone for analysing solute transport in a wide range of flow configurations and transport processes \citep{phillips1997initial,griffiths1999hydrodynamic,Mazumder_Das_1992, Wu_Chen_2014,Jiang2020}. Subsequent studies extended the classical theory to non-Newtonian fluids, and demonstrated that complex rheological behaviour can substantially modify hydrodynamic dispersion \citep{rana2016solute,singh2023significance,rana2026solute}. Among the various constitutive models, the couple-stress fluid theory proposed by  \cite{10.1063/1.1761925} extends classical continuum mechanics by incorporating the influence of material microstructure through couple stresses. This model has successfully been employed to describe polymeric lubricants, emulsions, suspensions and physiological fluids \citep{stokes2012theories,Devakar_Iyengar_2010,el1994couple,Lin_2007}. Among these models, Soundalgekar \citep{10.1063/1.1693276,soundalgekar1975effects,soundalgekar1980effects} first investigated Taylor dispersion in couple-stress fluids, demonstrating that microstructural stresses significantly influence reactive solute transport. Subsequently, \cite{rudraiah1986effect} extended these analyses to transient channel flows, showing that the classical Newtonian behaviour is recovered in the limit of vanishing couple stresses.}
\vspace{-1.2mm}

\DD{ Besides, reactive solute transport involving simultaneous homogeneous (bulk) and heterogeneous (catalytic wall) first-order reactions has attracted considerable attention owing to its importance in catalytic reactors, chemical engineering and microfluidic systems.  Foundational studies on convection--diffusion--reaction processes and catalytic wall reactions were reported in seminal works \citep{katz1959chemical,walker1992chemical,gupta1972effect,neely1976mathematical}. Aforementioned reported analyses were subsequently extended to complex fluids, oscillatory flows, adsorption--desorption mechanisms, buoyancy-driven transport and electro-osmotic systems \citep{ng2008convective,roy2019hydrodynamic,jiang2019solute,saha2023solute,poddar2023scalar,Das_Dhar_Kairi_Mondal_2025}. These developments, together with advances in multiscale homogenization, have considerably improved the understanding of reactive solute transport. Nevertheless, transport and disperison charactertics of reactive solute using analytical framework in confined microfluidic channels remain relatively limited despite extensive investigations in parallel-plate and tubular geometries \citep{wu2011multi,Wu_Chen_2014,jiang2022analytical,singh2023significance,gervais2006mass,ramon2011solute}. It may be mentioned here that the underlying transport characteristics are further enriched in the presence of magnetic fields, which provide an effective mechanism for controlling momentum and mass transport in electrically conducting fluids and have found widespread applications in chemical processing, metallurgy and related industrial technologies \citep{barman2025enhancement,davidson1999magnetohydrodynamics,bojarevics2023mhd}. Following the pioneering work of \cite{cowling1957magnetohydrodynamics}, Taylor dispersion in magnetized channel flows was first investigated by  \cite{gupta1968dispersion} and later extended by  \cite{annapurna1979exact} using Gill's generalized formulation \citep{Gillparam}. More recently, multiscale homogenization \citep{mei1996some,mei2002dispersion,doi:10.1142/7427} has been employed to investigate solute transport under magnetic fields \citep{poddar2021exact,dhar2021dispersion}, while \cite{pratt2022reynolds} examined dispersion in turbulent MHD flows. Most of these studies focused on externally applied magnetic fields, whereas \cite{doi:10.1098/rspa.2024.0091} demonstrated that induced magnetic fields can also significantly influence reactive solute dispersion and solute removal efficiency.
}

\DD{Despite these advances, a comprehensive theoretical framework describing reactive Taylor dispersion in electrically conducting couple-stress fluids under the combined influence of pressure-driven flow, applied and induced magnetic fields, gravitational body forces, and first-order heterogeneous catalytic wall reactions remains unexplored.} To bridge this critical gap, the present work develops a rigorous mathematical framework that simultaneously accounts for the complex interplay of these dynamic forces alongside the linearized micro-rotational kinematics of the electrically conducting medium, explicitly utilizing a couple-stress fluid model. We semi-analytically derive the fully developed, unidirectional velocity profile by integrating the momentum transport equations with the governing principles of induced electromagnetism. Subsequently, Mei’s multiscale homogenization technique is employed to resolve the coupled species transport equation, comprehensively accounting for the couple-stress rheology. This approach enables a rigorous evaluation of the longitudinal concentration gradients, precisely quantifying how key physical parameters modulate the underlying dispersive mechanics. Furthermore, to gain a deeper insight into the macroscopic transport behavior, statistical inferences of the concentration field have been rigorously explored using the moments equations. These theoretical and statistical findings are subsequently synthesized to establish concrete optimization guidelines for microfluidic devices, typically deployed in hydrometallurgy, chemical sensing, and diverse biotechnological applications.
\vspace{-1.2mm}

\DD{The remainder of the paper is organized as follows. Section \ref{sec:mathmodel} introduces the physical model, flow geometry, and the underlying assumptions. Section \ref{sec:cal} presents the governing equations of electromagnetism and hydrodynamics, together with the  solution for the velocity field, its validation, and the associated discussion. Section \ref{sec:sol} develops the multiscale framework for reactive solute transport, derives the Taylor dispersion model, and concentration distribution. Also, a comprehensive analysis of the transport characteristics through analytical, stochastic, and numerical approaches has been presented in Section \ref{sec:sol} as well. The analytical predictions are validated against stochastic Brownian dynamics simulations, and available experimental results, followed by a detailed discussion of the transport mechanisms and the main findings of the study.}

\section{Physical description of the system}
\label{sec:mathmodel}
\begin{figure}
	\centering
	\includegraphics[width=0.9\linewidth]{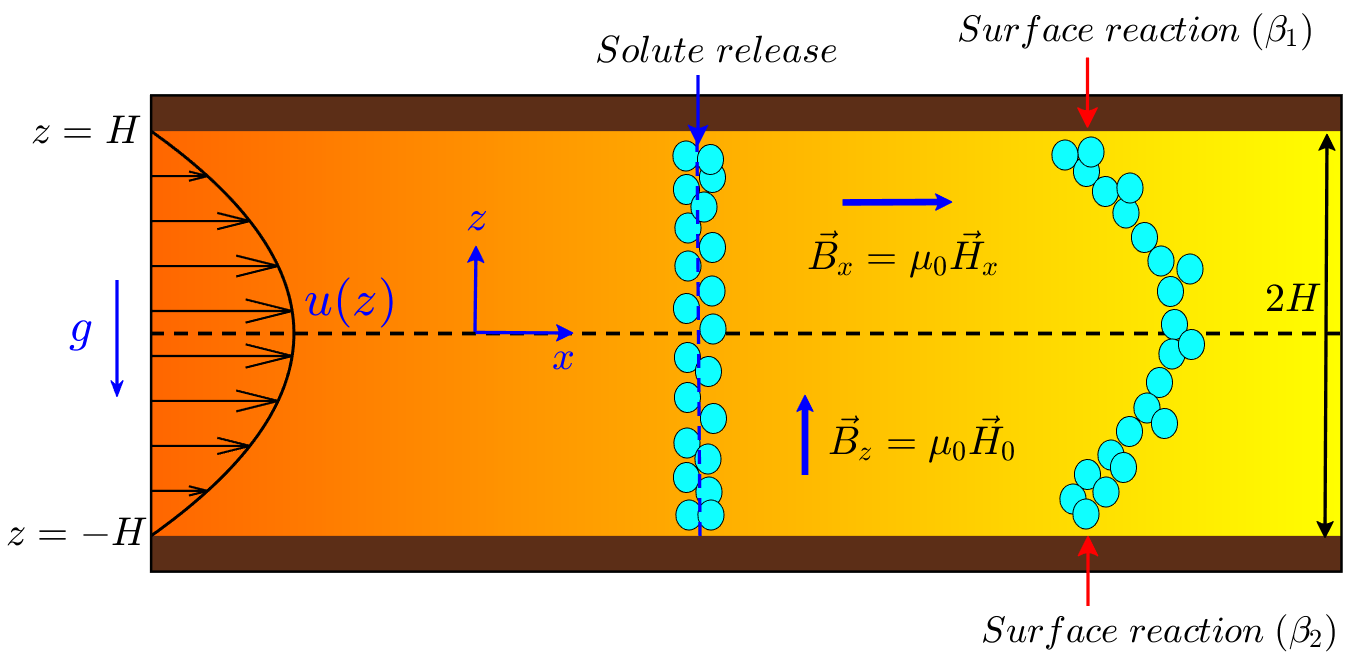}
	\caption{Schematic diagram of a combined pressure-driven unidirectional flow of a viscous electrically conducting couple-stress fluid in presence of an induced magnetic field ($\vec{B}$) through a microchannel. The walls are of non-conducting nature and hence insulated. The $x$ and $z$ axes aligns with the streamwise and the wall-normal direction of the fluid motion respectively. 
	}
	\label{fig:schematic}
\end{figure}

\DD{In the present study, we consider solute dispersion in electrically conducting reactive fluid as shown schematically in \textcolor{blue}{figure} \ref{fig:schematic}. In this analysis, fluid is driven by an applied pressure gradient, while combined effect of buoyancy and magnetic induction influnces the underlying transport. The rheological behaviour of the complex polar fluid is modeled using \DD{the} couple-stress model \citep{stokes2012theories}. In \textcolor{blue}{figure} \ref{fig:schematic}, we present a schematic of the flow configuration together with the relevant dimensions. The coordinate system is attached to the geometric center of the channel. Along the streamwise direction\DD{, we consider the $x$-axis, whereas the $z$-axis is oriented in the} wall-normal direction. Throughout this study, we assume that the channel length is much larger than both its height and width, consistent with the characteristics of a long, slender microfluidic channel.}

The externally imposed constant magnetic field of strength $H_0$ \DD{is applied} along the $z$-axis\DD{, while a weak induced magnetic field of strength $H_x$ is generated in the streamwise direction owing to fluid motion. Under the assumption of a low magnetic Reynolds number, $Re_m \ll 1$, the induced magnetic field remains weak relative to the imposed field.} Under the standard assumption of a linear, non-dispersive medium\DD{,} the magnetic induction follows the constitutive relation $\vec{B}=\mu_0\vec{H}$ \citep{Griffiths_2017} \DD{and is expressed as} $\vec{B}=\left(\mu_0H_x,0,\mu_0H_0\right)$.

\DD{The solute-dispersion process is initiated} at $t=0$ \DD{by an instantaneous pulse that is initially distributed uniformly across the channel cross-section}. Once introduced, the solute is transported  by the \DD{pressure-driven flow}, which is further modulated by \DD{the combined effects of gravity-induced buoyancy and the magnetically induced Lorentz force}. The interplay of these driving mechanisms governs the subsequent spatiotemporal evolution of the solute concentration within the microfluidic channel. \DD{The magnetic properties of the fluid and the permeability of free space are assumed to remain constant throughout the analysis.} Furthermore, the spatiotemporal variation of the concentration\DD{, including the effects of chemical reaction, is investigated using the multiscale homogenization framework of Mei and Vernescu \citep{doi:10.1142/7427}. It is worth to mention here that we develop a theoretical framework to obtain closed-form solutions for the transport equations semi-analytically, as discussed in the forthcoming sections. }

\section{Formulation and description of mathematical model}\label{sec:cal}
\subsection{Governing equations of electromagnetism}
The electromagnetic field is governed by Ampère’s circuital law \citep{Griffiths_2017} and the magnetic induction equation, \DD{which we invoke under the standard assumptions of a quasi-static, linear, and non-magnetic medium in which displacement currents are negligible}. Accordingly, Ampère’s law reduces to the following form \citep{bojarevics2023mhd}. 
\begin{equation}
\boldsymbol{\nabla} \times \mathbf{B} = \mu_{0}\,\mathbf{J}
\label{eq:amplaw}
\end{equation}
where $\mathbf{J}$ denotes the electric current density and $\mu_0$ is the permeability of free space. \DD{Using the magnetic-field distribution} $\mathbf{B} = \left(\mu_0 H_x,0,\mu_0 H_0\right)$, \DD{we obtain the current density and subsequently evaluate the Lorentz force as}
\begin{equation}
\mathbf{J} \times \mathbf{B} = \frac{1}{\mu_0}((\boldsymbol{\nabla} \times \mathbf{B}) \times \mathbf{B}) = \left(\mu_0H_0\frac{dH_x}{dz},0,-\mu_0H_x\frac{dH_x}{dz} \right) 
\label{eq:lorentzforce}
\end{equation}
\DD{To determine the induced magnetic field, we employ the magnetic induction equation} \citep{Griffiths_2017,jackson2012classical}, \DD{which takes the form as given below}
\begin{equation}
\frac{\partial \mathbf{B}}{\partial t}
= \boldsymbol{\nabla} \times \left( \mathbf{u} \times \mathbf{B} \right)
+ \eta \boldsymbol{\nabla}^{2}\mathbf{B}
\label{eq:maginduc}
\end{equation}
where $\mathbf{u}$ represents the fluid velocity vector and $\eta = (\mu_{0}\sigma)^{-1}$ is the magnetic diffusivity, \DD{with $\sigma$ denoting the electrical conductivity}. \DD{Since the magnetic field is steady, we set $\partial\mathbf{B}/\partial t=\mathbf{0}$ in (\ref{eq:maginduc}), which yields}
\begin{equation}
\boldsymbol{\nabla} \times (\mathbf{u} \times \mathbf{B}) + \eta \boldsymbol{\nabla} \times (\boldsymbol{\nabla} \times \mathbf{B})\ = 0
\label{eq:redmaginduc}
\end{equation}
\DD{In this study, we consider unidirectional flow} $\mathbf{u} = (u(z),0,0)$. \DD{Substituting this velocity field together with the prescribed magnetic-field distribution into (\ref{eq:redmaginduc}), we obtain}
\begin{equation}
\frac{d^2 H_x}{dz^2}
= -\sigma\mu_0H_0\frac{du}{dz}
\label{eq:mag_dimension}
\end{equation}
\subsection{Governing equations of hydrodynamics}
\vspace{-0.7mm}
The flow field is governed by the continuity and momentum transport equation \citep{Devakar_Iyengar_2010}, \DD{which are given by}
\begin{equation}
	\boldsymbol{\nabla} \cdot \textit{\textbf{u}} = 0
	\label{eq:cont}
\end{equation}
\begin{equation}
\rho \left( \frac{\partial \mathbf{u}}{\partial t} + \left( \mathbf{u} \cdot \boldsymbol{\nabla} \right)\mathbf{u} \right) = \frac{1}{2}\boldsymbol{\nabla} \times (\rho \mathbf{f_c}) + \boldsymbol{\nabla} \boldsymbol{\tau} + \frac{1}{2}\boldsymbol{\nabla} \times (\boldsymbol{\nabla}\cdot \boldsymbol{M}) + \boldsymbol{J} \times \boldsymbol{B} - \rho g \hat{k}
\label{eq:linearmoment}
\end{equation}
where  $\boldsymbol{\tau}$, $\boldsymbol{M}$ and $\mathbf{f_c}$ \DD{denote the deviatoric force-stress tensor, the couple-stress tensor and the body-couple vector per unit mass, respectively}. Consistent with earlier investigations \citep{10.1063/1.1693276,Lin_2007}, \DD{we neglect the body couple by setting} $\mathbf{f_c}=\mathbf{0}$. \DD{This assumption is commonly adopted for couple-stress fluids because the body-couple contribution is generally negligible compared with the force stresses and other applied body forces. Accordingly, Mindlin and Tiersten} \citep{Mindlin1962} \DD{proposed the following linear constitutive relations for} $\boldsymbol{\tau_{ij}}$ \DD{and} $\boldsymbol{M_{ij}}$\DD{:} 
\begin{equation}
\boldsymbol{\tau_{ij}} = -p \delta_{ij} + \lambda \boldsymbol{D_{rr}} \delta_{ij} + 2\mu \boldsymbol{D_{ij}} \quad \quad \boldsymbol{M_{ij}} = 4\eta \boldsymbol{\Omega_{ij}} + 4\eta' \boldsymbol{\Omega_{ji}}
\label{eq:constitutive}
\end{equation}
where $p$ denotes the fluid pressure, $\boldsymbol{D_{ij}}$ is the strain rate tensor, $\boldsymbol{\Omega_{ij}}$ denotes the curvature tensor(or spin tensor which is defined as $\boldsymbol{\Omega_{ij}} = \frac{1}{2}\boldsymbol{\nabla} \times \boldsymbol{u}$). Note that $\lambda$, $\mu$ represent the viscosity coefficients and $\eta$, $\eta'$ represent the couple stress viscosity coefficients, which are constrained by the following inequalities,
\begin{equation}
\mu \ge 0, \quad 3\lambda + 2\mu \ge 0, \quad \eta \ge 0, \quad |\eta'| < \eta
\end{equation}
The couple-stress formulation introduces a characteristic material length scale $\left(l = \sqrt{\frac{\eta}{\mu}}\right)$, which quantifies the intrinsic polarity of the fluid. This parameter distinguishes polar fluid medium from classical non-polar fluids, for which $\ell \equiv 0$ and the theory reduces to the conventional Newtonian framework. \\
For an incompressible, linear, and isotropic fluid medium, the general moment balance and constitutive relations (\ref{eq:linearmoment}–\ref{eq:constitutive}) simplify considerably. Under these assumptions, \DD{the governing equations reduce to the canonical couple-stress formulation reported in the literature} \citep{radhaprf,Das_Dhar_Kairi_Mondal_2025}\DD{, as  follows}:
\begin{equation}
\rho \left( \frac{\partial \mathbf{u}}{\partial t} + \left( \mathbf{u} \cdot \boldsymbol{\nabla} \right)\mathbf{u} \right) = - \boldsymbol{\nabla} P + \mu \boldsymbol{\nabla}^2 \boldsymbol{u} - \eta \boldsymbol{\nabla}^4 \boldsymbol{u} + \boldsymbol{J} \times \boldsymbol{B} - \rho g \hat{k}
\label{eq:mhdfinal_gov}
\end{equation}
\DD{Substituting (\ref{eq:lorentzforce}) and employing the unidirectional flow assumption} $\mathbf{u}=(u(z),0,0)$ into (\ref{eq:mhdfinal_gov}) yields the ordinary differential equations (ODEs) as written below:
\begin{equation}
-\frac{\partial p}{\partial x} + \mu \frac{d^2 u}{dz^2}  - \eta \frac{d^4 u}{dz^4} + \mu_0H_0\frac{dH_x}{dz} = 0 
\label{eq:ode1}
\end{equation}
\begin{equation}
-\frac{\partial p}{\partial z} - \rho g -\mu_0H_x\frac{dH_x}{dz} = 0 
\label{eq:ode2}
\end{equation}
At this point, we invoke the Boussinesq approximation which assumes that the fluid density varies linearly with temperature as given below:
\begin{equation}
\rho = \rho_{0}\left[1-\alpha\left(\theta^{}-\theta_{0}^{}\right)\right]
\label{eq:bousieq1}
\end{equation}
where $\alpha$ denotes the coefficient of thermal expansion and $\rho_0$, $\theta_0$ represent the reference density and temperature, respectively. \textcolor{black}{From literature it is evident that the  dimensionless Péclet number ($Pe = u^* H / \alpha$), is generally considered as quite low \citep{KIYASATFAR2026112324,XIE2018355} 
for hydromagnetic flow in microchannels. In the present study, the channel walls experiences   a uniform linear axial heating/cooling condition, which means that the temperature on the  walls varies linearly with the axial coordinate $x$ at a constant rate $\mathbb{N}$.  At a sufficiently large distance downstream of the entrance of the channel, where the effects of the entrance become negligible,  the bulk fluid temperature attains the same constant axial gradient as the walls (i.e., $\partial \theta/\partial x = \mathbb{N}$), while the transverse  distribution of temperature becomes invariant with $x$. Following this well-established decomposition \citep{MAZUMDER1976285}, the temperature distribution  in the bulk can thus be represented as a linear axial variation superposed on a transverse perturbation $\phi(z)$, and written below.}
\begin{equation}
\theta^{}-\theta_{0}^{} = \mathbb{N} x + \phi(z)
\label{eq:bousieq2}
\end{equation}
Where $\mathbb{N}$ is a constant temperature gradient, and both confining plates are maintained at uniform, linearly varying thermal conditions. \\
\DD{To eliminate the pressure term, we combine (\ref{eq:ode1}) and (\ref{eq:ode2}) under the Boussinesq approximation, which yields}
\begin{equation}
\mu \frac{d^3 u}{dz^3}  - \eta \frac{d^5 u}{dz^5} + \mu_0H_0\frac{d^2 H_x}{d z^2} + g\frac{d \rho}{d x} = 0
\label{eq:moment1}
\end{equation}
The density gradient along the axial direction, appearing in (\ref{eq:moment1}), can be replaced by using Boussinesq approximation with the help of (\ref{eq:bousieq1}) and (\ref{eq:bousieq2}). We recast (\ref{eq:moment1}) it in the following form:
\begin{equation}
\mu \frac{d^3 u}{dz^3}  - \eta \frac{d^5 u}{dz^5} + \mu_0H_0\frac{d^2 H_x}{d z^2} - \rho_0g\alpha\mathbb{N} = 0
\label{eq:moment2}
\end{equation}
Upon integrating (\ref{eq:moment2}) with respect to $z$, we express the same as gradient of a scalar potential $\mathbf{E}(x)$ along axial direction \citep{doi:10.1098/rspa.2024.0091} to obtain the following equation.
\begin{equation}
\mu \frac{d^2 u}{dz^2}  - \eta \frac{d^4 u}{dz^4} + \mu_0H_0\frac{d H_x}{d z} - \rho_0g\alpha\mathbb{N}z - \frac{d\mathbf{E}(x)}{dx} = 0
\label{eq:moment3}
\end{equation}
\DD{We introduce the following non-dimensional variables and parameters to rewrite (\ref{eq:mag_dimension}) and (\ref{eq:moment3}) in dimensionless form:}
\[
z^* = \frac{z}{H}, \quad \nu = \frac{\mu}{\rho}, \quad P_x = -\frac{H^3}{\rho_0 \nu^2}\frac{d\mathbf{E}(x)}{dx}, \quad \delta = \sqrt{\frac{\mu H^{2}}{\eta}}
\]
\[
 u^* = \frac{uH}{\nu P_x}, \quad H_x^* =  \frac{H_x}{\sigma \mu_0 H_0 \nu P_x}, \quad Ha = \mu_{0} H H_{0} \left( \frac{\sigma}{\mu} \right)^{1/2}, \quad Gr =  \frac{\rho_{0} g \alpha \mathbb{N} H^{4}}{\mu \nu P_x}
\]
Use the above dimensionless parameters, we obtain the dimensionless form of (\ref{eq:mag_dimension}) and  (\ref{eq:moment3}) as following
\begin{equation}
\frac{d^{2}H_x^{*}}{dz^{*2}} = -\frac{d u^{*}}{dz^*}
\label{eq:mhdfinal}
\end{equation}
\begin{equation}
\frac{d^{2}u^{*}}{dz^{*2}} - \frac{1}{\delta^{2}}\frac{d^{4}u^{*}}{dz^{*4}}
+ Ha^{2}\frac{d H_{x}^{*}}{dz^{*}}
- Gr\, z^{*} = -1
\label{eq:momentumfinal}
\end{equation}
For completeness, we note that in the limiting case $\delta^{-1} \to 0$, the coupled system of equations reduces to the classical MHD formulation, thereby confirming the asymptotic consistency of the present model. In order to obtain a well-posed boundary value problem for the coupled ordinary differential equations governing the velocity ($u^*$) and magnetic field distributions ($H_x^*$) (\ref{eq:mhdfinal}-\ref{eq:momentumfinal}), the classical no-slip condition and vanishing couple stress are imposed at the confining walls \citep{Devakar_Iyengar_2010} together with homogeneous boundary conditions on the magnetic field intensity, such that the wall is electrically insulating and carries no surface currents. The prescribed boundary conditions are written as follows:
\begin{equation}
u^*(z^* = \pm 1) = 0, \quad \left.\frac{d^2 u^*}{dz^{*2}}\right|_{z^* = \pm 1} = 0, \quad  H_x^*(z^* = \pm 1) = 0
\label{eq:odebc}
\end{equation}
\begin{figure}[htbp]
    \centering    \includegraphics[width=1.02\textwidth]{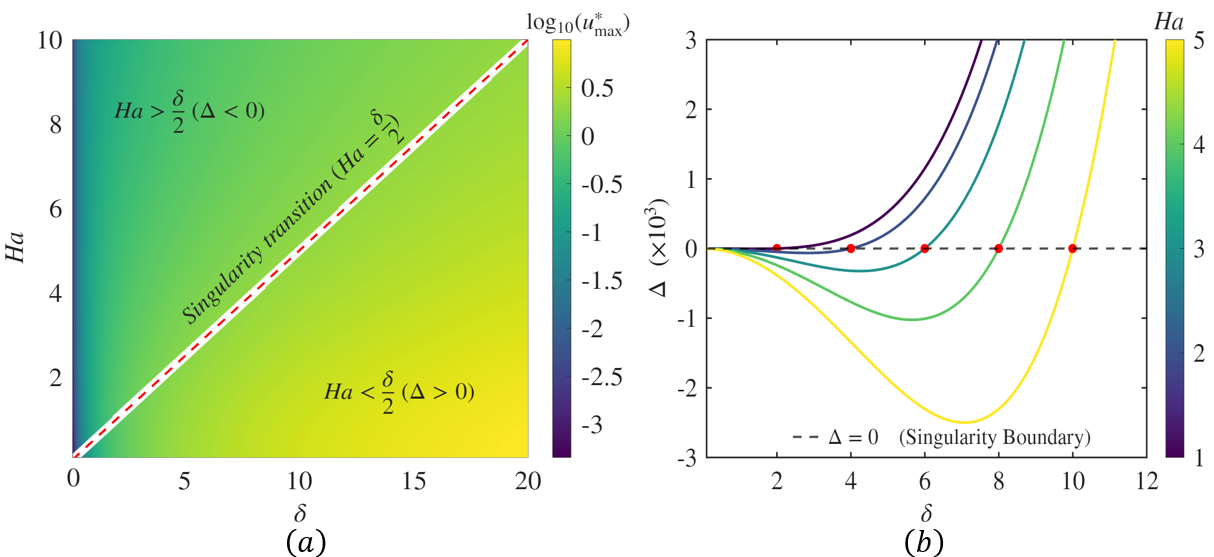}
    \caption{Illustrates the breakdown of the analytical solution. (a) Parametric distribution of maximum velocity magnitude in logarithmic scale depicting the singularity boundary at $Ha = \delta/2$ ($\Delta = 0$), and (b) discriminant profiles $\Delta(\delta)$ indicating critical zero-crossings. Due to the analytical failure along $\Delta = 0$, robust numerical integration of the underlying ODE is mandatory to evaluate the velocity field across the complete parametric spectrum.}
    \label{fig:velsol_fail}
\end{figure}
\DD{We integrate} (\ref{eq:mhdfinal}) once with respect to $z^*$, express the magnetic-field derivative in terms of the velocity field, and \DD{substitute the resulting expression into  (\ref{eq:momentumfinal}) to obtain the following ordinary differential equation:}
\begin{equation}
\frac{d^2 u^*}{dz^{*2}} - \frac{1}{\delta^2} \frac{d^4 u^*}{dz^{*4}} + Ha^2 (-u^* + c_1) - Gr z^* = -1
\end{equation}
where $c_1$ is an arbitrary constant of integration. \DD{We then solve the characteristic equation associated with the constant-coefficient ordinary differential equation and apply the boundary conditions given in  (\ref{eq:odebc}) to obtain the following closed-form analytical solution.}
\begin{equation}
  \begin{split}
    u^*(z^*) &= \frac{-m_1 m_2^3}{Ha^2 \bigl[ m_2^3 \sinh(m_1) - m_1^3 \cosh(m_1) \tanh(m_2) \bigr]} \times\\
    &\quad \biggl[ \cosh(m_1 z^*) - \cosh(m_1) - \frac{m_1^2 \cosh(m_1)}{m_2^2 \cosh(m_2)} \bigl( \cosh(m_2 z^*) - \cosh(m_2) \bigr) \biggr] \\
    &\quad + \frac{Gr}{Ha^2 (m_2^2 - m_1^2)} \biggl[ m_2^2 \frac{\sinh(m_1 z^*)}{\sinh(m_1)} - m_1^2 \frac{\sinh(m_2 z^*)}{\sinh(m_2)} \biggr] - \frac{Gr}{Ha^2} z^*
  \end{split}
\end{equation}
where the $m_1$ and $m_2$ are distinctly expressed in terms of the roots of the characteristics equation as 
\begin{equation}
m_1 = \sqrt{\frac{\delta^2 + \sqrt{\Delta}}{2}}, \quad m_2 = \sqrt{\frac{\delta^2 - \sqrt{\Delta}}{2}} \quad \left[\Delta = \delta^4 - 4\delta^2 Ha^2 \right]
\end{equation}
The analytical solution collapses due to the emergence of a singularity for $\delta = 2Ha$, particularly for specific combination of $(Ha,\delta)$ values, as evident from \textcolor{blue}{figure} \ref{fig:velsol_fail}\textcolor{black}{(a)}. The discriminant vanishes, yielding identical roots as apparent from \textcolor{blue}{figure} \ref{fig:velsol_fail}\textcolor{black}{(b)}. Consequently, to accommodate a wide range of parametric values encompassing both real distinct and repeated root cases, the governing ODE is computed numerically. 
\subsection{Selection of model parameters}
To establish an order of magnitude analysis of the pertinent physical parameters, we consider representative values of the physical properties for a conducting and reactive fluid. We undertake this endeavor essentially to ensure that the fundamental dimensionless variables fall precisely within the physical regimes as delineated in \textcolor{blue}{table \ref{tab:datatab}}. For a typical microfluidic configuration, the half-height of the channel is taken as $H \sim \mathcal{O}(10)\ \mu\text{m}$ \citep{CHAKRABORTY20131151, siva}.  Following the experiments of \citep{tan_rheology}, we consider an effective dynamic viscosity of $\mu \sim \mathcal{O}(10^{-2})$ $\text{Pa}\cdot\text{s}$ and fluid density of $\rho \sim \mathcal{O}(10^3)\ \text{kg}\cdot\text{m}^{-3}$ \citep{kharisov} to account for the situation of a dense microsuspension that is susceptible to induced magnetic fields. The baseline molecular diffusivity  ($\mathcal{D}$) of additives based on silicate derivatives is typically of the order $\mathcal{O}\left(10^{-9}\,\mathrm{m^2\,s^{-1}}\right)$\citep{yamaguchi}. Considering these values, the Péclet number is obtained in the range ($0.1 \le Pe \le 10$), which falls within the range considered in Mei's homogenization methodology \citep{mei1996some}. To faithfully capture the induced magnetic field interactions with the reactive couple stress fluid, we consider the magnetic permeability of $\mu_r = 1.1$ ($\mu_0 \approx 1.38 \times 10^{-6}\ \text{H}\cdot\text{m}^{-1}$) \citep{suwamagsus} and the operating magnetic flux density of $H_0 \sim \mathcal{O}(10^4 - 10^5)\ \text{A}\cdot\text{m}^{-1}$ \citep{wenmagnetic}, which cleanly bounds the Hartmann number within $0 < Ha \le 10$. Similarly, the balance coefficient of volumetric expansion $\alpha \mathbb{N} \sim \mathcal{O}(10^3)\ \text{m}^{-1}$ \citep{ZHOU202352} against the dimensionless axial pressure gradient $P_x \sim \mathcal{O}(10^{-2})\ \text{Pa}\cdot\text{m}^{-1}$ naturally spans the range of Grashof number in $-4 \le Gr \le 4$. For the heterogeneous first order surface reaction the bulk rate constant is considered in accordance with the experimentally verified kinetic data of \citep{kianidaniyal} $\hat{\mathcal{K}}_f \approx 0.1$.  Based on these practical considerations, we develop the present model to ensure that our numerical results and analytical solutions correspond to physically consistent microscale transport phenomena. 
\begin{table}
  \centering
  \caption{Summary of the dimensionless parameters and their investigated ranges.}
  \label{tab:parameters}
  \begin{tabular}{lcc}
    \toprule
    Parameters & Symbols & Values \& ranges \\
    \midrule
    P\'eclet number \citep{mei1996some}                           & $Pe$                          & $0.1 \le Pe \le 10$ \\
    Hartmann number \citep{Mazumder_Das_1992}                      & $Ha$                            & $0 < Ha \le 10$ \\
    Grashof number \citep{MAZUMDER1976285}                       & $Gr$                            & $-4 \le Gr \le 4$ \\
    dimensionless absorption parameters \citep{Gillparam}   & $\hat{\beta}_1, \hat{\beta}_2$ & $0 < \hat{\beta}_1, \hat{\beta}_2 \le 1.5$ \\
    dimensionless bulk chemical reaction \citep{doi:10.1098/rspa.2024.0091}      & $\hat{\mathcal{K}}_f$          & $0.1$ \\
    \bottomrule
  \end{tabular}
  \label{tab:datatab}
\end{table}
\subsection{Velocity Field: Validation and Analysis}
\begin{figure}[htbp]
\centering
\includegraphics[width=1.01\textwidth]{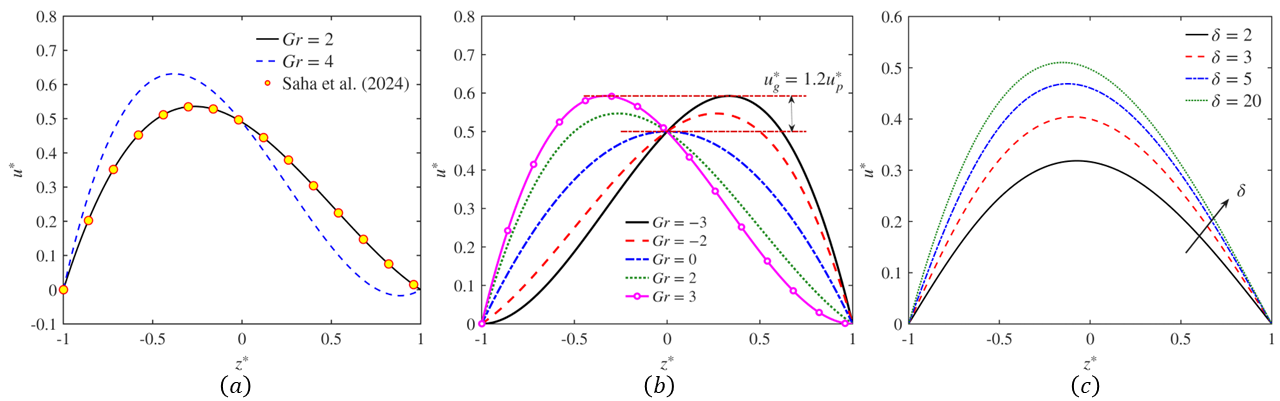}
\caption{Flow profiles for a couple stress fluid under the combined influence of induced magnetic forcing ($Ha = 0.5$) and buoyancy ($Gr$). \DD{The Hartmann number is fixed throughout to isolate the effects of buoyancy and couple stresses.}(\textit{a}) Validation of the present model against the results of \citep{doi:10.1098/rspa.2024.0091} for varying Grashof numbers, $Gr = 2$ and $Gr = 4$, exhibiting flow reversal near the wall $(z^* =1)$. (\textit{b}) Evolution of the dimensionless axial velocity $u^*$ across the transverse coordinate $z^*$ under opposing ($Gr < 0$), purely forced ($Gr = 0$), and aiding ($Gr > 0$) buoyancy regimes. The buoyancy effects induce pronounced flow asymmetry and shift the location and magnitude of the peak velocity. (\textit{c}) Effect of the couple stress parameter $\delta$ on the transverse velocity distribution. Increasing $\delta$ weakens the micro-rotational resistance within the fluid, which systematically enhances the flow magnitude and illustrates the asymptotic transition towards purely Newtonian behaviour ($\delta \to \infty$).}
\label{fig:velprofile}
\end{figure}

\DD{We first examine the dimensionless axial velocity distribution under the combined effects of couple stresses, buoyancy, and induced magnetic forcing,} as shown in \textcolor{blue}{figure} \ref{fig:velprofile}. \DD{For the plots depicted in \textcolor{blue}{figure} \ref{fig:velprofile}, the Hartmann number is fixed at $Ha=0.5$, allowing the individual effects of buoyancy and couple stresses to be isolated while maintaining a constant level of magnetic damping.} To validate the analytical \DD{solution}, we undertake an effort in \textcolor{blue}{figure} \ref{fig:velprofile}\textcolor{black}{(a)} \DD{to compare our results in the limiting case of Newtonian fluid} with the results reported by Saha et al. \citep{doi:10.1098/rspa.2024.0091} for $Gr = 2$. 

It is evident that the \DD{positive buoyancy accelerates the flow near} the lower wall ($z^* < 0$) \DD{while retarding} near the upper wall ($z^* > 0$). \DD{Consequently,}  this effect breaks the classic Poiseuille symmetry, shifting the velocity maximum towards $z^* \approx -0.4$ and skewing the velocity gradients at the boundaries. Additionally, in \textcolor{blue}{figure} \ref{fig:velprofile}\textcolor{black}{(a)}, as the buoyancy parameter is amplified to $Gr = 4$, a  localized momentum deficit \DD{develops near the upper wall}. \DD{Near} the lower wall $z^* = -0.4$, the peak velocity is nearly $u^* \approx 0.631$.\DD{This behaviour arises because} the buoyancy forces asymmetrically aid the pressure-driven flow on one side of the channel while opposing it on the other, constructively superimposing to shift and amplify the velocity peak. It is noteworthy to mention that the opposing thermal buoyancy forces completely overcome the driving axial pressure gradient and local viscous diffusion. This observation is graphically evident from the inflection point at $z^* = 0.88$ and flow reversal with $u^*_{min} \approx -0.017$, \DD{substantially modifies the wall shear stress} ($\tau_{rz} \propto du^*/dz^*$) at $z^* = 1$. \textcolor{blue}{Figure} \ref{fig:velprofile}\textcolor{black}{(b)} \DD{illustrates} the transition between opposing ($Gr < 0$), and aiding ($Gr > 0$) regimes. The isothermal baseline corresponds to $Gr = 0$ that recovers the symmetric Poiseuille flow, with a centralized peak velocity of $u^* = 0.5$ at $z^* = 0$. The system exhibits a strict morphological symmetry based on the sign of $Gr$. The profile for $Gr = 3$ is a perfect reflection of $Gr = -3$ across the channel centerline ($z^* = 0$). A specific enhancement factor is highlighted, denoting $u_g^* = 1.2 u_p^*$. This indicates that the superposition of strong buoyancy ($\vert{}Gr\vert{} = 3$) does not merely shift the peak but amplifies the maximum velocity by $20 \%$ compared to the purely pressure-driven case. \DD{The increased peak velocity is accompanied by steeper velocity gradients near the corresponding wall}, thinning the local momentum boundary layer. \textcolor{blue}{Figure} \ref{fig:velprofile}\textcolor{black}{(c)} depicts the effect of fluid's rheological behavior on the flow velocity, obtained by varying the couple stress parameter, $\delta$, which is inversely proportional to the fluid's micro-rotational resistance. For low values of $\delta$ (e.g., $\delta = 2$), the fluid strongly \DD{resists} rotational gradients, and this behavior aligns with the reported results as well \citep{radhaprf}. \DD{The enhanced microstructural resistance effectively increases the apparent viscosity, thereby flattening the velocity profile,} and suppresses the peak velocity to $u^* \approx 0.32$. As $\delta$ increases ($\delta \to \infty$), the length scale of the couple stresses vanishes, and the micro-rotational resistance weakens, \DD{indicating} an asymptotic convergence towards the standard parabolic Newtonian profile. 

\begin{figure}[htbp]
\centering
\includegraphics[width=1.03\textwidth]{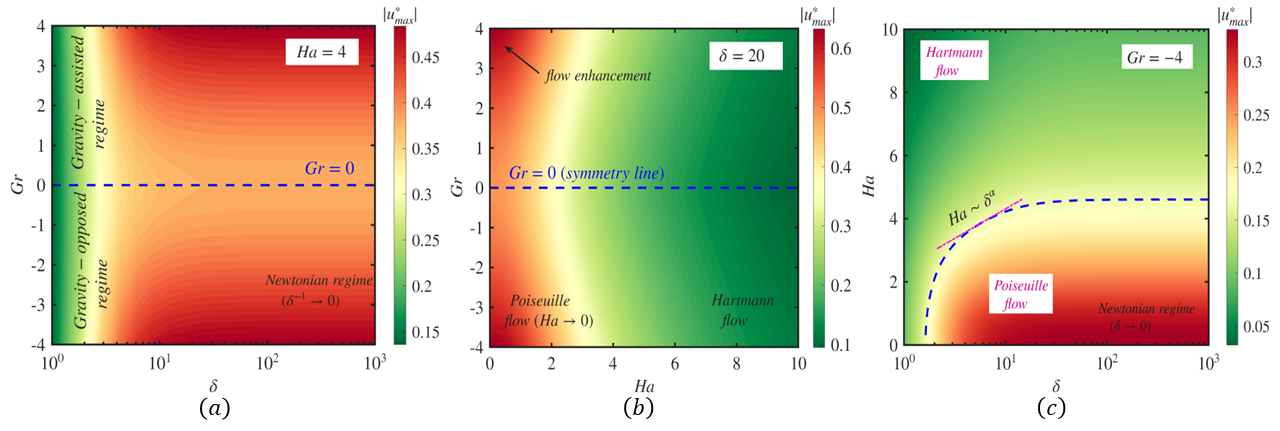}
\caption{Surface plots of the maximum absolute dimensionless velocity $\vert{}u_{max}^*\vert{}$, for a couple stress fluid subjected to transverse magnetic forcing and buoyancy. (a) Variation in the $(\delta, Gr)$ plane at a fixed Hartmann number $Ha = 4$, displaying the $Gr = 0$ symmetry line and the asymptotic recovery to Newtonian behaviour as the couple stress parameter $\delta \to \infty$. (b) Transitions in the $(Ha, Gr)$ plane at $\delta = 20$, illustrating the shift from a buoyancy-modified Poiseuille limit ($Ha \to 0$) to a Lorentz force dominated Hartmann flow. (c) Regime transitions in the $(\delta, Ha)$ plane for a gravity-opposed flow ($Gr = -4$). The parameter space illustrates the distinct competition between micro-rotational viscous diffusion and electromagnetic retardation. The boundary separating the couple-stress-dominated Poiseuille flow from the magnetically damped Hartmann flow is governed by the explicitly derived algebraic scaling law $Ha \approx \delta^{1/4}$. This scaling relation effectively bridges the highly couple-stress-dependent domain before the system ultimately converges to the classical, Newtonian limit ($\delta^{-1} \to 0$) at large $\delta$.}
\label{fig:velcmap}
\end{figure}

\DD{\textcolor{blue}{Figure} \ref{fig:velcmap} provides a global view of the maximum dimensionless velocity over the governing parameter space, highlighting the transitions between buoyancy, couple-stress, and magnetically dominated flow regimes.} \textcolor{blue}{Figure} \ref{fig:velcmap}\textcolor{black}{(a)} delineates the interplay between micro-rotational viscous resistance and thermal buoyancy at a fixed moderate magnetic field ($Ha = 4$). The horizontal axis, in logarithmic scale of $\delta$, clearly captures the asymptotic transition from a highly retarded couple-stress fluid ($\delta \to 10^0$) to a classical Newtonian fluid ($\delta^{-1} \to 0$). The velocity magnitude systematically increases as $\delta \to 10^3$. This is because of the suppression of micro-rotational gradients that reduces the apparent macro-viscosity of the fluid. Additionally, it is evident from \textcolor{blue}{figure} \ref{fig:velcmap}\textcolor{black}{(a)} that it maintains a strict symmetry across the $Gr = 0$ line. \DD{The maximum velocity therefore depends primarily on the magnitude of buoyancy, \(|Gr|\), rather than its direction. Consequently, both aiding (\(Gr>0\)) and opposing (\(Gr<0\)) buoyancy produce the same maximum velocity magnitude while shifting its location to opposite channel walls. } In both the gravity-assisted and gravity-opposed regimes, the magnitude of the peak velocity, \(|u_{\max}^*|\), increases with increasing \(|Gr|\), underscoring that buoyancy strengthens the flow irrespective of its direction. \textcolor{blue}{Figure} \ref{fig:velcmap}\textcolor{black}{(b)} \DD{illustrates the competition between Lorentz-force damping and buoyancy-induced acceleration.} As in $Ha \to 0$, the underlying scenario recovers the Poiseuille flow limit with modified buoyancy, which is characterized by high peak velocities. As the magnetic field augments ($Ha \to 10$), the induced Lorentz force flattens the velocity profile, severely damping and forcing the system into the Hartmann regime. \DD{The largest values of \(|u_{\max}^{*}|\) occur for weak magnetic fields \((Ha \rightarrow 0)\) combined with strong buoyancy \((|Gr| \approx 4)\), where Lorentz damping is minimal. As the Hartmann number increases, the induced Lorentz force progressively suppresses the velocity throughout the parameter space, demonstrating the dominant braking effect of the magnetic field.}
 \\
In \textcolor{blue}{figure} \ref{fig:velcmap}\textcolor{black}{(c)}, we depict the competition between two distinct damping mechanisms: internal couple-stress friction and electromagnetic damping, evaluated for a strongly opposed buoyancy regime ($Gr = -4$). \DD{The dashed curve separates regions where the dominant flow retardation arises from couple-stress effects from those where magnetic damping governs the flow dynamics.} The transition boundary in the semi-logarithmic plane is characterized by a power-law scaling. We can derive the scaling exponent by connecting this geometric slope to the semi-logarithmic derivative. Let us assume a generalized power-law relationship between the Hartmann number and the couple stress parameter at the boundary, $Ha = A \delta^{\alpha}$. Using the chain rule of derivative we obtain the following 
\begin{equation}
\frac{d(Ha)}{d(\log_{10} \delta)} = \frac{d}{d(\log_{10} \delta)} \left( A e^{\alpha (\log_{10} \delta) \ln 10} \right) = \alpha \ln(10) Ha
\end{equation}
Hence, from semi-logarithmic derivative we can estimate the characteristic transition value of $\alpha$ to approximately equal to $0.252$. Therefore, the boundary separating the rheologically damped flow from the magnetically damped flow \DD{is well approximated by} the algebraic scaling law: $Ha \sim \delta^{1/4}$. \DD{The scaling law provides a quantitative criterion for the transition from couple-stress-dominated to magnetically dominated flow,} dictating exactly how much magnetic forcing is required to overcome a specific degree of micro-rotational fluid resistance.
\subsection{\DD{Equations governing Solute Transport}}
\DD{To investigate reactive solute-dispersion characteristics, we introduce solute particle} through a continuous line source. The resulting solute concentration field, $C(x,z,t)$, is subsequently governed by the  \DD{following convection--diffusion--reaction equation \citep{Das_Dhar_Kairi_Mondal_2025}:}
\begin{equation}
    \frac{\partial C}{\partial t} + \left( \mathbf{u} \cdot \nabla \right) C = \mathcal{D} \nabla^2 C - \mathcal{K}_f C, \quad -H < z < H
    \label{eq:chemreac}
\end{equation}
where $\mathcal{D}$ represents the constant \DD{molecular} diffusivity and $\mathcal{K}_f$ represents the reaction rate of the first-order bulk irreversible chemical reaction. 
At $t=0$, \DD{we prescribe} the dispersion problem \DD{by introducing} an impulsive release of solute from an continuous line source. \DD{The corresponding initial concentration field is given by,}

\begin{equation}
C(x,z,t) = C_0\,\delta\!\left(\frac{x}{H}\right)
\label{eq:IC_dispersion}
\end{equation}
where $\delta(\cdot)$ denotes the Dirac delta distribution and $C_0$ is the initial concentration of the solute. \\
At the channel walls, the solute undergoes heterogeneous first-order reactions as discussed by the equations written below. Note that the following reaction equations form the boundary conditions for the present analysis.
\begin{subequations}
\begin{align}
-\,\mathcal{D}\,\frac{\partial C}{\partial z} - \beta_{1}\, C &= 0, 
\qquad z = H, 
\label{eq:BC_upper}
\\[4pt]
\phantom{-}\,\mathcal{D}\,\frac{\partial C}{\partial z} - \beta_{2}\, C &= 0, 
\qquad z = -H,
\label{eq:BC_lower}
\end{align}
\end{subequations}
In equations (\ref{eq:BC_upper},\ref{eq:BC_lower}), $\beta_{1}$ and $\beta_{2}$ denote the heterogeneous first-order reaction rate parameters at the upper and lower walls, respectively. Since solute particles are continuously released from a continuous line source, the transport process involves large scale chemical reactions within the bulk fluid inside the channel and simultaneous absorption at the bounding walls. However, at sufficiently large distances from the injection point, the effect of the source becomes negligible. Consequently, the concentration field satisfies the far-field condition as given below.  
\begin{equation}
\left.C(x,z,t)\right|_{z \longrightarrow \pm \infty} = 0
\label{eq:BC_farfield}
\end{equation}

\section{Description of solute concentration field}\label{sec:sol}

\DD{We here write the solute transport equation and the associated conditions, i.e., initial condition, boundary conditions, and far-field condition, in their dimensionless counter parts. Note that the dimensionless reactive solute transport equation forms the basis for the subsequent multiscale homogenization analysis. To render the convection-diffusion reaction equation together with its associated initial and boundary conditions dimensionless, we introduce the following non-dimensional variables and parameters:}

\[
x^* = \frac{x}{L}, \quad z^* = \frac{z}{H}, \quad t^* = \frac{t}{H^{2}/\mathcal{D}}, \quad 
C^* = \frac{C}{C_0}, \quad 
\mathcal{K}_f^* = \frac{\mathcal{K}_F L^{2}}{\mathcal{D}}, \quad
\beta^*_{1} = \frac{\beta_{1} L}{\mathcal{D}}, \quad
\beta^*_{2} = \frac{\beta_{2} L}{\mathcal{D}}
\]
Using these dimensionless variables and parameters, \DD{we recast} the governing convection diffusion  equation together with the associated \DD{initial and boundary  conditions into the following dimensionless form:}

\begin{equation}
\frac{\partial C^*}{\partial t^*} + \varepsilon Pe u^* \frac{\partial C^*}{\partial x^*} = \varepsilon^2 \frac{\partial^2 C^*}{\partial x^{*2}} + \frac{\partial^2 C^*}{\partial z^{*2}} - \varepsilon^2 \mathcal{K}^*_f C^*, \qquad -1 \le z^* \le 1
\label{eq:conc_nd}
\end{equation}
\begin{subequations}
\begin{align}
C^*(x^*, z^*, t^*) &= \delta \left( \frac{x^*}{\varepsilon} \right), \quad \text{for } t^* = 0, \label{eq:2.25a} \\
- \frac{\partial C^*}{\partial z^*} - \varepsilon\beta_1^* C^* &= 0, \quad \text{at } z^* = 1 \label{eq:2.25b} \\
\frac{\partial C^*}{\partial z^*} - \varepsilon\beta_2^* C^* &= 0, \quad \text{at } z^* = -1 \label{eq:2.25c} \\
C^*(x^*, z^*, t^*) &= 0, \quad \text{for } x^* \to \pm \infty \label{eq:2.25d}
\end{align}
\label{eq:bcicconc}
\end{subequations}
In (\ref{eq:conc_nd}), $\varepsilon \left(= \frac{H}{L} \ll 1\right)$ and $Pe \left(= \frac{\nu P_x}{\mathcal{D}}\right)$ denotes the aspect ratio of the channel and characteristics P\'eclet number respectively. The role of $\varepsilon$ is far from incidental; rather it fundamentally shapes the dispersion characteristics, and its effects have been systematically examined across a wide body of literature \citep{radhaprf,poddar2021exact,saha2023solute,Das_Dhar_Kairi_Mondal_2025}. 
\vspace{-5.5mm}
\subsection{Multiscale homogenization analysis}
\DD{To solve the transport equation  (\ref{eq:conc_nd}),} we adopt the asymptotic multiscale homogenization framework proposed by Mei \citep{mei1996some}, which exploits the strong separation between the advective and diffusive time scales in slender microchannels. \DD{Accordingly, we introduce the following three temporal scales as follows:}
\begin{subequations}
\begin{align}
T_0 = \frac{H^{2}}{\mathcal{D}} \\
T_1 = \frac{L}{\nu P_x/\mathcal{D}} \\
T_2 = \frac{L^{2}}{\mathcal{D}}
\end{align}
\end{subequations}
These scales represent, respectively, cross--stream diffusion, longitudinal
advection, and axial diffusion. The ratio of their relative magnitudes satisfies the proportion
\citep{doi:10.1142/7427}:
\begin{equation}
T_0 : T_1 : T_2 = 1 : \frac{1}{\epsilon} : \frac{1}{\epsilon^{2}}
\end{equation}
where \( \epsilon (\ll 1 \)) is the perturbation parameter.
Since temporal scales and temporal variables are reciprocal, we define
the corresponding fast, intermediate, and slow time variables :
\begin{equation}
t^*_0 = t^*, \qquad  
t^*_1 = \epsilon t^*, \qquad  
t^*_2 = \epsilon^{2} t^* 
\end{equation}
Using the chain rule, one can get the total time derivative as written below. 
\begin{equation}
\frac{\partial}{\partial t^*}
= \frac{\partial}{\partial t^*_0}
+ \epsilon \frac{\partial}{\partial t^*_1}
+ \epsilon^2 \frac{\partial}{\partial t^*_2}
\label{eq:chainrule}
\end{equation}
In accordance with the standard asymptotic regular perturbation framework for multiscale transport \citep{fife1975dispersion}, \DD{we expand the dimensionless solute concentration as an asymptotic series in the small parameter $\epsilon$ }\citep{bender2013advanced}:
\begin{equation}
\begin{split}
C^*(x^*,z^*,t^*) = C^*_0(x^*,z^*,t^*_0,t^*_1,t^*_2)
+ \epsilon\, C^*_1(x^*,z^*,t^*_0,t^*_1,t^*_2) 
& + \epsilon^2\, C^*_2(x^*,z^*,t^*_0,t^*_1,t^*_2) + \\\epsilon^3\, C^*_3(x^*,z^*,t^*_0,t^*_1,t^*_2) + \mathcal{O}(\epsilon^4) 
\end{split}
\label{eq:csolperturb}
\end{equation}
where $C^*_0$, $C^*_1$, $C^*_2$ and $C^*_3$ denotes the zeroth, first, second and third order concentration of solute. Substituting (\ref{eq:chainrule}-\ref{eq:csolperturb}) in (\ref{eq:conc_nd} - \ref{eq:bcicconc}) we obtain the following partial differential equations and boundary conditions
\begin{equation}
\begin{aligned}
    &\left( \frac{\partial C_{0}^{*}}{\partial t_{0}^{*}} - \frac{\partial^{2} C_{0}^{*}}{\partial z^{*2}} \right) 
    + \varepsilon \left( \frac{\partial C_{0}^{*}}{\partial t_{1}^{*}} + \frac{\partial C_{1}^{*}}{\partial t_{0}^{*}} + u^{*} Pe \frac{\partial C_{0}^{*}}{\partial x^{*}} - \frac{\partial^{2} C_{1}^{*}}{\partial z^{*2}} \right) + \varepsilon^{2} \biggl( \frac{\partial C_{0}^{*}}{\partial t_{2}^{*}} + \frac{\partial C_{1}^{*}}{\partial t_{1}^{*}} + \frac{\partial C_{2}^{*}}{\partial t_{0}^{*}} + u^{*} Pe \frac{\partial C_{1}^{*}}{\partial x^{*}} \\
    &\qquad\qquad - \frac{\partial^{2} C_{0}^{*}}{\partial x^{*2}} - \frac{\partial^{2} C_{2}^{*}}{\partial z^{*2}} + \mathcal{K}_{f}^{*} C_{0}^{*} \biggr) + \varepsilon^{3} \biggl( \frac{\partial C_{1}^{*}}{\partial t_{2}^{*}} + \frac{\partial C_{2}^{*}}{\partial t_{1}^{*}} + \frac{\partial C_{3}^{*}}{\partial t_{0}^{*}} + u^{*} Pe \frac{\partial C_{2}^{*}}{\partial x^{*}} \\
    &\qquad\qquad - \frac{\partial^{2} C_{1}^{*}}{\partial x^{*2}} - \frac{\partial^{2} C_{3}^{*}}{\partial z^{*2}} + \mathcal{K}_f^* C_{1}^{*} \biggr) + \mathcal{O}(\varepsilon^{4}) = 0
\end{aligned}
\label{eq:conceps}
\end{equation}
\begin{equation}
\begin{aligned}
&\frac{\partial C_{0}^{*}}{\partial z^{*}}
+ \varepsilon
\left(-
\frac{\partial C_{1}^{*}}{\partial z^{*}}
- \beta_{1}^{*} C_{0}^{*}
\right)
+ \varepsilon^{2}
\left(
-\frac{\partial C_{2}^{*}}{\partial z^{*}}
- \beta_{1}^{*} C_{1}^{*}
\right) + 
\varepsilon^{3}
\left(
-\frac{\partial C_{3}^{*}}{\partial z^{*}}
- \beta_{1}^{*} C_{2}^{*}
\right)
+ \mathcal{O}(\varepsilon^{4}) = 0,
\qquad z^{*} = 1
\end{aligned}
\label{eq:bc1}
\end{equation}
\begin{equation}
\begin{aligned}
&\frac{\partial C_{0}^{*}}{\partial z^{*}}
+ \varepsilon
\left(
\frac{\partial C_{1}^{*}}{\partial z^{*}}
- \beta_{2}^{*} C_{0}^{*}
\right)
+ \varepsilon^{2}
\left(
\frac{\partial C_{2}^{*}}{\partial z^{*}}
- \beta_{2}^{*} C_{1}^{*}
\right) + 
\varepsilon^{3}
\left(
\frac{\partial C_{3}^{*}}{\partial z^{*}}
- \beta_{2}^{*} C_{2}^{*}
\right) +
\mathcal{O}(\varepsilon^{4}) = 0,
\qquad z^{*} = -1
\end{aligned}
\label{eq:bc2}
\end{equation}
\DD{Collecting terms at successive orders of $\varepsilon$ produces the following hierarchy of governing equations. We solve these equations sequentially, beginning with the leading-order problem.}
\subsubsection{Zeroth-order perturbation $\Big(\mathcal{O}(\varepsilon^{0})\Big)$}
At the leading order, $\mathcal{O}(\varepsilon^{0})$, (\ref{eq:conceps}), (\ref{eq:bc1}), and (\ref{eq:bc2}) reduce to the classical one-dimensional diffusion problem governing the zeroth-order concentration field, yielding
\begin{equation}
\frac{\partial C_{0}^{*}}{\partial t_{0}^{*}}
=
\frac{\partial^{2} C_{0}^{*}}{\partial z^{*2}},
\qquad -1 \le z^{*} \le 1
\label{eq:pdezero}
\end{equation}
subject to homogeneous Neumann boundary conditions,
\begin{equation}
\frac{\partial C_{0}^{*}}{\partial z^{*}} = 0,
\qquad z^{*} = \pm 1
\label{eq:bczero}
\end{equation}
\DD{We solve (\ref{eq:pdezero}) subject to the boundary condition in (\ref{eq:bczero}) and obtain the general solution.} Note that the general solution given below consists of a depth-averaged mode and a set of exponentially decaying higher-order Fourier modes:
\begin{equation}
C_{0}^{*}(x^{*},z^{*},t_{0}^{*},t_{1}^{*},t_{2}^{*})
=
C_{0}^{*(0)}(x^{*},t_{1}^{*},t_{2}^{*})
+
\sum_{n=1}^{\infty}
\mathrm{Re}\!\left[
C_{0}^{*(n)}(x^{*},t_{1}^{*},t_{2}^{*})
\, e^{i n \pi z^{*}}
\right]
e^{-n^{2}\pi^{2} t_{0}^{*}}
\label{eq:zerosol}
\end{equation}
Since $t_{0}^{*}$ represents the fastest diffusive time scale, \DD{the solution in} (\ref{eq:zerosol}) becomes asymptotically large in comparison with $t_{1}^{*}$ and $t_{2}^{*}$. Consequently, all higher Fourier modes decay exponentially, and the zeroth-order concentration rapidly diminishes over long-time evolution across the depth $\left(e^{-n^{2}\pi^{2} t_{0}^{*}} \to 0\right)$. \DD{The leading-order concentration therefore reduces to the  form}:
\begin{equation}
C_{0}^{*}
=
C_{0}^{*(0)}(x^{*},t_{1}^{*},t_{2}^{*})
\label{eq:perturbsol0}
\end{equation}
which is independent of the transverse coordinate $z^{*}$.
\subsubsection{First-order perturbation $\Big(\mathcal{O}(\varepsilon^{1})\Big)$}
\DD{Having established that the zeroth-order concentration,} \(C_{0}^{*}\), is independent of the transverse coordinate \(z^{*}\), \DD{we now consider} the  first order problem that governs  non-uniform correction to the solute \DD{concentration}. Collecting all terms of order \(\varepsilon^{1}\) from (\ref{eq:conceps}) yields the following.
\begin{equation}
\frac{\partial C_{0}^{*}}{\partial t_{1}^{*}}
+
\frac{\partial C_{1}^{*}}{\partial t_{0}^{*}}
+
u^{*} Pe\, \frac{\partial C_{0}^{*}}{\partial x^{*}}
=
\frac{\partial^{2} C_{1}^{*}}{\partial z^{*2}},
\qquad -1< z^{*} < 1
\label{eq:pdeone}
\end{equation}
\DD{subject to} the following first-order boundary conditions: 
\begin{subequations}
\begin{align}
-\frac{\partial C_{1}^{*}}{\partial z^{*}} - \beta_{1}^{*} C_{0}^{*} = 0,
\quad z^{*} = 1 \\
\frac{\partial C_{1}^{*}}{\partial z^{*}} - \beta_{2}^{*} C_{0}^{*} = 0,
\quad z^{*} = -1
\end{align}
\label{eq:bcone}
\end{subequations}
Following the long-time homogenization \DD{analysis} of  \citep{teng2023diffusioosmotic}, \DD{\DD{we consider the asymptotic regime in which} }the fast diffusive time scale satisfies \(t_{0}^{*}\gg 1\), \DD{while} the intermediate \DD{time} scale \(t_{1}^{*}\) remains finite. \DD{Under these conditions}, the term 
\(\partial C_{1}^{*}/\partial t_{0}^{*}\)
becomes asymptotically negligible and \DD{is therefore neglected}. Consequently, (\ref{eq:pdeone}) reduces to form given below.
\begin{equation}
\frac{\partial C_{0}^{*}}{\partial t_{1}^{*}}
+
u^{*} Pe\, \frac{\partial C_{0}^{*}}{\partial x^{*}}
=
\frac{\partial^{2} C_{1}^{*}}{\partial z^{*2}}
\label{eq:pdeonesimplify}
\end{equation}
For \DD{subsequent analysis}, we  define \DD{cross-sectional averaging}  operator  as \citep{monin2007statistical}
\begin{equation}
\langle f^* \rangle
=
\frac{1}{2}\int_{-1}^{1} f^*\, dz^{*}
\label{eq:statavg}
\end{equation}
\DD{which represents the cross-sectional average of any dimensionless physical quantity} \(f^*\). \DD{Applying the averaging operator to }\( C_0^* \) and the transverse diffusion term 
\(\partial^2 C_n^*/\partial z^{*2}\), \DD{we obtain}
\begin{subequations}
\begin{align}
\langle C_0^* \rangle &= C_0^* \\
\left\langle \frac{\partial^2 C_n^*}{\partial z^{*2}} \right\rangle
&= -\frac{1}{2}\!\left(\beta_1^*\,C_{n-1}^*\big|_{z^*=1}
+\beta_2^*\,C_{n-1}^*\big|_{z^*=-1}\right),
\qquad n=1,2,\dots
\end{align}
\end{subequations}

Applying the \DD{cross-sectional averaging} operator to (\ref{eq:pdeone}), \DD{together with}  boundary conditions described \DD{in} (\ref{eq:bcone}), yields
\begin{equation}
\frac{\partial C_0^*}{\partial t^*_1}
+ \langle u^* \rangle Pe\, \frac{\partial C_0^*}{\partial x^*}
= -\frac{1}{2}(\beta_1^*+\beta_2^*) C_0^*
\label{3.45_new}
\end{equation}
Subtracting (\ref{3.45_new}) from  (\ref{eq:pdeonesimplify}), \DD{while noting that} \(C_0^*\) is uniform in \(z^*\), leads to
\begin{equation}
Pe\, (u^* - \langle u^* \rangle) 
\frac{\partial C_0^*}{\partial x^*}
- \frac{1}{2}(\beta_1^* + \beta_2^*) C_0^*
= \frac{\partial^2 C_1^*}{\partial z^{*2}}
\label{3.46_new}
\end{equation}
The above (\ref{3.46_new}) suggests the following ansatz that has been prevalent in many literature studies \citep{radhaprf} to \DD{represent} the  first-order \DD{concentration field}:
\begin{equation}
C_1^*
= Pe\, A_1(z^*)\, \frac{\partial C_0^*}{\partial x^*}
+ \frac{1}{2}(\beta_1^*+\beta_2^*) A_2(z^*)\, C_0^*
\label{C1_expansion}
\end{equation}
which separates the contributions associated with streamwise gradients and
reaction kinetics. \DD{Equating}  the coefficients of 
\(\partial C_0^*/\partial x^*\), and \(C_0^*\) in \eqref{C1_expansion}, and 
\eqref{3.46_new} yields two independent boundary-value problems for the transverse functions \(A_1(z^*)\) and \(A_2(z^*)\), which are presented in \textcolor{blue}{Appendix A}.
\subsubsection{Second-order perturbation $\Big(\mathcal{O}(\varepsilon^{2})\Big)$}
For the second order of the perturbation expansion \((\mathcal{O}(\varepsilon^2))\), 
(\ref{eq:conceps} -- \ref{eq:bc2}) yield
\begin{equation}
\frac{\partial C_0^*}{\partial t^*_2}
+ \frac{\partial C_1^*}{\partial t^*_1}
+ \frac{\partial C_2^*}{\partial t^*_0}
+ u^* Pe \frac{\partial C_1^*}{\partial x^*}
= \frac{\partial^2 C_0^*}{\partial x^{*2}}
+ \frac{\partial^2 C_2^*}{\partial z^{*2}}
- \mathcal{K}_f^* C_0^*,
\qquad -1 < z^* < 1,
\label{eq:second_order_local}
\end{equation}
with \DD{the following } boundary conditions
\begin{equation}
\left. \left(- \frac{\partial C_2^*}{\partial z^*} - \beta_1^* C_1^* \right) \right|_{z^*=1} = 0,
\qquad
\left. \left( \frac{\partial C_2^*}{\partial z^*} - \beta_2^* C_1^* \right) \right|_{z^*=-1} = 0
\label{eq:second_order_bc}
\end{equation}
Under the long-time assumption \( t^*_0 \gg 1 \), the fast-time derivative 
\( \partial C_2^*/\partial t^*_0 \) becomes negligible \citep{teng2023diffusioosmotic}, reducing 
(\ref{eq:second_order_local}) to the form as given below.
\begin{equation}
\frac{\partial C_0^*}{\partial t^*_2}
+ \frac{\partial C_1^*}{\partial t^*_1}
+ u^* Pe \frac{\partial C_1^*}{\partial x^*}
= \frac{\partial^2 C_0^*}{\partial x^{*2}}
+ \frac{\partial^2 C_2^*}{\partial z^{*2}}
- \mathcal{K}_f^* C_0^*
\label{eq:second_order_reduced}
\end{equation}
Applying \DD{cross-sectional averaging} operator to  (\ref{eq:second_order_reduced}), \DD{together with the }boundary conditions in Eq. \ref{eq:second_order_bc}, \DD{we obtain the following:}
\begin{equation}
\frac{\partial C_0^*}{\partial t^*_2}
+ \frac{\partial \langle C_1^*\rangle}{\partial t^*_1}
+ Pe\, \left\langle u^* \frac{\partial C_1^*}{\partial x^*}\right\rangle = \frac{\partial^2 C_0^*}{\partial x^{*2}}
- \frac{1}{2} \left( \beta_1^* C_1^*(1) + \beta_2^* C_1^*(-1) \right)- \mathcal{K}_f^* C_0^*
\label{eq:meanvarint}
\end{equation}
Subtracting (\ref{eq:meanvarint}) from (\ref{eq:second_order_local}) yields
\begin{equation}
\frac{\partial C_2^*}{\partial t^*_0}
+ \frac{\partial C_1^*}{\partial t^*_1}
+ Pe\, u^* \frac{\partial C_1^*}{\partial x^*}
- Pe \left\langle u^* \frac{\partial C_1^*}{\partial x^*} \right\rangle
- \frac{\partial \langle C_1^* \rangle}{\partial t^*_1}
= \frac{\partial^2 C_2^*}{\partial z^{*2}}
+ \frac{1}{2} \left( \beta_1^* C_1^*(1) + \beta_2^* C_1^*(-1) \right)
\label{eq:second_order_transverse}
\end{equation}
(\ref{eq:second_order_transverse}) \DD{governs} the second-order transverse correction \(C_2^*(z^*)\), which contributes to the higher-order
concentration corrections discussed subsequently. Using  (\ref{3.45_new}) and (\ref{C1_expansion}), \DD{together with the results present in} \textcolor{blue}{Appendix~B}, \DD{we evaluate each term in  }
(\ref{eq:second_order_transverse})  explicitly.  
The height-averaged first--order correction satisfies the condition given below.
\begin{equation}
\frac{\partial \langle C_1^* \rangle}{\partial t^*_1} = 0
\label{eq:c1mean}
\end{equation}
\DD{whereas the temporal evolution of \(C_1^*\)  is  gievn by,}
\begin{equation}
\frac{\partial C_1^*}{\partial t^*_1}
= -Pe^2 \langle u^* \rangle A_1 
\frac{\partial^2 C_0^*}{\partial x^{*2}}
- \frac{Pe}{2}(\beta_1^*+\beta_2^*)
\left( A_1 + \langle u^* \rangle A_2 \right)
\frac{\partial C_0^*}{\partial x^*}
- \frac{1}{4}(\beta_1^*+\beta_2^*)^2 A_2 C_0^*
\label{eq:c1temporal}
\end{equation}
The streamwise \DD{advection term is gievn by,}
\begin{equation}
u^* \frac{\partial C_1^*}{\partial x^*}
=
Pe\, u^* A_1 
\frac{\partial^2 C_0^*}{\partial x^{*2}}
+ \frac{1}{2}(\beta_1^*+\beta_2^*) 
u^* A_2 
\frac{\partial C_0^*}{\partial x^*}
\label{eq:c1streamwise}
\end{equation}
\DD{whose cross-sectional average form is written below.}
\begin{equation}
\left\langle u^* \frac{\partial C_1^*}{\partial x^*} \right\rangle
=
Pe\, \langle u^* A_1 \rangle \frac{\partial^2 C_0^*}{\partial x^{*2}}
+ \frac{1}{2}(\beta_1^*+\beta_2^*)
\langle u^* A_2 \rangle
\frac{\partial C_0^*}{\partial x^*}.
\label{eq:c1uaverage}
\end{equation}
Substituting (\ref{eq:c1mean}-\ref{eq:c1uaverage}) into (\ref{eq:second_order_transverse}) \DD{and collecting like terms, we obtain the below written equation.}
\begin{align}
\frac{\partial C_2^*}{\partial t^*_0}
&+ Pe^2 \left[ u^* A_1 - \langle u^* \rangle A_1 - \langle u^* A_1 \rangle \right]
\frac{\partial^2 C_0^*}{\partial x^{*2}}
\notag \\[4pt]
&+ \frac{Pe}{2}(\beta_1^*+\beta_2^*)
\left[ u^* A_2 - \langle u^* A_2 \rangle
- A_1 - \langle u^* \rangle A_2
- \frac{\beta_1^* A_1(1) + \beta_2^* A_1(-1)}{\beta_1^*+\beta_2^*} \right]
\frac{\partial C_0^*}{\partial x^*}
\notag \\[4pt]
&- \frac{1}{4}(\beta_1^*+\beta_2^*)^2 
\left[ A_2 
+ \frac{\beta_1^* A_2(1) + \beta_2^* A_2(-1)}{\beta_1^*+\beta_2^*} \right]
C_0^*
= \frac{\partial^2 C_2^*}{\partial z^{*2}} 
\label{eq:c1final}
\end{align}
Following the methodology of \citep{Das_Dhar_Kairi_Mondal_2025}, \DD{we express the second-order concentration} $C^*_2$ in  (\ref{eq:c1final}) \DD{using the following}  ansatz:
\begin{equation}
C_2^{*}
=
Pe^{2}\,A_{3}(z^{*})\,
\frac{\partial^{2} C_{0}^{*}}{\partial x^{*2}}
+
\frac{Pe}{2}\left(\beta_{1}^{*}+\beta_{2}^{*}\right)
A_{4}(z^{*})\,
\frac{\partial C_{0}^{*}}{\partial x^{*}}
+
\frac{1}{4}\left(\beta_{1}^{*}+\beta_{2}^{*}\right)^{2}
A_{5}(z^{*})\,C_{0}^{*}
\label{eq:C2sol}
\end{equation}
\DD{Equating the coefficients of} \(\partial^2 C_0^*/\partial x^{*2}\),\(\partial C_0^*/\partial x^*\) and \(C_0^*\), we obtain the set of boundary-value problems for the transverse functions $A_{3}(z^{*})$,$A_{4}(z^{*})$, and $A_{5}(z^{*})$, as mentioned in \textcolor{blue}{Appendix~B}.
\subsubsection{Third-order perturbation $\Big(\mathcal{O}(\varepsilon^{3})\Big)$}
\DD{Collecting the $\mathcal{O}(\varepsilon^{3})$ terms, we obtain the following governing transport equation.} 
\begin{equation}
    \frac{\partial C_{1}^{*}}{\partial t_{2}^{*}} +\frac{\partial C_{2}^{*}}{\partial t_{1}^{*}} +\frac{\partial C_{3}^{*}}{\partial t_{0}^{*}} +u^{*}Pe\,\frac{\partial C_{2}^{*}}{\partial x^{*}} = \frac{\partial^{2}C_{1}^{*}}{\partial {x^{*}}^{2}} +\frac{\partial^{2}C_{3}^{*}}{\partial {z^{*}}^{2}} -\mathcal{K}_f^* C_{1}^{*}
    \label{eq:C3_gov}
\end{equation}
The aforementioned equation is subject to the corresponding boundary conditions at the upper and lower channel walls:
\begin{equation}
    \left[ \frac{\partial C_{3}^{*}}{\partial z^{*}} +\beta_{1}^{*}C_{2}^{*} \right]_{z^{*}=1} = 0, \qquad
    \left[ \frac{\partial C_{3}^{*}}{\partial z^{*}} -\beta_{2}^{*}C_{2}^{*} \right]_{z^{*}=-1} = 0
    \label{eq:C3_bc}
\end{equation}
For the third-order approximation, after neglecting the term $\partial C_{3}^{*}/\partial t_{0}^{*}$, the governing equation takes the form as follows:
\begin{equation}
    \frac{\partial C_{1}^{*}}{\partial t_{2}^{*}} +\frac{\partial C_{2}^{*}}{\partial t_{1}^{*}} +u^{*}Pe\,\frac{\partial C_{2}^{*}}{\partial x^{*}} = \frac{\partial^{2}C_{1}^{*}}{\partial {x^{*}}^{2}} +\frac{\partial^{2}C_{3}^{*}}{\partial {z^{*}}^{2}} -\mathcal{K}_{f}^{*}C_{1}^{*}
    \label{eq:third_order_governing}
\end{equation}
Furthermore, taking the cross-sectional average to (\ref{eq:third_order_governing}), \DD{together with the boundary conditions in (\ref{eq:C3_bc}), we obtain the following equation.}
\begin{equation}
    \frac{\partial \langle C_{1}^{*}\rangle}{\partial t_{2}^{*}} +\frac{\partial \langle C_{2}^{*}\rangle}{\partial t_{1}^{*}} +Pe\left\langle u^{*}\frac{\partial C_{2}^{*}}{\partial x^{*}} \right\rangle = \frac{\partial^{2}\langle C_{1}^{*}\rangle}{\partial {x^{*}}^{2}} -\frac{1}{2} \Bigl( \beta_{1}^{*}C_{2}^{*}(1) +\beta_{2}^{*}C_{2}^{*}(-1) \Bigr) -\mathcal{K}_{f}^{*}\langle C_{1}^{*}\rangle
    \label{eq:third_order_average}
\end{equation}
Subtracting (\ref{eq:third_order_average}) from  (\ref{eq:third_order_governing}), \DD{yields the governing equation for the third-order} fluctuation equation as written below.
\begin{equation}
    \begin{aligned}
        \frac{\partial C_{1}^{*}}{\partial t_{2}^{*}} +\frac{\partial C_{2}^{*}}{\partial t_{1}^{*}} &+ Pe\left( u^{*}\frac{\partial C_{2}^{*}}{\partial x^{*}} - \left\langle u^{*}\frac{\partial C_{2}^{*}}{\partial x^{*}} \right\rangle \right) -\frac{\partial \langle C_{1}^{*}\rangle}{\partial t_{2}^{*}} -\frac{\partial \langle C_{2}^{*}\rangle}{\partial t_{1}^{*}} \\
        &= \frac{\partial^{2}C_{1}^{*}}{\partial {x^{*}}^{2}} +\frac{\partial^{2}C_{3}^{*}}{\partial {z^{*}}^{2}} -\frac{\partial^{2}\langle C_{1}^{*}\rangle}{\partial {x^{*}}^{2}} +\frac{1}{2} \Bigl( \beta_{1}^{*}C_{2}^{*}(1) +\beta_{2}^{*}C_{2}^{*}(-1) \Bigr) +\mathcal{K}_{f}^{*}\langle C_{1}^{*}\rangle -\mathcal{K}_{f}^{*}C_{1}^{*}
    \end{aligned}
    \label{eq:third_order_fluctuation}
\end{equation}
\DD{As in the preceding orders, the averaged derivatives vanish identically because the higher-order concentration corrections are defined to have zero cross-sectional mean, thereby ensuring the uniqueness of the multiscale expansion. Consequently, we write the following equation.}
\begin{equation}
    \frac{\partial \langle C_{1}^{*}\rangle}{\partial t_{2}^{*}} = 0, \qquad
    \frac{\partial \langle C_{2}^{*}\rangle}{\partial t_{1}^{*}} = 0, \qquad
    \frac{\partial^{2}\langle C_{1}^{*}\rangle}{\partial {x^{*}}^{2}} = 0
    \label{eq:derivative_relations_zero}
\end{equation}
\DD{Employing the same procedure outlined in \citep{Das_Dhar_Kairi_Mondal_2025} , we express the third-order concentration correction using the following ansatz:}
\begin{equation}
    \begin{aligned}
        C^*_{3} &= Pe^{3} A_{6}(z^*) \frac{\partial^{3}C^*_{0}}{\partial x^{3}} 
        + \frac{Pe^{2}}{2} (\beta_{1}^{*} + \beta_{2}^{*})^{2} A_{7}(z^*) \frac{\partial^{2}C^*_{0}}{\partial x^{2}} 
        + \frac{Pe}{4} (\beta_{1}^{*} + \beta_{2}^{*})^{2} A_{8}(z^*) \frac{\partial C^*_{0}}{\partial x} + \frac{(\beta_{1}^{*} + \beta_{2}^{*})^{3}}{8} A_{9}(z^*)C^*_{0}
    \end{aligned}
    \label{eq:C3_ansatz}
\end{equation}
By equating the coefficients of the derivative terms $\partial^3 C_0^*/\partial x^{*3}$, $\partial^2 C_0^*/\partial x^{*2}$, $\partial C_0^*/\partial x^*$, and $C_0^*$, we obtain a system of boundary-value problems for the transverse functions $A_{6}(z^*)$, $A_{7}(z^*)$, $A_{8}(z^*)$, and $A_{9}(z^*)$. These equations are detailed in \textcolor{blue}{Appendix~C}.

\subsubsection{Taylor dispersion coefficient}
\DD{Following the classical analysis of \citep{taylor1953dispersion}, the evolution of the } cross-sectionally averaged solute concentration \DD{is governed by the balance} between axial shear dispersion and \DD{transverse} molecular diffusion. Substituting (\ref{eq:c1mean}) and (\ref{eq:c1uaverage}) into (\ref{eq:meanvarint}), we obtain the equation written below:
\begin{equation}\label{eq:macro_dispersion}
\frac{\partial C_0^*}{\partial t_2^*} + \frac{Pe}{2} \left(\beta_1^* + \beta_2^*\right) \langle u^* A_2 \rangle \frac{\partial C_0^*}{\partial x^*} = \left(1 - Pe^2 \langle u^* A_1 \rangle \right) \frac{\partial^2 C_0^*}{\partial x^{*2}} - \frac{1}{2} \left[\beta_1^* C_1^*(1) + \beta_2^* C_1^*(-1)\right] - \mathcal{K}_f^* C_0^*
\end{equation}
\DD{To recover the effective macroscale transport equation, we combine the governing equations associated with the different temporal scales.} Multiplying  (\ref{3.45_new}) by  $\varepsilon$ and  (\ref{eq:macro_dispersion}) by $\varepsilon^2$, we sum the equations to eliminate intermediate derivatives. This combination provides a single, unified governing equation describing the effective advection-diffusion-reaction dynamics:
\begin{equation}\label{eq:unified_multiscale}
\begin{aligned}
\frac{\partial C_0^*}{\partial t_0^*} &+ \varepsilon \frac{\partial C_0^*}{\partial t_1^*} + \varepsilon^2 \frac{\partial C_0^*}{\partial t_2^*} + \varepsilon Pe \left[ \langle u^* \rangle + \frac{\varepsilon}{2}\left(\beta_1^* + \beta_2^*\right) \langle u^* A_2 \rangle + \frac{\varepsilon}{2}\beta_1^* A_1(1) + \frac{\varepsilon}{2}\beta_2^* A_1(-1) \right] \frac{\partial C_0^*}{\partial x^*} \\
&= \varepsilon^2 \left[1 - Pe^2 \langle u^* A_1 \rangle \right] \frac{\partial^2 C_0^*}{\partial x^{*2}} - \left[ \frac{\varepsilon}{2}\left(\beta_1^* + \beta_2^*\right) + \frac{\varepsilon^2}{4}\left(\beta_1^* + \beta_2^*\right) \left\{ \beta_1^* A_2(1) + \beta_2^* A_2(-1) \right\} + \varepsilon^2 \mathcal{K}_f^* \right] C_0^*
\end{aligned}
\end{equation}
\DD{Using the chain-rule relation for the multiple time scales, we recover the original time variable} $t^*$, and (\ref{eq:unified_multiscale}) \DD{becomes}:
\begin{equation}\label{eq:original_time}
\begin{aligned}
\frac{\partial C_0^*}{\partial t^*} &+ \varepsilon Pe \left[ \langle u^* \rangle + \frac{\varepsilon}{2}\left(\beta_1^* + \beta_2^*\right) \langle u^* A_2 \rangle + \frac{\varepsilon}{2}\beta_1^* A_1(1) + \frac{\varepsilon}{2}\beta_2^* A_1(-1) \right] \frac{\partial C_0^*}{\partial x^*} \\
&= \varepsilon^2 \left[1 - Pe^2 \langle u^* A_1 \rangle \right] \frac{\partial^2 C_0^*}{\partial x^{*2}} - \left[ \frac{\varepsilon}{2}\left(\beta_1^* + \beta_2^*\right) + \frac{\varepsilon^2}{4}\left(\beta_1^* + \beta_2^*\right) \left\{ \beta_1^* A_2(1) + \beta_2^* A_2(-1) \right\} + \varepsilon^2 \mathcal{K}_f^* \right] C_0^*
\end{aligned}
\end{equation}
To simplify the governing equation, we introduce the following scaled dimensionless parameters:
\begin{equation}\label{eq:new_vars}
\hat{\beta}_1 = \varepsilon \beta_1^*, \quad \hat{\beta}_2 = \varepsilon \beta_2^*, \quad \hat{\mathcal{K}_f} = \varepsilon^2 \mathcal{K}_f^*
\end{equation}
Substituting (\ref{eq:new_vars}) into (\ref{eq:original_time}) yields the transport equation in compact form and is given as:
\begin{equation}\label{eq:final_transport}
\begin{aligned}
\frac{\partial C_0^*}{\partial t^*} &+ \varepsilon Pe \left[ \langle u^* \rangle + \frac{1}{2}\left(\hat{\beta}_1 + \hat{\beta}_2\right) \langle u^* A_2 \rangle + \frac{1}{2}\hat{\beta}_1 A_1(1) + \frac{1}{2}\hat{\beta}_2 A_1(-1) \right] \frac{\partial C_0^*}{\partial x^*} \\
&= \varepsilon^2 \left[1 - Pe^2 \langle u^* A_1 \rangle \right] \frac{\partial^2 C_0^*}{\partial x^{*2}}- \left[ \frac{1}{2}\left(\hat{\beta}_1 + \hat{\beta}_2\right) + \frac{1}{4}\left(\hat{\beta}_1 + \hat{\beta}_2\right) \left\{ \hat{\beta}_1 A_2(1) + \hat{\beta}_2 A_2(-1) \right\} + \hat{\mathcal{K}_f} \right] C_0^*
\end{aligned}
\end{equation}
Grouping the effective macroscale transport terms and rewriting (\ref{eq:final_transport}) in compact form we arrive at the following form:
\begin{equation}\label{eq:advection_diff_rxn}
\frac{\partial C_0^*}{\partial t^*} + \varepsilon Pe \chi \frac{\partial C_0^*}{\partial x^*} = \varepsilon^2 D^{\text{Taylor}} \frac{\partial^2 C_0^*}{\partial x^{*2}} - \Gamma C_0^*
\end{equation}
where the effective dimensionless advection ($\chi$), Taylor dispersion ($D^{\text{Taylor}}$), and heterogenous reaction ($\Gamma$) coefficients are defined explicitly as:
\begin{equation}\label{eq:chi_definition}
\chi = \langle u^* \rangle + \frac{1}{2}(\hat{\beta}_1 + \hat{\beta}_2) \langle u^* A_2 \rangle + \frac{1}{2}\hat{\beta}_1 A_1(1) + \frac{1}{2}\hat{\beta}_2 A_1(-1)
\end{equation}
\begin{equation}\label{eq:dispersion_definition}
D^{\text{Taylor}} = 1 - Pe^2 \langle u^* A_1 \rangle
\end{equation}
and
\begin{equation}\label{eq:reaction_definition}
\Gamma = \frac{1}{2}(\hat{\beta}_1 + \hat{\beta}_2) + \frac{1}{4}(\hat{\beta}_1 + \hat{\beta}_2) \left\{\hat{\beta}_1 A_2(1) + \hat{\beta}_2 A_2(-1)\right\} + \hat{\mathcal{K}_f}
\end{equation}
Introducing the moving coordinate frame defined by $\tau = t^*$ and $\xi = (x^*/\varepsilon) - Pe \chi t^*$, solving (\ref{eq:advection_diff_rxn}) under the given initial and boundary conditions yields the following longitudinal Gaussian concentration distribution:
\begin{equation}\label{eq:gaussian_distribution}
C_0^* = \frac{1}{\sqrt{4\pi \tau D^{\text{Taylor}}}} \exp\left( -\frac{\xi^2}{4\tau D^{\text{Taylor}}} - \Gamma \tau\right)
\end{equation}
This transformed coordinate system $(\xi, \tau)$ \DD{facilitates the analysis of the dispersion dynamics in a} reference frame co-moving with the mean flow at a constant speed of $Pe \chi$.

\begin{figure}[htbp]
    \centering
    \begin{subfigure}[b]{0.48\textwidth}
        \centering
        \includegraphics[width=\textwidth]{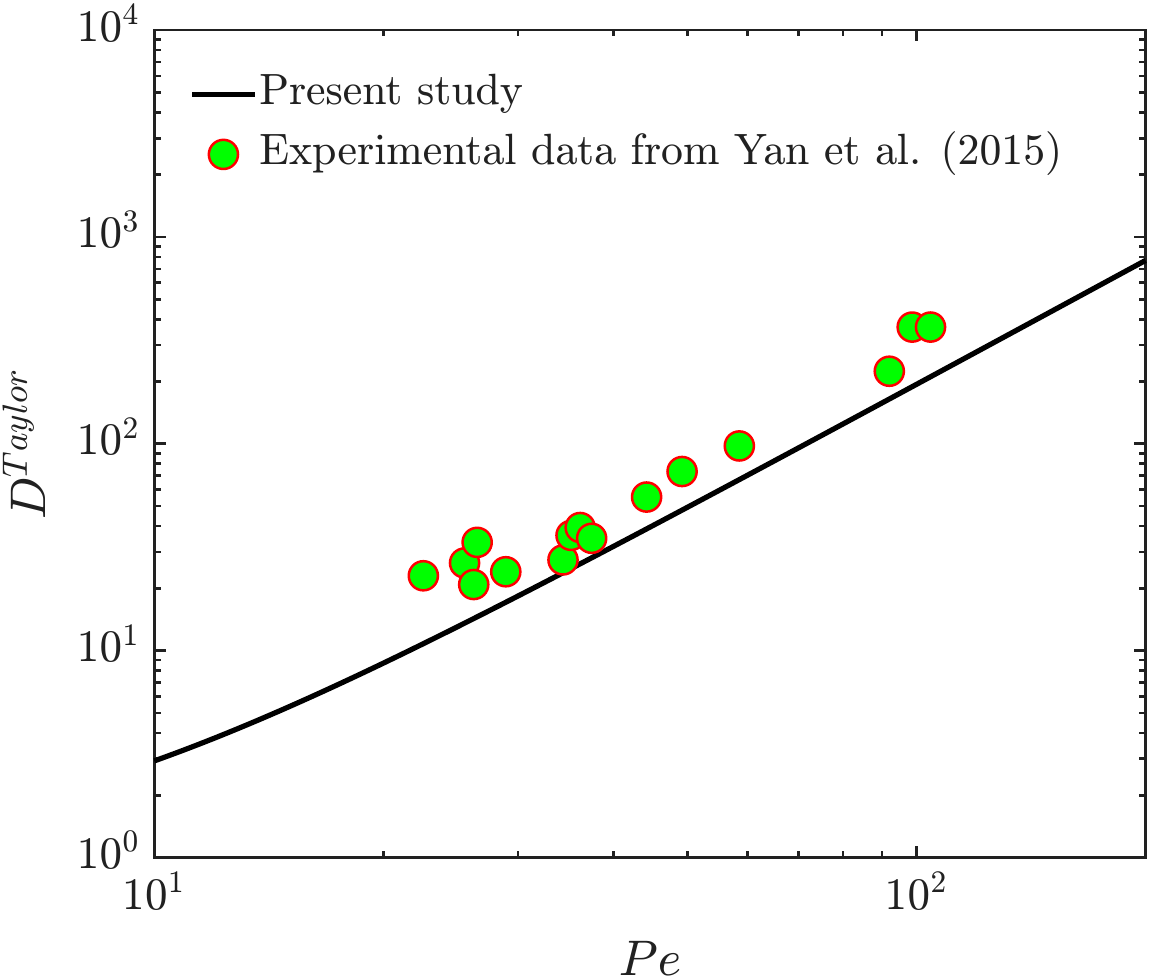}
        \vspace{18mm} 
        \caption{}
        \label{fig:5a}
    \end{subfigure}
    \hfill
    \begin{subfigure}[b]{0.49\textwidth}
        \centering
        \includegraphics[width=\textwidth]{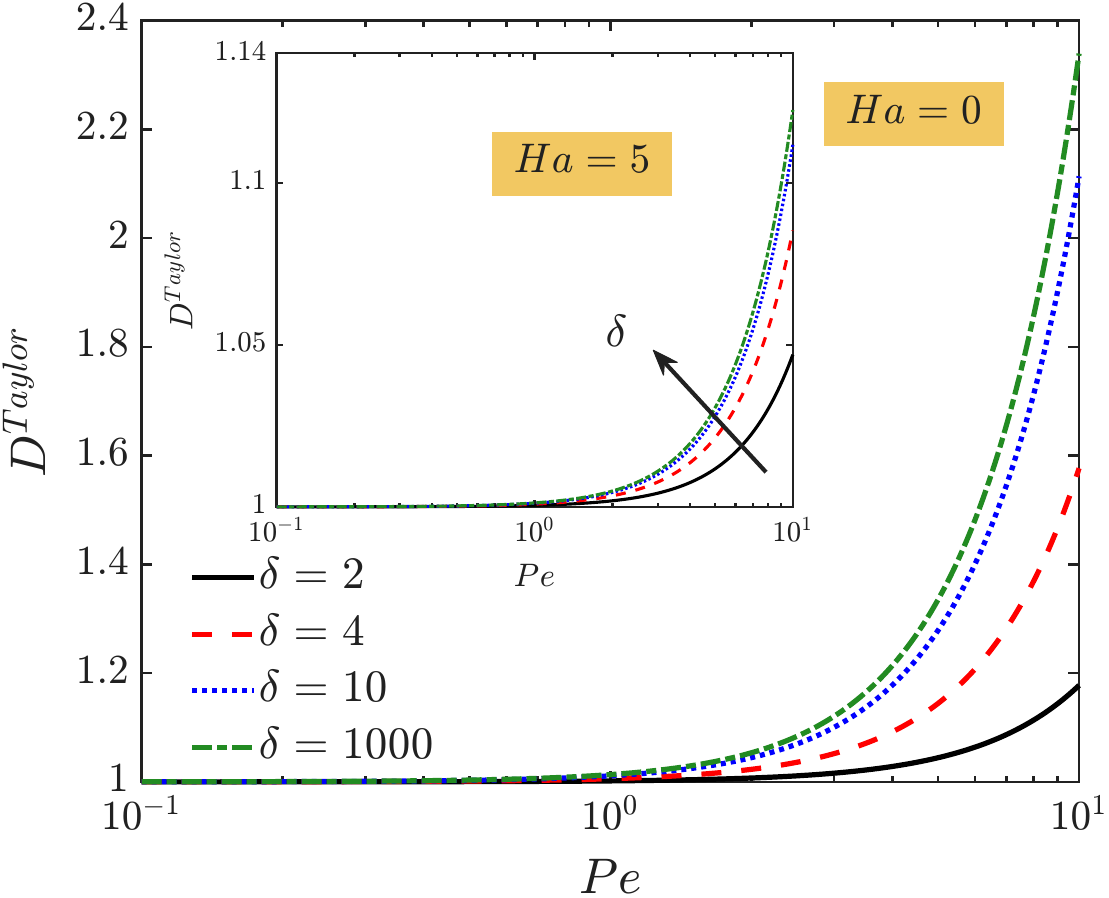}
        \vspace{18mm} 
        \caption{}
        \label{fig:5b}
    \end{subfigure}
  
    \begin{subfigure}[b]{0.48\textwidth}
        \centering        \includegraphics[width=\textwidth,height=6.65cm]{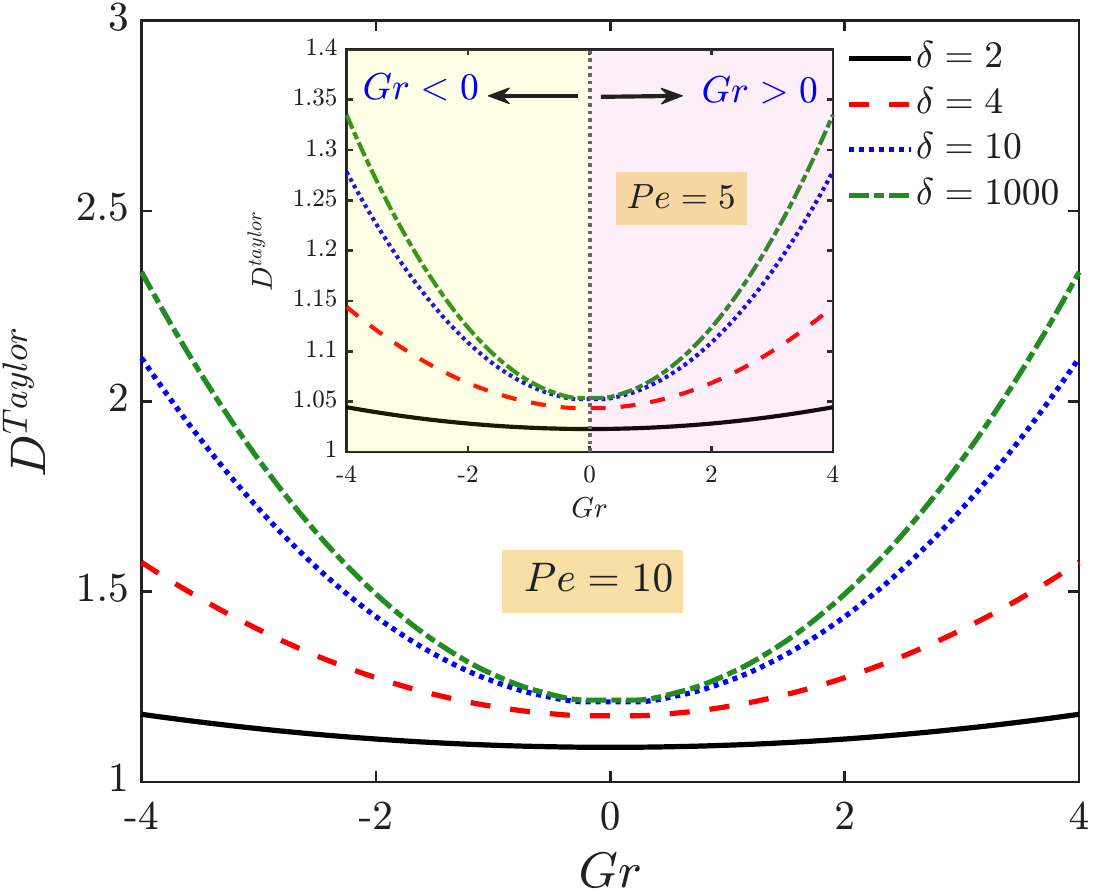}
        \vspace{18mm} 
        \caption{}
        \label{fig:5c}
    \end{subfigure}
    \hspace{0.02\textwidth}
    \begin{subfigure}[b]{0.48\textwidth}
        \centering
      \includegraphics[width=\textwidth,height=6.65cm]{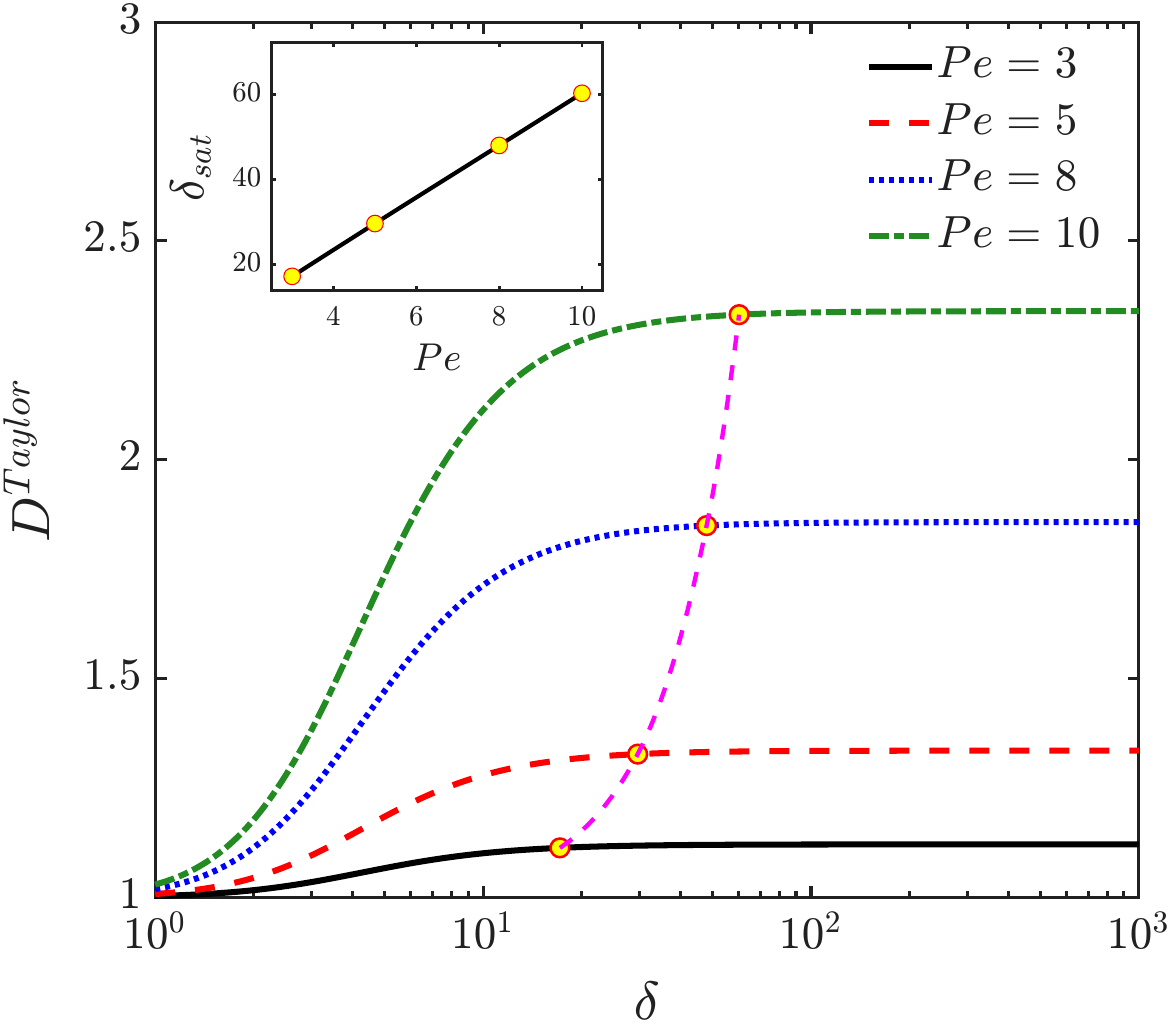}
        \vspace{18mm} 
        \caption{}
        \label{fig:5d}
    \end{subfigure}
    \vspace{0.4mm}
    \caption{The four figure panels showing  characteristics of the Taylor dispersion coefficient, $D^{\text{Taylor}}$, for a couple stress fluid flow. (a) Theoretical prediction (solid line) demonstrates excellent quantitative agreement of $D^{\text{Taylor}}$ against experimental data from \citep{yan2015chip} recorded in the presence of gravitational effects with no absorption ($\hat{\beta}_1 = \hat{\beta}_2 \to 0$) and magnetic effects ($Ha = 0$). (b) Monotonic growth of dispersion with Péclet number ($Pe$), where stronger couple stresses (decreasing $\delta$) flatten the core velocity and inhibit shear-driven dispersion (magnetic damping shown at $Ha = 5$ in the inset). (c) Symmetric parabolic variation with Grashof number ($Gr$), illustrating identical dispersion enhancements in gravity-assisted ($Gr > 0$) and gravity-opposed ($Gr < 0$) modes contribute identically to the enhanced shear dispersion. (d) Detailed saturation dynamics of $D^{\text{Taylor}}$ as a function of the couple stress parameter $\delta$ for representative values of $Pe$. As $\delta$ increases, micropolar effects diminish and $D^{\text{Taylor}}$ monotonically approaches an upper asymptotic plateau representing the Newtonian limit. The onset of saturation across different Péclet numbers follows a distinct critical locus (magenta line). The corresponding linear dependence of the threshold couple stress parameter, $\delta_{\text{sat}}$, on the Péclet number ($Pe$) is detailed in the inset. }
    \label{fig:tayloredisp}
\end{figure}

\DD{To validate the present analytical model, we compare the Taylor dispersion coefficient,} ($D^{\text{Taylor}}$), \DD{predicted for the limiting case of a non-reactive Newtonian fluid under buoyancy ($Gr = 0.2$), with the reported experimental results} \citep{yan2015chip}, \DD{as shown in} \textcolor{blue}{figure} \ref{fig:tayloredisp}\textcolor{black}{(a)}. It is important to mention that the authors of referred experimental study \citep{yan2015chip} investigated \DD{fluorescein dispersion} in pressure-driven laminar flow within rectangular PDMS microchannels under gravitational effects using time-resolved fluorescence measurements. In the present work, \DD{we employ the classical } Taylor–Aris \citep{aris_statement,taylor1953dispersion} \DD{over the same range of Péclet numbers considered experimentally} \citep{yan2015chip}. As evident from \textcolor{blue}{figure} \ref{fig:tayloredisp}\textcolor{black}{(a)}, that our analytical results has good agreement with the experimental data. The slight deviation from the experimental results is due to the fact of using a two-dimensional model for a three dimensional PDMS microchannel due to which the spanwise component effect is missing as perviously reported by \citep{Das_Dhar_Kairi_Mondal_2025}. Furthermore, it confirms the recovery of the classical Taylor-Aris scaling regime in absence of magnetic forcing ($Ha = 0$) and chemical reactions ($\hat{\beta}_1 = \hat{\beta}_2 \to 0$), wherein the shear-induced dispersion dominates longitudinal diffusion, yielding $D^{\text{Taylor}} \propto Pe^2$. \textcolor{blue}{Figure} \ref{fig:tayloredisp}\textcolor{black}{(b)} maps the monotonic growth of $D^{\text{Taylor}}$ with $Pe$ across varying couple stress parameters ($\delta$). As $\delta$ decreases from 1000 (near-Newtonian) to 2 (strongly non-Newtonian), $D^{\text{Taylor}}$ drops significantly. For instance, at $Pe = 10$, the dispersion coefficient is roughly halved. 
This is a direct consequence of the couple stresses increasing the apparent macroscopic viscosity, which flattens the parabolic velocity core and severely diminishes the transverse velocity gradients responsible for stretching the solute cloud. The inset in \textcolor{blue}{figure} \ref{fig:tayloredisp}\textcolor{black}{(b)} quantitatively demonstrates the profound damping effect at $Ha = 5$. The maximum dispersion at $Pe = 10$ and $\delta = 1000$ drops drastically from approximately $2.4$ to roughly $1.14$ (cf. inset figure). The Hartmann damping strongly flattens the velocity profile, eradicating the high-shear zones required for robust hydrodynamic dispersion. 

In \textcolor{blue}{figure} \ref{fig:tayloredisp}\textcolor{black}{(c)} isolates the effect of mixed convection on the dispersion coefficient, revealing a distinct parabolic symmetry centered at $Gr = 0$. \DD{The minimum dispersion occurs at $Gr=0$, corresponding to the purely pressure-driven Poiseuille flow.}  The identical dispersion enhancement for both $Gr > 0$ and $Gr < 0$ indicates that longitudinal dispersion is strictly sensitive to the magnitude of the transverse velocity gradients, not their sign. \DD{Both gravity-assisted and gravity-opposed buoyancy strengthen the transverse shear with equal intensity, thereby producing identical increases in $D^{\text{Taylor}}$.} Additionally, in \textcolor{blue}{figure} \ref{fig:tayloredisp}\textcolor{black}{(c)}, upon comparing the main plot ($Pe = 10$) with the inset ($Pe = 5$), it is evident that the relative enhancement of dispersion due to buoyancy is amplified at higher advection rates. \DD{Increasing the couple stress effect (smaller $\delta$) uniformly suppresses the dispersion coefficient over the entire range of $Gr$, while preserving the symmetric parabolic variation.} The couple stress parameter $\delta$ acts as a consistent scaling factor, uniformly dampening the parabolas without altering their fundamental symmetric topology. \textcolor{blue}{Figure} \ref{fig:tayloredisp}\textcolor{black}{(d)}, on the other hand, \DD{illustrates the} asymptotic behavior of the dispersion coefficient as the fluid transitions from a highly micro-structured state to a Newtonian continuum. For any given $Pe$, $D^{\text{Taylor}}$ grows with $\delta$ but eventually saturates at an upper horizontal plateau, representing the classical Newtonian limit where internal micro-rotational friction is indeed negligible. \DD{An important quantitative feature is the onset of saturation, characterized by the threshold couple stress parameter}($\delta_{\text{sat}}$). The explicit quadratic dependence of $D_{\text{sat}}$ on $\log_{10}\delta$ is evaluated by using a logarithmic fit of the data points yielding the following empirical relation:
\begin{equation}
D_{\text{sat}} = 4.5473 (\log_{10}\delta)^2 - 11.5350 \log_{10}\delta + 8.4291
\label{eq:satcurve}
\end{equation}
\DD{(\ref{eq:satcurve}) indicates that the maximum attainable dispersion increases nonlinearly as the influence of couple stresses weakens and the flow approaches the Newtonian limit.} The inset in \textcolor{blue}{figure} \ref{fig:tayloredisp}\textcolor{black}{(d)} extracts a strict linear relationship between the saturation threshold and the advective strength: $\delta_{\text{sat}} \propto Pe$. This indicates that as the flow velocity increases, the shear rates amplify, making the dispersion coefficient far more sensitive to even minor microstructural rheology effects.
\subsubsection{Total dispersion coefficient and concentration distribution}
\begin{figure}[htbp] 
    \centering
    \begin{subfigure}[b]{0.48\textwidth}
        \centering
        \includegraphics[width=\textwidth]{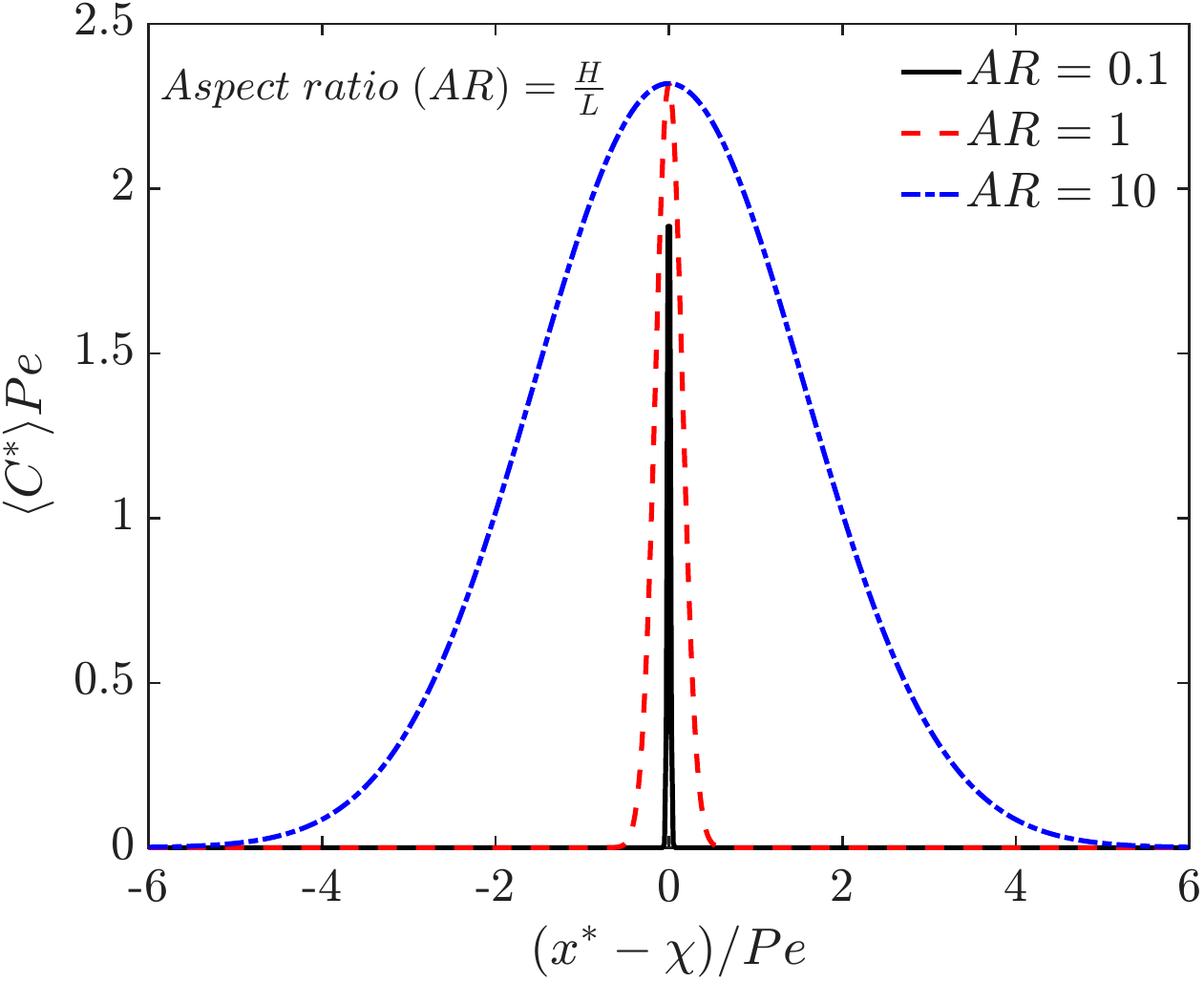}
        \vspace{18mm} 
        \caption{}
        \label{fig:6a}
    \end{subfigure}
    \hfill
    \begin{subfigure}[b]{0.48\textwidth}
        \centering
        \includegraphics[width=\textwidth]{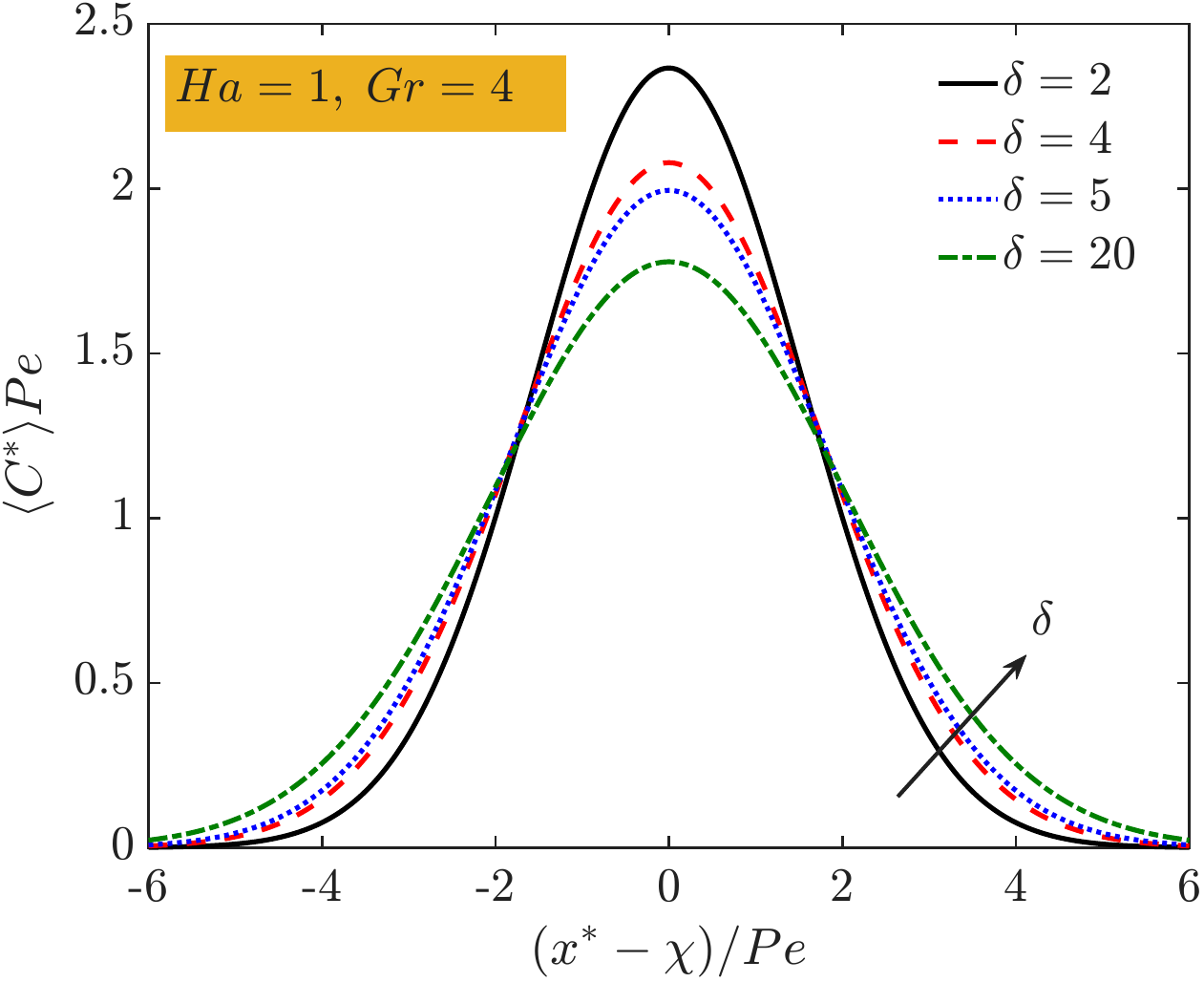}
        \vspace{18mm} 
        \caption{}
        \label{fig:6b}
    \end{subfigure}
  
    \begin{subfigure}[b]{0.48\textwidth}
        \centering        
        \includegraphics[width=\textwidth,height=6.65cm]{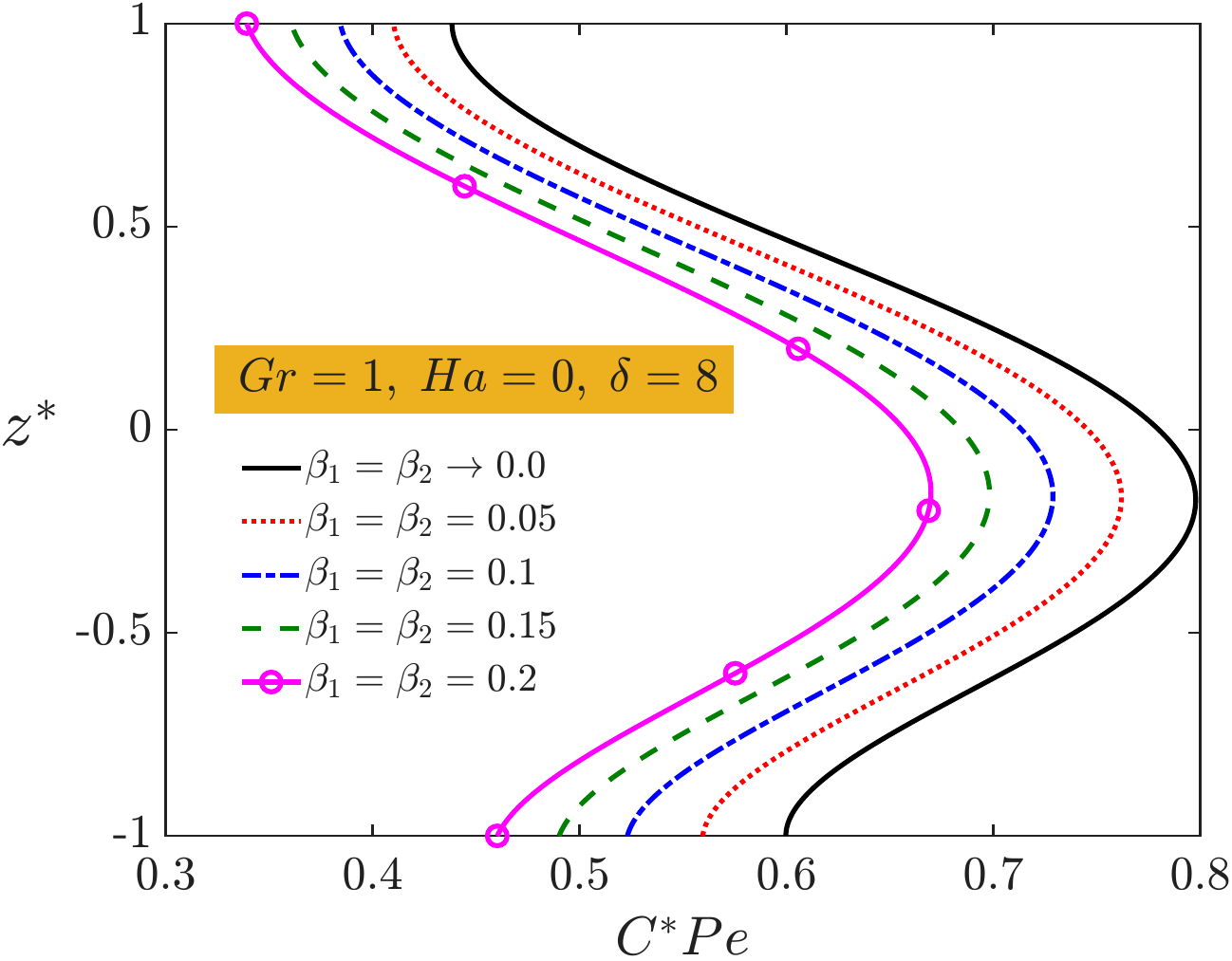}
        \vspace{18mm} 
        \caption{}
        \label{fig:6c}
    \end{subfigure}
    \hspace{0.02\textwidth}
    \begin{subfigure}[b]{0.48\textwidth}
        \centering
        \includegraphics[width=\textwidth,height=6.65cm]{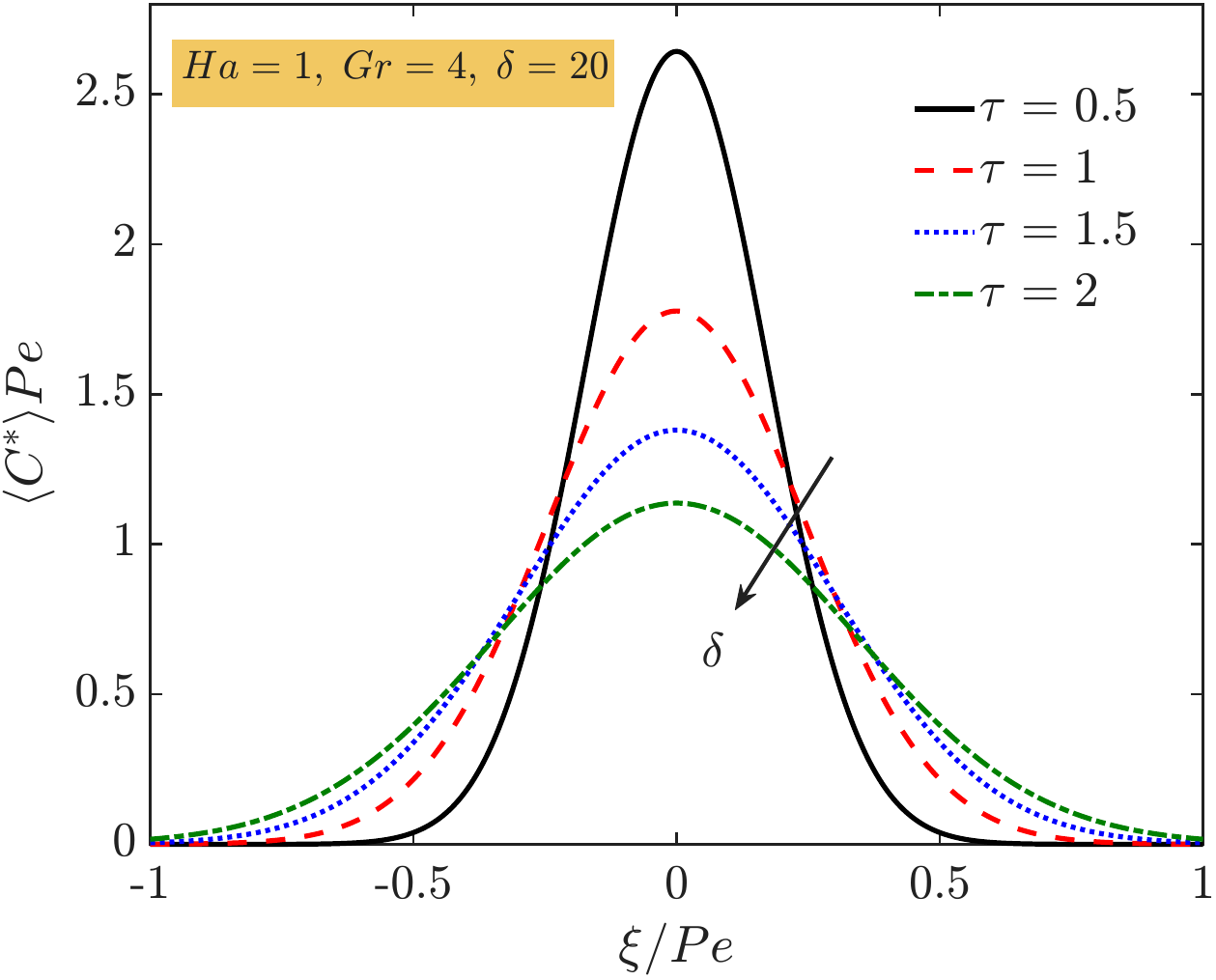}
        \vspace{18mm} 
        \caption{}
        \label{fig:6d}
    \end{subfigure}
    \caption{(a) Streamwise profile broadening of mean concentration, $\langle C^* \rangle Pe$, with increasing aspect ratio ($AR$) due to enhanced longitudinal dispersion. (b) Influence of couple stress parameter ($\delta$), showing core peak enhancement and tail attenuation with decreasing $\delta$. (c) Transverse solute concentration profiles, $C^* Pe(z^*)$, illustrating uniform mass depletion with increasing wall absorption ($\beta_1 = \beta_2$) while preserving buoyancy-induced asymmetry for $Gr = 1$. (d) Transient evolution across dimensionless time scales $\tau$, depicting peak attenuation alongside symmetric spreading.}
    \label{fig:concvary}
\end{figure}

Traditional frameworks for hydrodynamic transport typically rely on a taylor dispersion coefficient, $D^{\text{Taylor}}$, which isolates solute spreading strictly to the coupled mechanisms of fluid advection and molecular diffusion. However, when heterogeneous surface reactions occur at the boundaries, they distort the concentration profiles over the complete lateral extent, thereby fundamentally reshaping the macro-transport behavior. To quantify this interfacial influence, a total effective dispersion coefficient, $D^{\text{eff}}_T$, is formulated by employing higher-order asymptotic expansions with respect to the perturbation parameter $\varepsilon$. This modified coefficient specifically accounts for the convective-diffusive variations of solute induced by boundary absorption. It is worth noting that while both homogeneous bulk reactions and heterogeneous boundary kinetics introduce non-conservative effects, manifested as exponential decay in the total solute mass, their impacts on transport mechanics are distinct. Only reactions driven by the boundaries possess the spatial dependency required to alter transverse gradients and modulate the overall dispersion coefficient \citep{aris_statement}; homogeneous bulk processes uniformly deplete the solute without modifying the underlying dispersive dynamics. In order to deal with this non-conservative system, we introduce a coordinate transformation by inserting the variables $\tau = t^*$, $\xi = (x^*/\varepsilon) - Pe[\langle u^*\rangle + (1/2)(\hat{\beta}_1 + \hat{\beta}_2)\langle u^* A_2\rangle + (1/2)\hat{\beta}_1 A_1(1) + (1/2)\hat{\beta}_2 A_1(-1)]t^*$ into the governing equation. Employing this coordinate transformation and making use of the boundary conditions, the following dimensionless concentration equation (\ref{eq:conc_nd}) and its boundary conditions (\ref{eq:2.25b},\ref{eq:2.25c}) are obtained:
\begin{equation}
\label{eq:dimensionless_concentration}
\frac{\partial C^*}{\partial t^*} + Pe \, u_\chi \frac{\partial C^*}{\partial \xi} = \frac{\partial^2 C^*}{\partial \xi^2} + \frac{\partial^2 C^*}{\partial {z^*}^2} - \hat{\mathcal{K}_f} C^*
\end{equation}
and
\begin{subequations}
\label{eq:boundary_conditions}
\begin{align}
\label{eq:bc_top}
\frac{\partial C^*}{\partial z^*} + \hat{\beta}_1 C^* &= 0, \quad z^* = 1 \\
\label{eq:bc_bottom}
\frac{\partial C^*}{\partial z^*} - \hat{\beta}_2 C^* &= 0, \quad z^* = -1
\end{align}
\end{subequations}
Here, the velocity term $u_\chi$ is expressed as $u_\chi = u^* - \chi$, with the tracking parameter $\chi$ explicitly formulated in (\ref{eq:chi_definition}):
\begin{equation}
\label{eq:chi_definition}
\chi = \langle u^* \rangle + \frac{1}{2}(\hat{\beta}_1 + \hat{\beta}_2)\left[\langle u^* A_2 \rangle + \frac{u^* A_1(1) + \hat{\beta}_1 A_1(-1)}{\hat{\beta}_1 + \hat{\beta}_2}\right]
\end{equation}
On averaging the convection–diffusion equation with respect to the transverse variable $z^*$, and then applying the boundary conditions (\ref{eq:bc_top}) and (\ref{eq:bc_bottom}), we obtain the following equation:
\begin{equation}
\label{eq:averaged_transport}
\frac{\partial \langle C^* \rangle}{\partial t^*} + Pe \left\langle u^* \frac{\partial C^*}{\partial \xi} \right\rangle - Pe \, \chi \left\langle \frac{\partial C^*}{\partial \xi} \right\rangle = \frac{\partial^2 \langle C^* \rangle}{\partial \xi^2} - \frac{1}{2} \left[\hat{\beta}_1 C^*(1) + \hat{\beta}_2 C^*(-1)\right] - \hat{\mathcal{K}}_f \langle C^* \rangle = 0
\end{equation}
The asymptotic form of the solute concentration, up to the third order, can be expressed as
\begin{equation}
\label{eq:asymptotic_concentration}
C^* = C_0^* + \varepsilon C_1^* + \varepsilon^2 C_2^* + \varepsilon^3 C_3^*
\end{equation}
or, equivalently,
\begin{align}
\label{eq:expanded_concentration}
C^* = {} & \left[ 1 + \frac{\hat{\beta}_1 + \hat{\beta}_2}{2} A_2(z^*) + \frac{(\hat{\beta}_1 + \hat{\beta}_2)^2}{4} A_5(z^*) + \frac{(\hat{\beta}_1 + \hat{\beta}_2)^3}{8} A_9(z^*) \right] C_0^* \nonumber \\
& + Pe \left[ A_1(z^*) + \frac{\hat{\beta}_1 + \hat{\beta}_2}{2} A_4(z^*) + \frac{(\hat{\beta}_1 + \hat{\beta}_2)^2}{4} A_8(z^*) \right] \frac{\partial C_0^*}{\partial \xi} \nonumber \\
& + Pe^2 \left[ A_3(z^*) + \frac{\hat{\beta}_1 + \hat{\beta}_2}{2} A_7(z^*) \right] \frac{\partial^2 C_0^*}{\partial \xi^2} + Pe^3 A_6(z^*) \frac{\partial^3 C_0^*}{\partial \xi^3}
\end{align}
where $A_6, A_7, A_8, A_9$ are transverse functions, and $C_0^*$ denotes the mean concentration (since $\langle A_i \rangle = 0$ for $i \ge 1$).
Finally, we plug in the expression of expanded concentration $C^*$, given in (\ref{eq:expanded_concentration}), into the dimensionless transport equation, Eq. (\ref{eq:dimensionless_concentration}), and collect the coefficient adhering to the second-order derivative $\partial^2 C_0^* / (\partial \xi^2)$. Following this task, we identify the total dispersion coefficient  in terms of the wall absorption parameters $\hat{\beta_1}$ and $\hat{\beta_2}$ as:
\begin{equation}
\label{eq:total_dispersion}
\begin{split}
D_T^{\text{eff}} = 1 - Pe^2 \Biggl\{ & \langle u^* A_1 \rangle + \frac{\hat{\beta}_1 + \hat{\beta}_2}{2} \left[ \langle u^* A_4 \rangle + \frac{\hat{\beta}_1 A_3(1) + \hat{\beta}_2 A_3(-1)}{\hat{\beta}_1 + \hat{\beta}_2} \right] \\
& + \frac{(\hat{\beta}_1 + \hat{\beta}_2)^2}{4} \left[ \langle u^* A_8 \rangle + \frac{\hat{\beta}_1 A_7(1) + \hat{\beta}_2 A_7(-1)}{\hat{\beta}_1 + \hat{\beta}_2} \right] \Biggr\}
\end{split}
\end{equation}
Following the formulation established by the researchers \citep{radhaprf,Das_Dhar_Kairi_Mondal_2025}, the total dispersion coefficient in (\ref{eq:total_dispersion}) is defined as: $D_T^{eff} = D^{\text{Taylor}} + D^{\text{abs}}$, where $D^{\text{abs}}$ is the component of the boundary absorption dependent dispersion coefficient. It is worth adding here that when the absorption parameters $\hat{\beta}_1$ and $\hat{\beta}_2$ tend to zero, the expression for the total dispersion coefficient reduces to the classical Taylor dispersion coefficient. \DD{Accordingly, the mean concentration distribution reads as:}
\begin{equation}
\label{eq:mean_concentration_solution}
C_0^* = \frac{1}{\sqrt{4\pi \tau D_T^{\text{eff}} }} \times \exp \left( -\frac{\xi^2}{4\tau D_T^{\text{eff}} } - \left\{ \frac{1}{2}(\hat{\beta}_1 + \hat{\beta}_2) + \frac{1}{4}(\hat{\beta}_1 + \hat{\beta}_2) \left\{ \hat{\beta}_1 A_2(1) + \hat{\beta}_2 A_2(-1) \right\} + \hat{\mathcal{K}}_f \right\} \tau \right)
\end{equation}
In \textcolor{blue}{figure} \ref{fig:concvary}, we analyse the concentration distribution along both the longitudinal and
transverse directions. Before analyzing the plots, it is crucial to establish the physical implications of the coordinate transformation $\xi$ and effective velocity $\chi$. The expression  \DD{for the transformed coordinate $\xi$ and the corresponding effective cloud velocity $\chi$ (see Eq. \ref{eq:chi_definition})} reveals a profound coupling between heterogeneous reactions and advective transport. This indicates that the center of mass of the solute cloud does not merely translate at the cross-sectional mean fluid velocity, $\langle u^* \rangle$. Instead, the \DD{heterogeneous wall absorption} ($\hat{\beta}_1, \hat{\beta}_2 > 0$) induces an apparent advective acceleration. Physically, because the solute is preferentially depleted in the slow-moving viscous sublayers near the walls, the surviving mass distribution is weighted toward the faster-moving central core. \DD{Consequently, the transformed coordinate $\xi$ tracks the effective motion of the solute cloud rather than the bulk fluid, enabling the intrinsic effects of longitudinal dispersion and wall reactions to be examined independently of the mean advective transport.}
Within this framework, \textcolor{blue}{figure} \ref{fig:concvary}\textcolor{black}{(a)} illustrates the sensitivity of the mean concentration, $\langle C^* \rangle Pe$, to the channel's aspect ratio in the moving frame of reference. It is evident from \textcolor{blue}{figure} \ref{fig:concvary}\textcolor{black}{(a)} that \DD{the peak concentration remains nearly unchanged, whereas the longitudinal width of the solute cloud increases substantially.} This observation can be attributed as follows:\DD{increasing the aspect ratio enlarges the effective streamwise diffusion length relative to the transverse channel height}. A larger AR amplifies the effective Taylor dispersion, causing the solute to \DD{evolve from a narrow plug-like distribution at $AR=0.1$ to a much broader Gaussian profile at $AR=10$.} \DD{These }results are consistent with the inferences reported in the literature \citep{ling2021macroscale} as well. It is worth mentioning here that the velocity field in the presesnt study is independent of the perturbation expansion that allows our theoretical predictions to remain accurate over a broader range of aspect ratios. \textcolor{blue}{Figure} \ref{fig:concvary}\textcolor{black}{(b)} illustrates the variation of \DD{the longitudinal} concentration distribution with varying couple stress parameters obtained for $Gr = 4$ and $Ha = 1$. \DD{As $\delta$ increases to $20$, the influence of couple stresses on underlying transport diminishes and the flow approaches the Newtonian limit, producing stronger velocity gradients and enhanced Taylor dispersion. Consequently, the concentration peak decreases to approximately $1.75$ (a reduction of about $27\%$), while the leading and trailing tails broaden owing to enhanced longitudinal spreading.} \textcolor{blue}{Figure} \ref{fig:concvary}\textcolor{black}{(c)}, depicts \DD{the effect of boundary absorption parameter on} the mean concentration profile along the transverse direction, obtained for fixed parameter values $Gr = 1$, $Ha = 0$, and $\delta = 8$. It is evident that the concentration peak is distinctly skewed towards the lower half of the channel because \DD{buoyancy shifts the region of maximum axial advection } away from the centerline. As the identical surface reaction parameters transition from purely reflective ($\beta_1 = \beta_2 \to 0$) to highly absorptive ($\beta_1 = \beta_2 = 0.2$), the peak concentration drops sequentially from $0.80$ to $ 0.65$ approximately. This occurs \DD{because stronger wall absorption steepens the concentration gradients near both channel walls, thereby increasing the diffusive mass flux from the fluid to the boundaries.} This observation is in strong agreement with previously reported results of solute dispersion in MHD transport \citep{dhar2021dispersion,doi:10.1098/rspa.2024.0091}. \textcolor{blue}{Figure} \ref{fig:concvary}\textcolor{black}{(d)}, illustrates the temporal \DD{evolution of the longitudinal concentration profile in the transformed coordinate system.} \DD{With increasing dimensionless time, the concentration peak progressively attenuates while the profile broadens symmetrically, reflecting the combined effects of longitudinal dispersion and molecular diffusion on the underlying transport. The accompanying growth of the leading and trailing tails indicates the continuous redistribution of solute over an increasingly wider spatial region.}

After \DD{obtaining} the concentration field $C^*(\xi, z^*, \tau)$, \DD{we next quantify}  the degree of cross-sectional non-uniformity  within the channel. To this end, we introduce the transverse variation \DD{rate}, denoted by $R_{\mathrm{tr}}(\xi)$, which \DD{measures  } the localized peak-to-peak \DD{concentration difference normalized by the centerline concentration}  \citep{Wu_Chen_2014}. This dimensionless \DD{quantities} is defined as,
\begin{equation}
    R_{\mathrm{tr}}(\xi) = \frac{\displaystyle\max_{-1 \le z^* \le 1} C^*(\xi, z^*, \tau) - \displaystyle\min_{-1 \le z^* \le 1} C^*(\xi, z^*, \tau)}{C^*(0, 0, \tau)}
    \label{eq:transverse_variation}
\end{equation}
where $z^* \in [-1, 1]$ is the dimensionless transverse coordinate. \DD{The index} $R_{\mathrm{tr}}(\xi)$ \DD{characterizes the cross-sectional non-uniformity of the solute cloud by quantifying the maximum concentration variation across the channel. This index therefore provides a direct measure of how shear-induced advection and transverse diffusion redistribute the solute over the channel cross section.} In addition to characterizing \DD{concentration non-uniformity,} it is \DD{also necessary}  to quantify the \DD{solute mass removed through reactive channel walls} . The mass removal rate within a specified streamwise interval $[\xi_a, \xi_b]$ is \DD{defined as:} 
\begin{equation}
    R_{\mathrm{rate}} = \frac{1}{2} \int_{\xi_a}^{\xi_b} \left[ \hat{\beta}_1 C^*(\xi, 1, \tau) + \hat{\beta}_2 C^*(\xi, -1, \tau) \right] \mathrm{d}\xi
\end{equation}
Finally, to assess the overall effectiveness of the reactive boundaries, we define the total removal efficiency, $\eta_R$, as the fraction of the initially injected solute mass removed from the system:
\begin{equation}
\eta^{R} = 1 - \frac{1}{2}\int_{-\infty}^{\infty}\int_{-1}^{1} C^*(\xi,z^*) \, \mathrm{d}z^* \, \mathrm{d}\xi
\end{equation}
\begin{figure}[htbp]
    \centering
        \begin{subfigure}[b]{0.48\textwidth}
        \centering        
        \includegraphics[width=\textwidth,height=6.65cm]{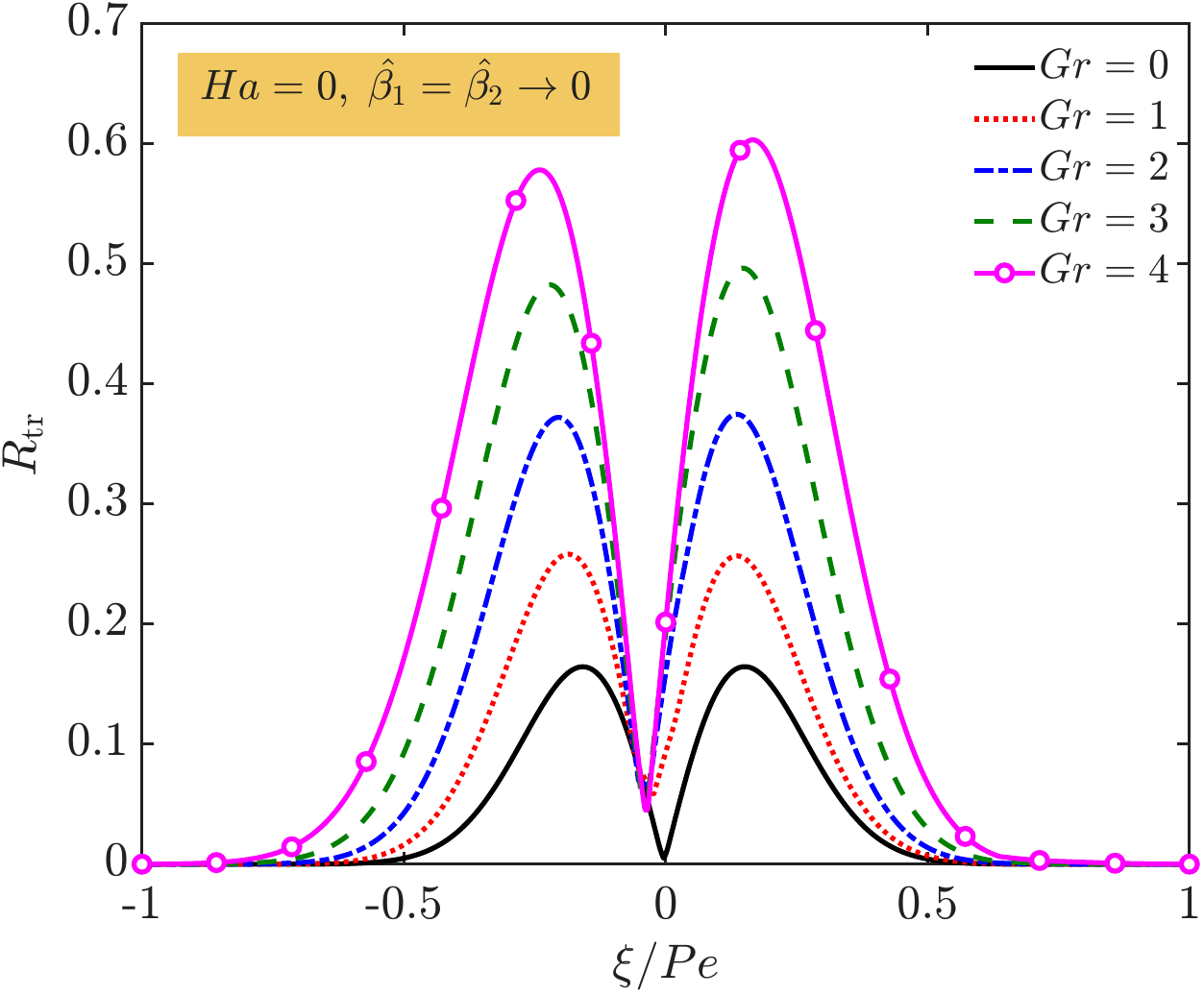}
        \vspace{18mm} 
        \caption{}
        \label{fig:6e}
    \end{subfigure}
    \hspace{0.02\textwidth}
    \begin{subfigure}[b]{0.48\textwidth}
        \centering
        \includegraphics[width=\textwidth,height=6.65cm]{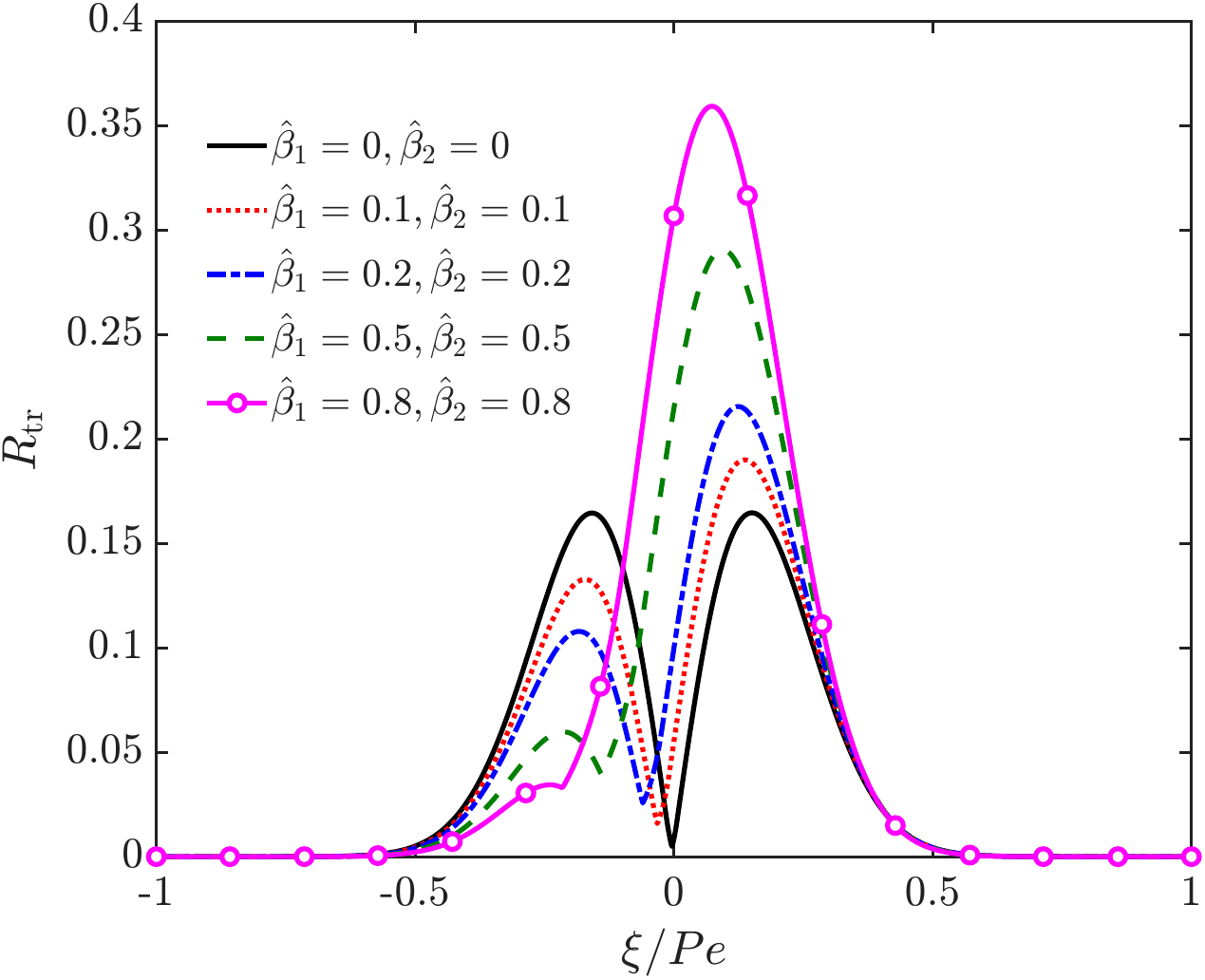}
        \vspace{18mm} 
        \caption{}
        \label{fig:6f}
    \end{subfigure}
    
    \vspace{0.4mm}
    \begin{subfigure}[b]{0.48\textwidth}
        \centering        
        \includegraphics[width=\textwidth,height=6.65cm]{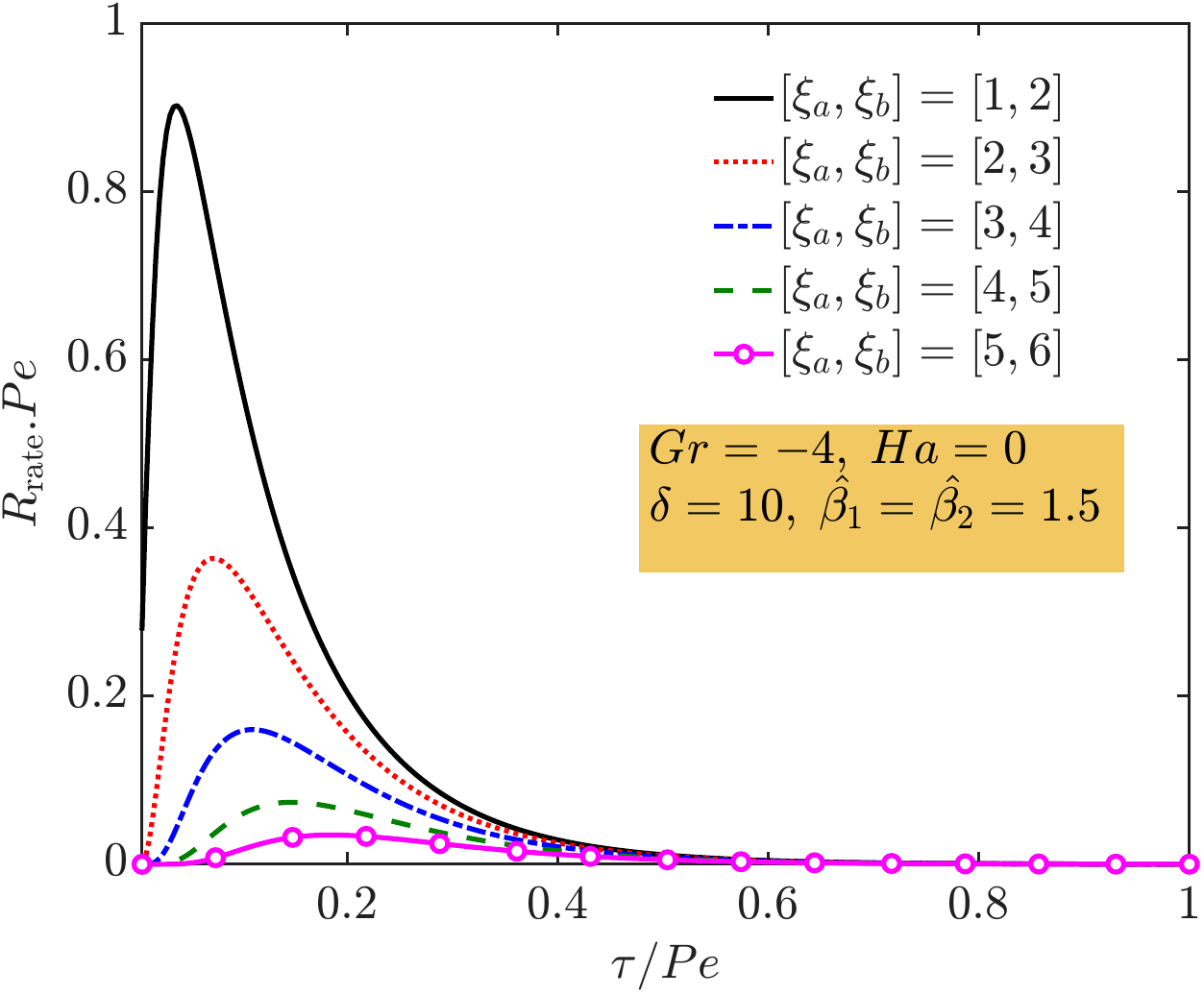}
        \vspace{18mm} 
        \caption{}
        \label{fig:6g}
    \end{subfigure}
    \hspace{0.02\textwidth}
    \begin{subfigure}[b]{0.48\textwidth}
        \centering
        \includegraphics[width=\textwidth,height=6.65cm]{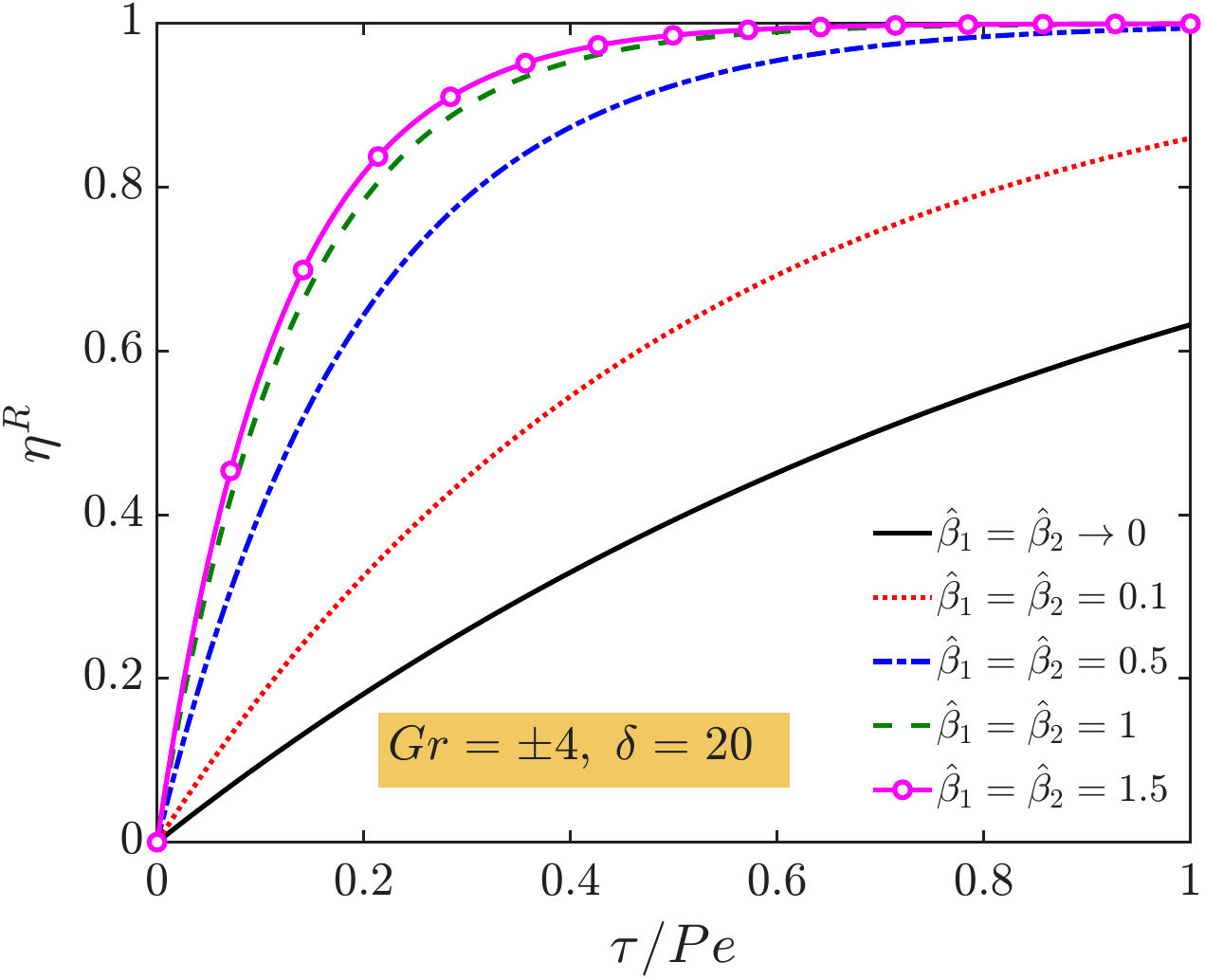}
        \vspace{18mm} 
        \caption{}
        \label{fig:6h}
    \end{subfigure}
    
    \vspace{0.4mm}
    \caption{The four figures panel depicting solute transport characteristics. (a) Longitudinal profiles of transverse removal rate ($R_{tr}$) at $\delta = 20$, showing symmetric bimodal enhancement with increasing Grashof number ($Gr$). (b) Variations in $R_{tr}$ for different uniform wall adsorption parameters ($\hat{\beta}_1 = \hat{\beta}_2$), depicting a transition in peak height dominance driven by boundary sinks. (c) Temporal evolution of inter-sectional mass removal rate ($R_{\text{rate}} Pe$) between consecutive axial positions $[\xi_a, \xi_b]$, displaying downstream peak decay and delay. (d) Temporal dynamics of total mass removal efficiency ($\eta^R$), illustrating accelerated saturation toward complete removal ($\eta^R \to 1$) with higher absorption rates.}
    \label{fig6}
\end{figure}
\DD{In \textcolor{blue}{figure} \ref{fig6}, we further show the effect of buoyancy and wall reactions on solute transport by examining the transverse variation index, $R_{\mathrm{tr}}$, the local mass removal rate, $R_{\mathrm{rate}}$, and the overall removal efficiency, $\eta^{R}$. These quantities provide complementary measures of cross-sectional concentration non-uniformity, localized wall uptake, and the cumulative efficiency of reactive solute removal. 
\textcolor{blue}{Figure} \ref{fig6}\textcolor{black}{(a)} depicts the longitudinal evolution of the transverse variation index, $R_{\mathrm{tr}}$, for different values of Grashof number in the absence of wall absorption $(\hat{\beta}_1=\hat{\beta}_2\rightarrow0)$. The index exhibits a symmetric bimodal distribution about the cloud centroid $(\xi=0)$, where $R_{\mathrm{tr}}$ attains its minimum value. Moving away from the centroid, the combined action of shear-induced advection and transverse diffusion generates increasing cross-sectional concentration differences, which in turn, produces two symmetric maxima. When $Gr=0$, the the transveerse variation rate is almost uniform. As the Grashof number increases, the peak magnitude of $R_{\mathrm{tr}}$ increases monotonically, whereas the peak locations remain nearly unchanged. This behaviour indicates that buoyancy intensifies the cross-sectional concentration non-uniformity by strengthening the transverse velocity gradients without significantly modifying the longitudinal extent of the concentration cloud.
\textcolor{blue}{Figure} \ref{fig6}\textcolor{black}{(b)} illustrates the influence of identical wall absorption parameters on the transverse variation index. For weak wall absorption, the two peaks remain nearly comparable. However, increasing the reaction parameters $(\hat{\beta}_1=\hat{\beta}_2)$ progressively suppresses the upstream peak while substantially enhancing the downstream peak, thereby breaking the fore--aft symmetry of $R_{\mathrm{tr}}$. Notably, continuous solute removal at the reactive walls reduces the solute concentration near wall region and leads to steeper transverse concentration gradients. Consequently, heterogeneous wall reactions modify not only the total solute mass but also the internal spatial structure of the concentration field.
\textcolor{blue}{Figure} \ref{fig6}\textcolor{black}{(c)} depicts the temporal evolution of the local mass removal rate, $R_{\mathrm{rate}}$, over successive streamwise intervals. For every observation interval, the removal rate decreases rapidly to a distinct maximum before gradually decaying towards zero. The initial peak corresponds to the arrival of the concentrated solute cloud within the respective axial segment, whereas the subsequent decay results from the combined effects of longitudinal dispersion and continuous wall absorption. Furthermore, the peak removal rate decreases and its occurrence is progressively delayed as the observation interval moves farther downstream, demonstrating that the dispersed solute cloud reaches successive channel locations later and with a reduced concentration.
Finally, \textcolor{blue}{figure} \ref{fig6}\textcolor{black}{(d)} shows the temporal evolution of the overall removal efficiency, $\eta^{R}$, for different wall reaction strengths. The removal efficiency increases monotonically with time and asymptotically approaches unity, indicating the progressive depletion of the injected solute from the fluidic channel. Increasing the wall absorption parameters markedly accelerates this process, allowing almost complete solute removal within a comparatively short transport time. In contrast, weakly reactive walls exhibit a much slower increase in $\eta^{R}$, underscoring that the boundary reaction kinetics primarily determine the characteristic time required for solute clearance. Overall, \textcolor{blue}{figure} \ref{fig6} demonstrates that buoyancy primarily governs the spatial non-uniformity of the solute cloud, whereas the wall reaction parameters control both the rate and the overall efficiency of reactive solute removal.}
\begin{figure}[htbp]
    \centering
    
    \includegraphics[width=1.05\textwidth]{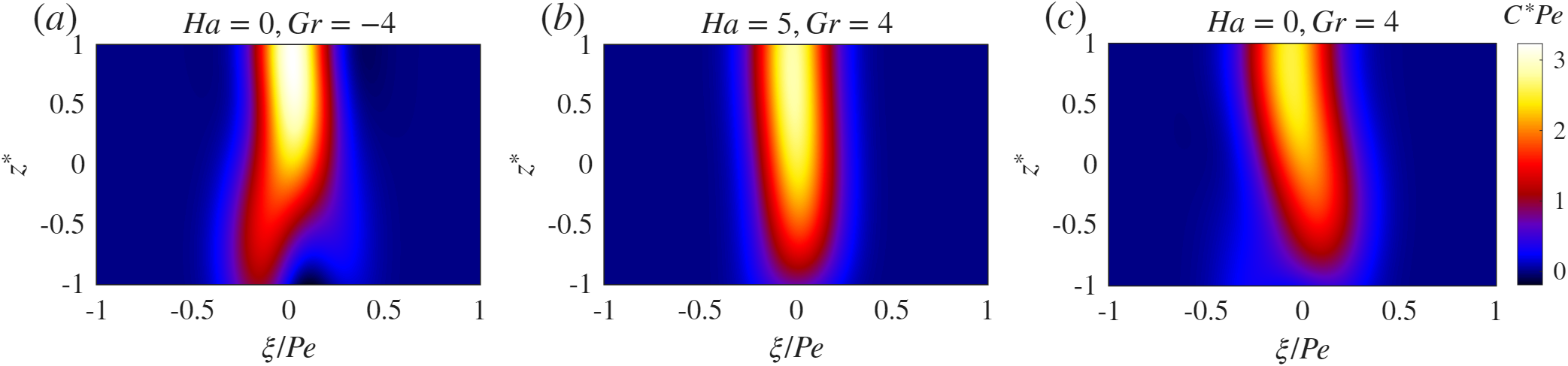}
    
    \vspace{0.5mm}
    
    \includegraphics[width=1.05\textwidth]{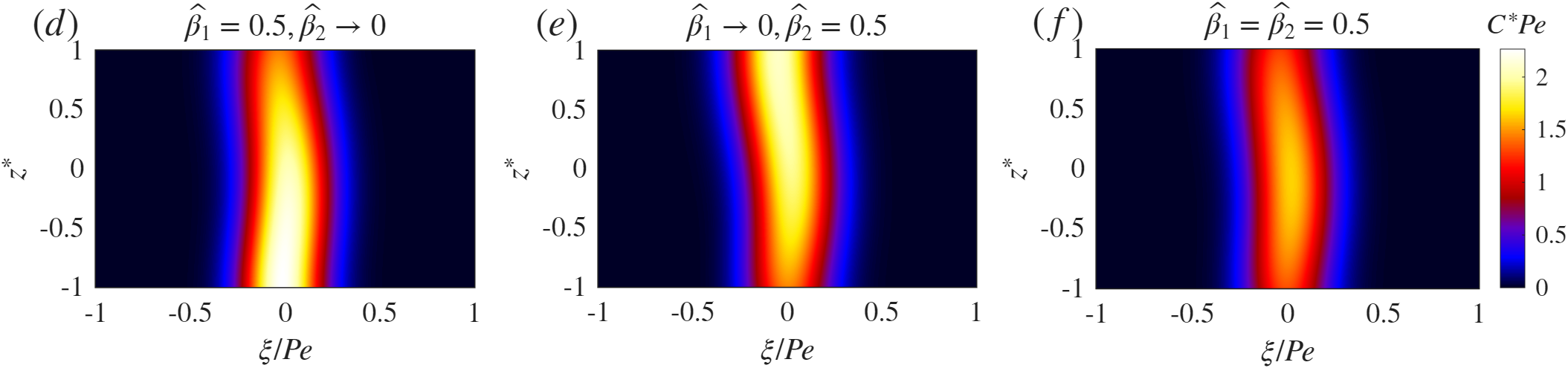}
    
    \vspace{0.5mm}
    
    \includegraphics[width=1.05\textwidth]{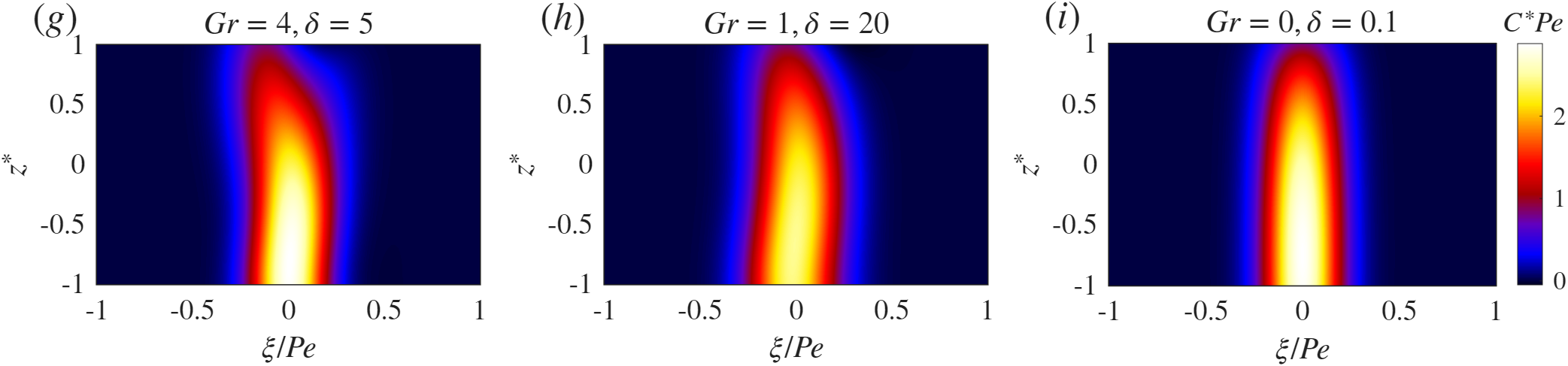}
    \caption{Concentration contour of solute in the $(\xi/Pe, z^*)$ plane, demonstrating the coupled effects of magnetic forcing, buoyancy, couple stress parameters, and differential wall adsorption. \textbf{(a--c)} Under asymmetric adsorption ($\hat{\beta}_1 = 0.1, \hat{\beta}_2 = 1.5, Ha = 0$), opposing buoyancy ($Gr = -4$) causes lower-wall distortion and backflow, while assisting buoyancy ($Gr = 4$) tilts the plume upward; introducing a magnetic field ($Ha = 5$) stabilizes this shear and restores core symmetry. \textbf{(d--f)} Varying the wall adsorption parameters ($\hat{\beta}_1, \hat{\beta}_2$) demonstrates how boundary mass locally and forces peak concentration toward the non-adsorbing wall, whereas uniform adsorption ($\hat{\beta}_1 = \hat{\beta}_2 = 0.5$) eliminates this bias to yield a perfectly symmetric, narrowed core. \textbf{(g--i)} With inverted adsorption ($\hat{\beta}_1 = 1.5, \hat{\beta}_2 = 0.1, Ha = 0$), high buoyancy and couple stresses ($\delta = 5, Gr = 4$) advectively skew the plume upward (mitigated at $\delta = 20, Gr = 1$), while nuetrally buoyant, micropolar fluid ($\delta = 0.1, Gr = 0$) maintains a blunt, symmetric core modified solely by the upper boundary sink.}
    \label{fig7p}
\end{figure}
\DD{\textcolor{blue}{Figure} \ref{fig7p}\textcolor{black}{(a--c)} illustrates the combined influence of buoyancy and magnetic forcing on the two-dimensional solute concentration field. In the absence of magnetic damping ($Ha=0$), buoyancy induces a pronounced inclination of the concentration plume. Opposing buoyancy ($Gr=-4$) produces a right-leaning plume, whereas assisting buoyancy ($Gr=4$) reverses the inclination, yielding a left-leaning plume. Increasing the Hartmann number to $Ha=5$ suppresses the buoyancy-induced transverse shear via the Lorentz force, thereby restoring an almost vertically aligned, more symmetric concentration plume.
Fig. \ref{fig7p}\textcolor{black}{(d--f)} examines the effect of wall adsorption on the concentration distribution. Selective adsorption at either wall $(\hat{\beta}_1\neq\hat{\beta}_2)$ continuously depletes solute near the reactive boundary, causing the concentration core to migrate towards the less reactive wall. In contrast, identical wall adsorption $(\hat{\beta}_1=\hat{\beta}_2)$ preserves the transverse symmetry of the concentration field while uniformly reducing the peak concentration due to balanced mass removal at both boundaries.
Fig. \ref{fig7p}\textcolor{black}{(g--i)} demonstrates the influence of couple stresses on concentration field for different buoyancy conditions. For moderate couple stresses $(\delta=5)$ and strong assisting buoyancy $(Gr=4)$, the plume remains noticeably skewed towards the upper wall, attributed primarily to the enhanced buoyancy-driven advection. Increasing the couple stress parameter to $\delta=20$, while reducing the assisting buoyancy to $Gr=1$, yields a smoother concentration distribution with reduced transverse distortion. Under neutrally buoyant conditions $(Gr=0)$, the concentration field becomes nearly symmetric, with negligible skewness towards the upper wall persists for $\delta=0.1$. This observation indicates that the plume skewness is governed predominantly by buoyancy effects rather than couple stresses. Therefore, in the absence of buoyancy, the plume evolution is governed primarily by pressure-driven advection and molecular diffusion.}

\subsection{Numerical solution of stochastic governing equations}

To validate the analytical solutions \DD{obtained from the multiscale asymptotic analysis}, we \DD{perform} Brownian dynamics simulations \DD{of} the governing advection-diffusion equation \DD{subject to the corresponding} initial and boundary conditions.  Brownian dynamics provides an efficient \DD{Lagrangian framework for simulating stochastic solute transport, and has been widely employed in seminal studies as well \citep{Jiang2020,Das_Dhar_Kairi_Mondal_2025}}. \DD{A key aspect of the numerical formulation is the implementation of the reactive absorbing boundary conditions at the channel walls. Accordingly,} the stochastic differential equations (SDEs) \DD{governing the particle trajectories, which are equivalent to the corresponding} Fokker-Planck equation, are given as follows:
\begin{equation}
d\xi(t^*) = Pe(u^* - \chi) dt^* + \sigma_{\xi} dW_{\xi}(t^*)
\end{equation}
\begin{equation}
dz^*(t^*) = \sigma_{z^*} dW_{z^*}(t)
\end{equation}
where $\xi(t^*)$ and $z^*(t^*)$ represent the  longitudinal and transverse stochastic positions of a particle, respectively. \DD{Consistent with the notation introduced in the previous section, the dimensionless time is denoted by} $t^* = \tau$. \DD{The stochastic processes}, $W_{\xi}$ and $W_{z^*}$ correspond to independent standard Wiener processes. The associated coefficients are defined by:
\begin{equation}
\sigma_\xi = \sqrt{2D_\xi}, \quad \sigma_{z^*} = \sqrt{2D_{z^*}}
\end{equation}
with molecular diffusivities \DD{taken as} $ D_\xi = 1$ and $D_{z^*} = 1$. The flow velocity profile, denoted as $u(z^*)$, incorporates several effects that stem from couple-stress, induced magnetic field conditions, and buoyancy induced effects.
\DD{The stochastic differential equations are integrated using the forward Euler--Maruyama scheme with a} time step of $\Delta t = 10^{-4}$. The boundary conditions at the channel walls are implemented \DD{ using a probabilistic reaction algorithm:} particles that exit the upper ($z^* > 1$) or lower ($z^* < -1$) boundaries during
\DD{each} time step are either absorbed with  probabilities $P_1 = \beta_1 \sqrt{\Delta t / D_{z^*}}$ and $P_2 = \beta_2 \sqrt{\Delta t / D_{z^*}}$, respectively, or reflected back into the domain. The spatial domain is \DD{discretized} into a $201 \times 201$ \DD{computational grid}. The solute concentration within each block is \DD{computed as}  $C_{Brownian} = n_{ij} / (N \Delta \xi \Delta z^*)$, where $n_{ij}$ represents the particle count in the $i,j$-th block, $N$ denotes the total particle population, and $\Delta \xi$, $\Delta z^*$ \DD{denote the grid spacings. The cross-sectionally averaged concentration is then evaluated by integrating over the transverse direction} using Simpson's one-third rule.
\begin{figure}[htbp]
    \centering
    
    \includegraphics[width=1.02\textwidth]{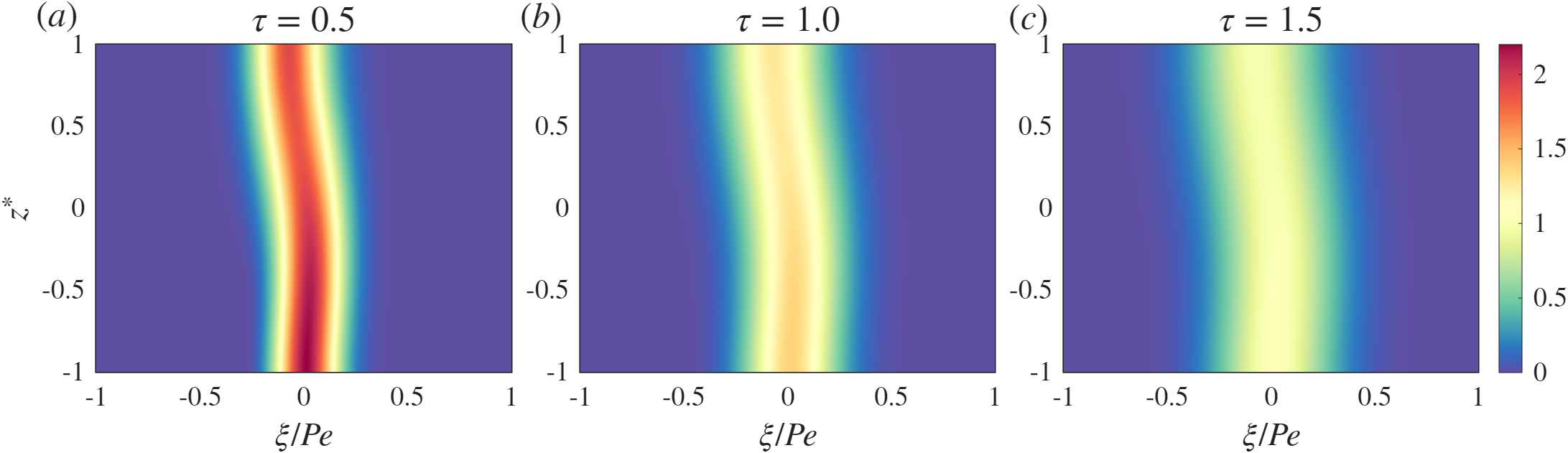}
    \vspace{0.5mm}
    
    \includegraphics[width=1.02\textwidth]{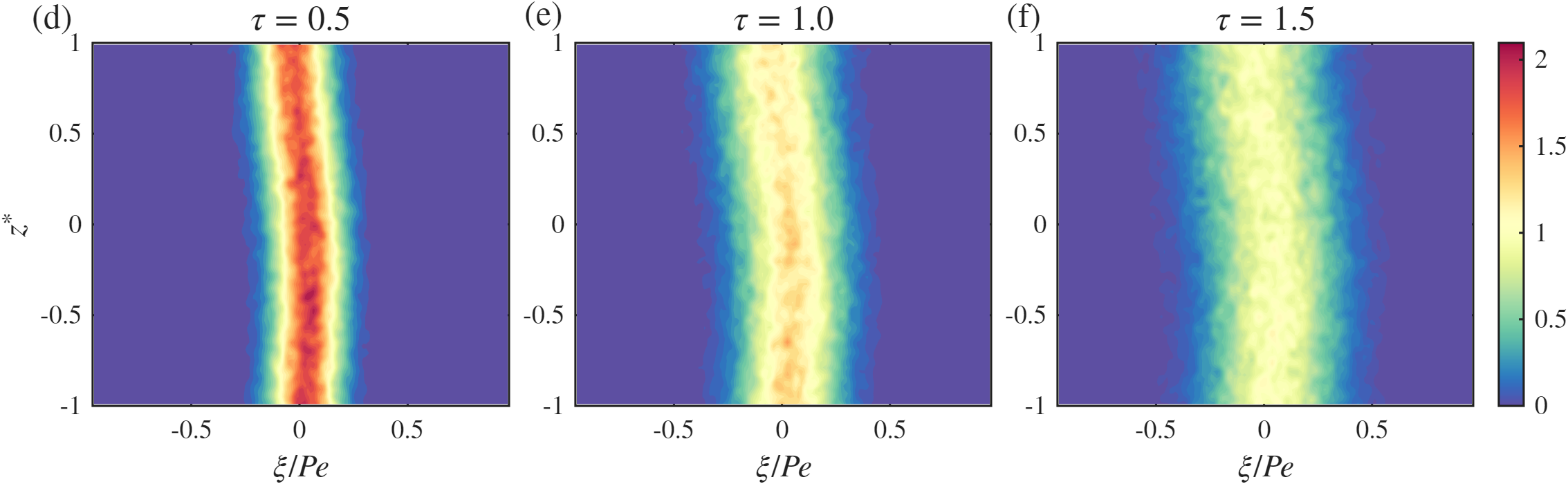}
    \caption{Benchmark comparisons between Brownian dynamics simulations and the analytical solution of the quasi-steady solute concentration field at three values of dimensionless time $\tau$, in the transformed coordinate space $(\xi/Pe, z^*)$. (a–c) Analytical predictions illustrating progressive peak attenuation alongside longitudinal plume dispersion over time. (d–f) Stochastic Brownian dynamics (BD) simulations showing excellent structural and quantitative agreement with the continuum theoretical model, accurately capturing cross-sectional plume bending due to buoyancy effects ($Gr = 2$, $Ha = 0.5$) and diffusive spreading kinetics over time ($(\hat{\beta_1},\hat{\beta_2}) \to 0$).}
    \label{fig:brownvalid}
\end{figure}
\textcolor{blue}{Figure} \ref{fig:brownvalid} plots the analytical solutions obtained from multiscale model and compares them with the results of Brownian dynamics simulations. The analytical continuum model (\textcolor{blue}{figure} \ref{fig:brownvalid}\textcolor{black}{(a,b,c)}) and stochastic Brownian dynamics simulations (\textcolor{blue}{figure} \ref{fig:brownvalid}\textcolor{black}{(d,e,f)}) \DD{exhibit excellent agreement} in capturing  longitudinal dispersion. Both methods \DD{consistently predict the temporal broadening of the solute accompanied by}  progressive peak attenuation as $\tau$ \DD{increases from} $0.5$ to $1.5$. However, \DD{minor differences are confined to the local concentration strength}, as evidenced by the colorbar intensities. The \DD{finite-particle statistical noise inherent} in the stochastic particle tracking method introduces minor, localized fluctuations in the maximum concentration magnitudes, yielding slight deviations from the perfectly smooth theoretical contours.

\begin{figure}[htbp]
    \centering
    \includegraphics[width=1\textwidth]{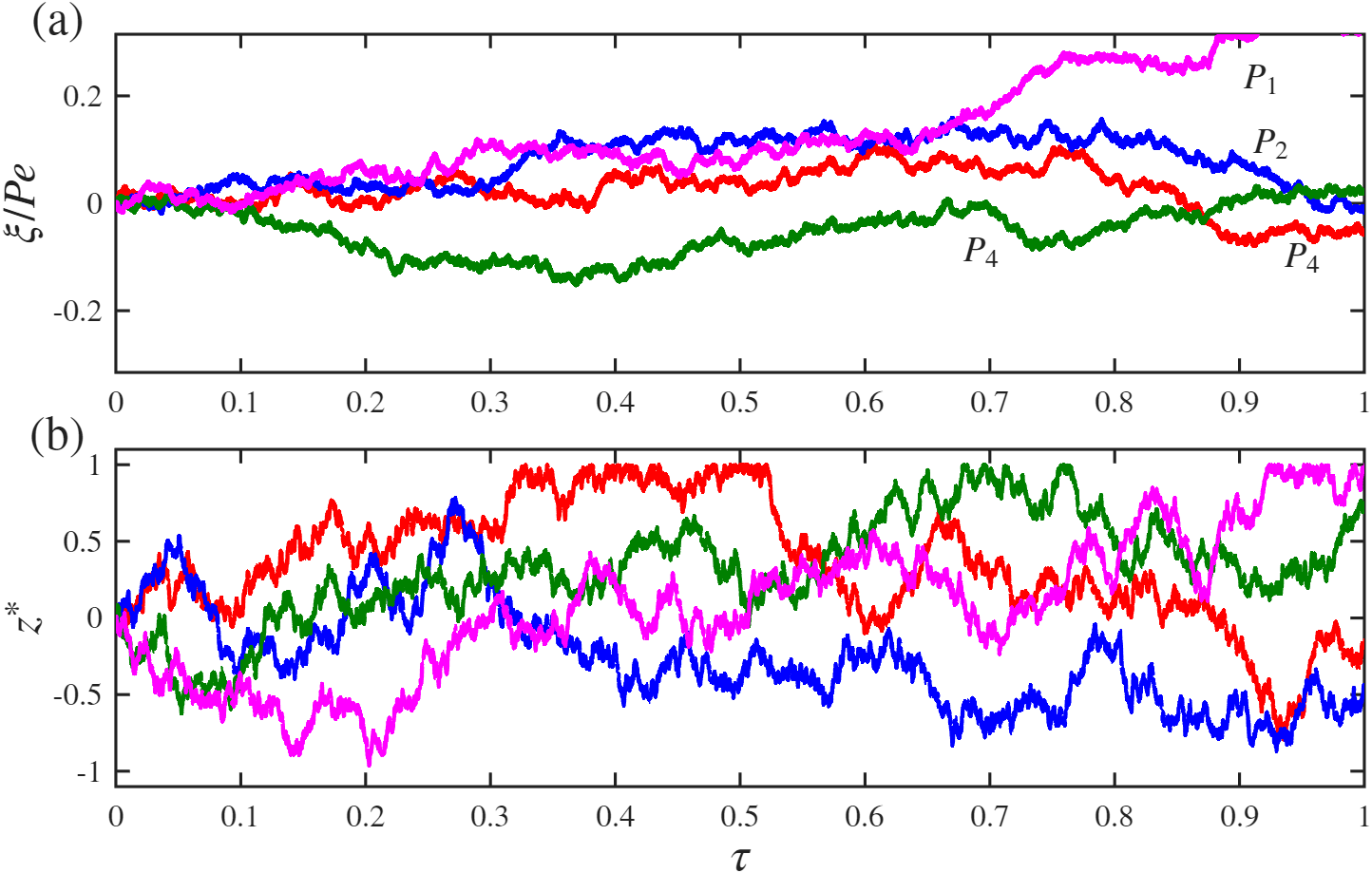}
    
    \vspace{0.5mm}
    
    \includegraphics[width=1\textwidth]{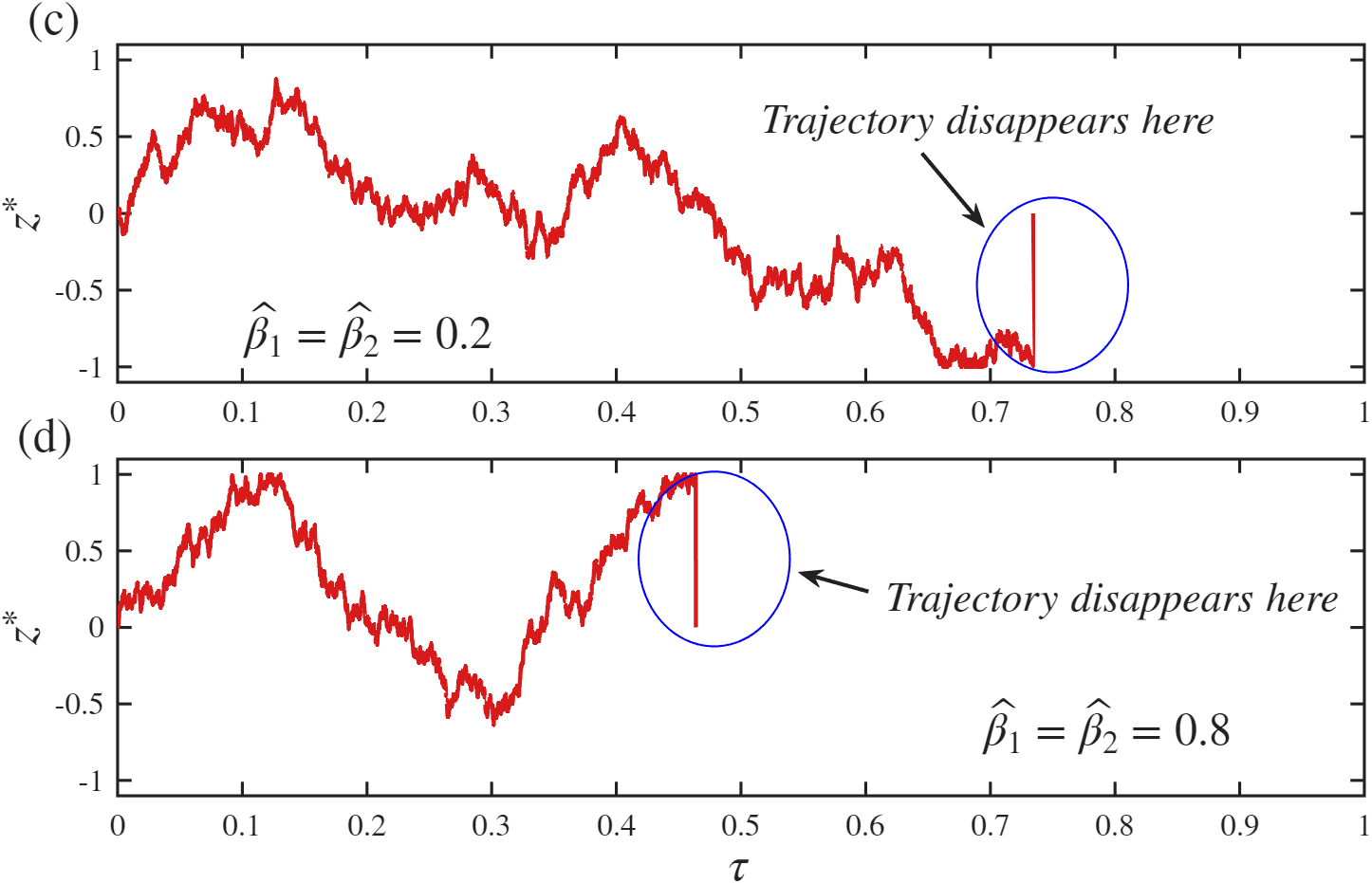}
    
    \vspace{0.5mm}

    \caption{Brownian dynamics (BD) particle trajectories highlighting stochastic dispersion and wall adsorption dynamics. (a) Longitudinal ($\xi/Pe$) and (b) transverse ($z^*$) random walks for four distinct solute particles ($P_1 - P_4$) illustrating spatiotemporal fluctuations across the channel.(c, d) Transverse trajectory ($z^*$) of a single solute particle under purely diffusive flow ($\delta = 20, Ha = Gr = 0$) for uniform wall adsorption values of weak absorption ($\hat{\beta}_1 = \hat{\beta}_2 = 0.2$) and strong absorption ($\hat{\beta}_1 = \hat{\beta}_2 = 0.8$) respectively. The abrupt disappearance of the trajectory signifies an irreversible boundary capture event at the wall ($z^* = \pm 1$) \citep{erban2007reactive}.}
    \label{fig:randomwalk}
\end{figure}

\textcolor{blue}{Figure} \ref{fig:randomwalk}\textcolor{black}{(a,b)}, depicts the random walks of four distinct non-reacting solute particles ($P_1$ through $P_4$) in the longitudinal ($\xi/Pe$) (see \textcolor{blue}{figure} \ref{fig:randomwalk}\textcolor{black}{(a)}) and transverse ($z^*$) (see \textcolor{blue}{figure} \ref{fig:randomwalk}\textcolor{black}{(b)}) directions, respectively. It is evident that, as time progresses, particles exhibit \DD{increasing} longitudinal dispersion away from the source \DD{while undergoing}  gradual transverse migration toward the walls. \DD{To isolate the influence of}  the reactive boundary conditions at the channel walls, we depict \textcolor{blue}{figure} \ref{fig:randomwalk}\textcolor{black}{(c,d)} to illustrate the transverse trajectory of a single solute particle. As mentioned previously, the probability of \DD{particle capture upon collision with the channel walls is a strong} function of the surface reaction parameters ($\hat{\beta}_1, \hat{\beta}_2$). In \textcolor{blue}{figure} \ref{fig:randomwalk}\textcolor{black}{(c)}, where absorption is relatively weak ($\hat{\beta}_1 = \hat{\beta}_2 = 0.2$), the particle survives longer in the domain (until $\tau \approx 0.73$), \DD{undergoing multiple reflections before being permanently captured. In contrast,} under strong absorption conditions ($\hat{\beta}_1 = \hat{\beta}_2 = 0.8$), the probability of capture per collision is much higher as witnessed in \textcolor{blue}{figure} \ref{fig:randomwalk}\textcolor{black}{(d)}. Consequently, the particle is permanently adsorbed much earlier in its trajectory (at $\tau \approx 0.46$) during one of its first encounters with the upper boundary. \DD{The abrupt termination of the trajectory signifies} an irreversible boundary capture event. Physically, this means \DD{that} the solute particles have collided with the adsorptive channel wall and undergone a heterogeneous chemical reaction. Once it reacts or \DD{or is adsorbed}, it is permanently removed from the bulk fluid flow. \DD{According to the forward Euler--Maruyama discretization scheme}\citep{kloeden1977numerical,erban2007reactive}, the computational tracking of that specific particle immediately terminates abruptly in the next pseudo time step. The complex interplay between the induced magnetic field and buoyancy resolved through Brownian dynamics simulations is illustrated in \textcolor{blue}{Supplementary Movie 1}. 
\vspace{-5mm}
\subsection{Numerical solution of deterministic governing equations}
To solve the governing  advection-diffusion-reaction equation (\ref{eq:conc_nd}) \DD{subject to} initial and boundary conditions (\ref{eq:bcicconc}), we employ \DD{a first-order upwind discretization for the advective term, second-order central finite differences for the diffusive terms, and a first-order explicit Euler scheme for time integration.} The \DD{coordinate transformation $\chi$ defined in}  (\ref{eq:chi_definition}) \DD{provides} a substantial numerical advantage by rendering the evolving solute cloud quasi-statically \DD{in the transformed reference frame.}  Consequently, it \DD{eliminates the need for explicitly track the translating concentration front in the physical domain, thereby improving the numerical stability and computational efficiency of the scheme. The first-order forward-time, two-point stencil for the explicit Euler discretization is given by,}
\begin{equation}
    \left. \frac{\partial C^*}{\partial \tau} \right|_{i,j}^k = \frac{C_{i,j}^{*k+1} - C_{i,j}^{*k}}{\Delta \tau} + \mathcal{O}(\Delta \tau)
\end{equation}
The advective flux is discretized using a first-order upwind finite-difference scheme, wherein the stencil direction is determined by the local velocity $u - \chi$, yielding 
\begin{equation}
    \begin{split}
        Pe (u - \chi) \left. \frac{\partial C^*}{\partial \xi} \right|_{i,j}^k 
        &= Pe \Bigg[ \frac{\max(u - \chi, 0)}{\Delta \xi} (C_{i,j}^{*k} - C_{i-1,j}^{*k}) \\
        &\quad + \frac{\max(-(u - \chi), 0)}{\Delta \xi} (C_{i+1,j}^{*k} - C_{i,j}^{*k}) \Bigg] + \mathcal{O}(\Delta \xi)
    \end{split}
\end{equation}
The diffusive terms in the $\xi$- and $z^*$-directions are discretized using second-order central  differences on a uniform grid, \DD{yielding}
\begin{equation}
    \left. \frac{\partial^2 C^*}{\partial \xi^2} \right|_{i,j}^k = \frac{C_{i+1,j}^{*k} - 2C_{i,j}^{*k} + C_{i-1,j}^{*k}}{(\Delta \xi)^2} + \mathcal{O}((\Delta \xi)^2)
\end{equation}
\begin{equation}
    \left. \frac{\partial^2 C^*}{\partial z^{*2}} \right|_{i,j}^k = \frac{C_{i,j+1}^{*k} - 2C_{i,j}^{*k} + C_{i,j-1}^{*k}}{(\Delta z^*)^2} + \mathcal{O}((\Delta z^*)^2)
\end{equation}
\DD{Neglecting the truncation error terms, the fully discrete concentration equation is given below}
\begin{equation}
    \begin{split}
        \frac{C_{i,j}^{*k+1} - C_{i,j}^{*k}}{\Delta \tau} 
        &+ \varepsilon Pe \Bigg[ \frac{\max(u - \chi, 0)}{\Delta \xi} (C_{i,j}^{*k} - C_{i-1,j}^{*k}) + \frac{\max(-(u - \chi), 0)}{\Delta \xi} (C_{i+1,j}^{*k} - C_{i,j}^{*k}) \Bigg] \\
        &= \varepsilon^2 \frac{C_{i+1,j}^{*k} - 2C_{i,j}^{*k} + C_{i-1,j}^{*k}}{(\Delta \xi)^2} 
        + \frac{C_{i,j+1}^{*k} - 2C_{i,j}^{*k} + C_{i,j-1}^{*k}}{(\Delta z^*)^2} - \varepsilon^2\mathcal{K}^*_f C^{*}_{i,j}
    \end{split}
\end{equation}
\DD{The governing equation is discretized using an explicit finite-difference scheme.} For the numerical simulations, the length of the domain along $\xi$ and $z^*$ is \DD{chosen} as $L_\xi = 20.0$ and $L_{z^*} = 2.0\;(L_{z^*} = \varepsilon L_\xi)$ as with the analytical work (since $\varepsilon = 0.1$). The number of points along the $\xi$ and $z^*$ axes is taken as $N_\xi = 201$ and $N_{z^*} = 51$. \DD{The corresponding grid spacings }  along $\xi$ and $z^*$ are taken as $\Delta\xi = L_\xi/(N_\xi - 1)$ and $\Delta z^* = L_{z^*}/(N_{z^*} - 1)$. \DD{To satisfy the stability requirement of the explicit scheme, the time step is chosen as} $\Delta \tau = \min(\Delta\xi, \Delta z^*)^2/4$ \citep{courant1967partial}. The ranges of $\xi$, $z^*$ are $(-L_\xi/2, L_\xi/2)$ and $(-L_{z^*}/2, L_{z^*}/2)$. The dimensionless initial condition is discretized using a Kronecker delta approximation to represent the initial Dirac-delta distribution, \DD{which is written as:} 
\begin{equation}
    C^* \left( \frac{N_\xi - 1}{2} + 1, j \right) = \frac{1}{\Delta \xi}
\end{equation}
The reactive boundary conditions at the transverse walls \DD{are discretized using first-order finite differences}, yielding the discrete expressions as written below.
\begin{subequations}
\begin{align}
    C^*(i, 1, k) &= C^*(i, 2, k) - \hat{\beta_2} \Delta z^* C^*(i, 2, k) \\
    C^*(i, N_{z^*}, k) &= C^*(i, N_{z^*}-1, k) - \hat{\beta_1} \Delta z^* C^*(i, N_{z^*}, k)
\end{align}
\end{subequations}
Here $N_{z^*}$ denotes the total number of discrete nodes along the $z^*$ direction.
\begin{figure}[htbp]
    \centering
    
    \includegraphics[width=1.03\textwidth]{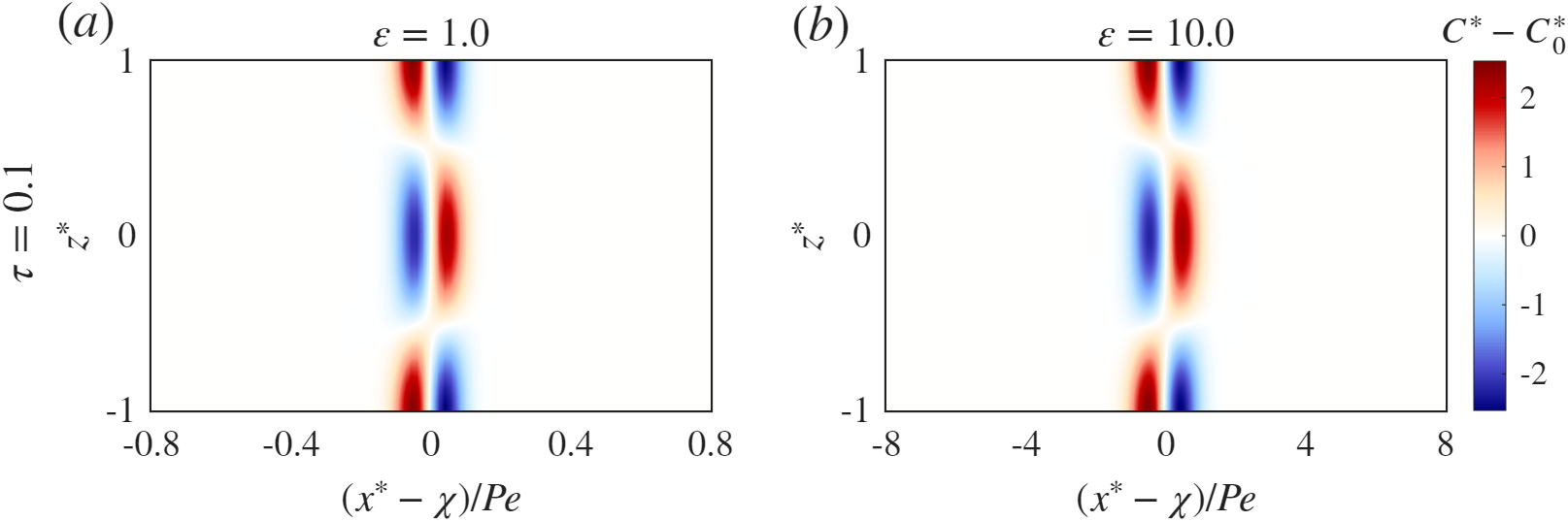}
    
    \vspace{0.5mm}
    \includegraphics[width=1.03\textwidth]{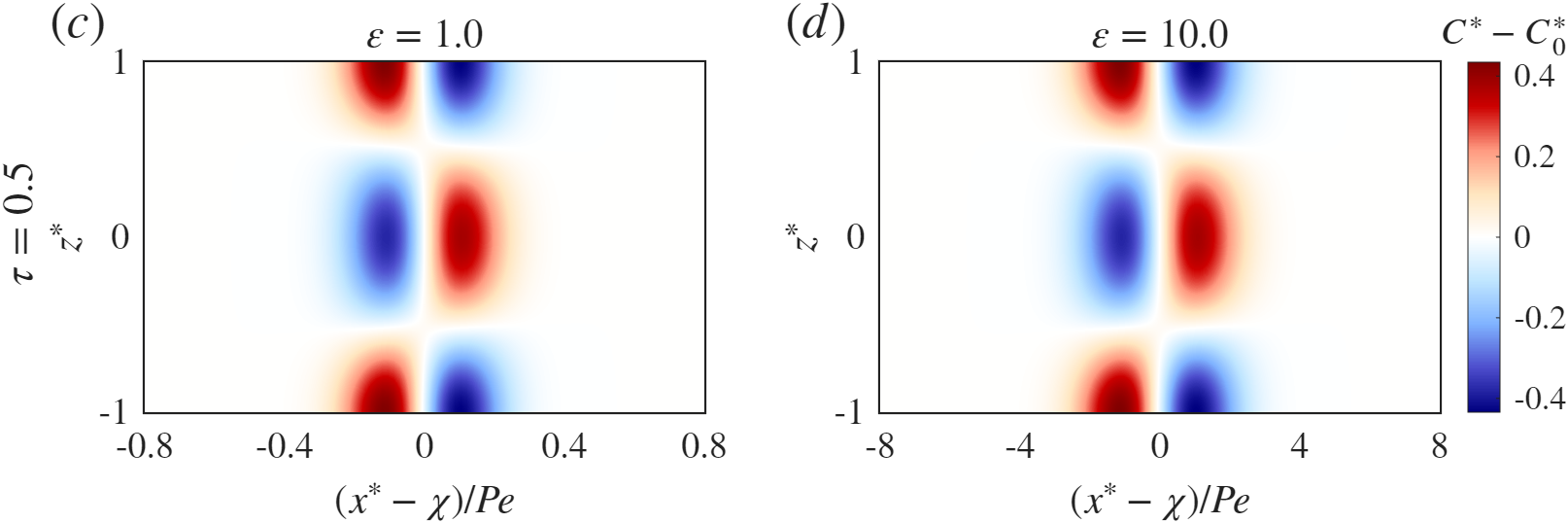}
    \vspace{0.5mm}
   \includegraphics[width=1.03\textwidth]{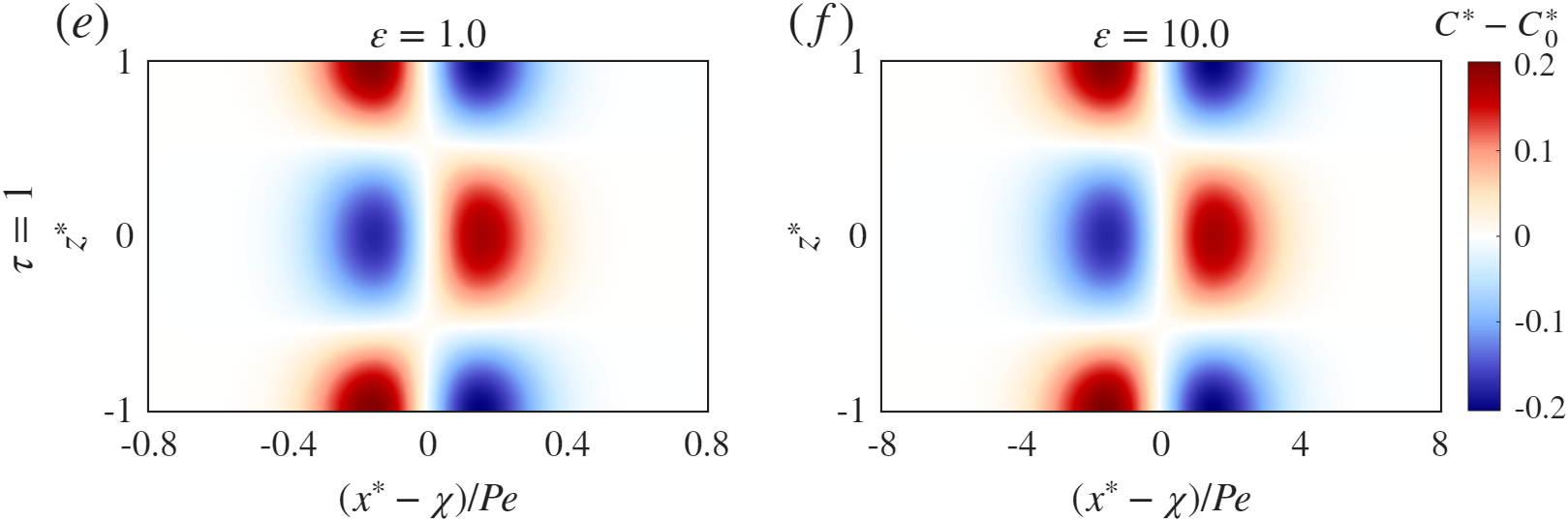}
    \caption{Numerical simulation results of the higher-order solute concentration perturbation ($C^* - C_0^*$) fields, in the transformed coordinate plane for $Pe = 10$, $\delta = 20$, $Gr = Ha = 0$ and first-order reaction rate $\hat{K}_f = 0.1$ using finite difference methodology.}
    \label{fig:fdmdiff}
\end{figure}

As a critical starting point, it is essential to recognize that the dispersion of solutes in a fluid \DD{system} is fundamentally governed by the geometrical aspect ratio, $\varepsilon$. As established in \DD{the} literature \citep{ling2021macroscale,teng2023diffusioosmotic}, standard analytical perturbation solutions often lose their predictive fidelity at higher aspect ratios ($\varepsilon \approx 2$). For values such as $\varepsilon = 10$, theoretical models tend to exhibit significant deviations, particularly \DD{in predicting the} uniform solute depletion near the centerline. Consequently, the simulated results presented in \textcolor{blue}{figure} \ref{fig:fdmdiff}  \DD{particularly become important as they} provide an accurate, non-approximate resolution of the higher-order concentration perturbation field ($C^* - C_0^*$) specifically in these sensitive geometric regimes. Upon comparing the left column for $\varepsilon = 1.0$ (see \textcolor{blue}{figure} \ref{fig:fdmdiff}\textcolor{black}{(a,c,e)}) \DD{with} the right column for $\varepsilon = 10.0$ (see \textcolor{blue}{figure} \ref{fig:fdmdiff}\textcolor{black}{(b,d,f)}), it is observed that the fundamental transverse pattern remains \DD{unchanged}. However, the longitudinal spatial footprint is stretched by \DD{approximately} an order of magnitude. This \DD{observation} confirms that narrow fluidic channels with high aspect ratio, the higher-order dispersive deviations are smeared over a \DD{much} larger streamwise domain, reinforcing why standard perturbation theories fail and full numerical solutions are required in this limit. \DD{As time increases, the perturbation amplitude decreases monotonically throughout the domain. This decay reflects the progressive relaxation of the concentration field during which the higher-order spatial asymmetry weakens and the bulk solute cloud approaches the classical symmetric Gaussian state.} It is worth adding here that the observed dependence of \DD{the} higher-order solute concentration on temporal evolution and channel aspect ratio agrees well with previous findings \citep{samuel2025taylor,Das_Dhar_Kairi_Mondal_2025}.
\subsection{Method of Moments}
\DD{To quantify the transient longitudinal transport of solutes, we employ the method of moments, in which} global moments characterize the spatial evolution and spreading of the cross-sectional mean concentration distribution. \DD{The $n$-th global} moment $M_n(t)$ \DD{is defined as} \citep{latini2001transient,Jiang_2022}:
\begin{equation}
    M_n(t) = \mathcal{\bar{M}}_n(t) \triangleq \int_{-\infty}^{\infty} \xi^n \langle{C^*}(\xi,z^*)\rangle \, d{\xi}, \quad n = 0, 1, \dots
    \label{eq:momentform}
\end{equation}
Here, $\langle .\rangle$ \DD{denotes} the cross-sectional mean concentration distribution \DD{and $\xi$ is the longitudinal} coordinate. The fundamental dispersion parameters, \DD{namely the} mean drift velocity ($U_d$) and \DD{effective Taylor dispersivity} ($D^{eff}_T$), are \DD{obtained from the} first- and second-order global moments as written below.
\begin{equation}
U_d(t) \triangleq \frac{\mathrm{d}\mu_x}{\mathrm{d}t} = \frac{\mathrm{d}M_1}{\mathrm{d}t} \label{eq:drift}
\end{equation}
\begin{equation}
D^{eff}_T(t) \triangleq \frac{1}{2} \frac{\mathrm{d}\sigma^2}{\mathrm{d}t} = \frac{1}{2} \frac{\mathrm{d}M_2}{\mathrm{d}t} - M_1 \frac{\mathrm{d}M_1}{\mathrm{d}t} \label{eq:dispersivity}
\end{equation}
where $\mu_x$ represents the mean displacement and $\sigma^2$ denotes the \DD{longitudinal} variance (mean square displacement). \DD{The temporal evolution of} $U_d(t)$ and $D^{eff}_T(t)$ \DD{provides quantitative insight into transient longitudinal transport prior to the establishment of the asymptotic Taylor dispersion regime} \citep{Gillparam, chatwin1970approach, latini2001transient, Wu_Chen_2014}. These statistical quantities ($\mu_x , \sigma^2$) are \DD{ given by}:
\begin{equation}
\mu_x \triangleq \frac{M_1}{M_0} = M_1, \quad \sigma^2 \triangleq \frac{M_2}{M_0} - \frac{M_1^2}{M_0^2} = M_2 - M_1^2 \label{eq:moments}
\end{equation}

\begin{figure}[htbp]
    \centering
    
    \includegraphics[width=1.02\textwidth]{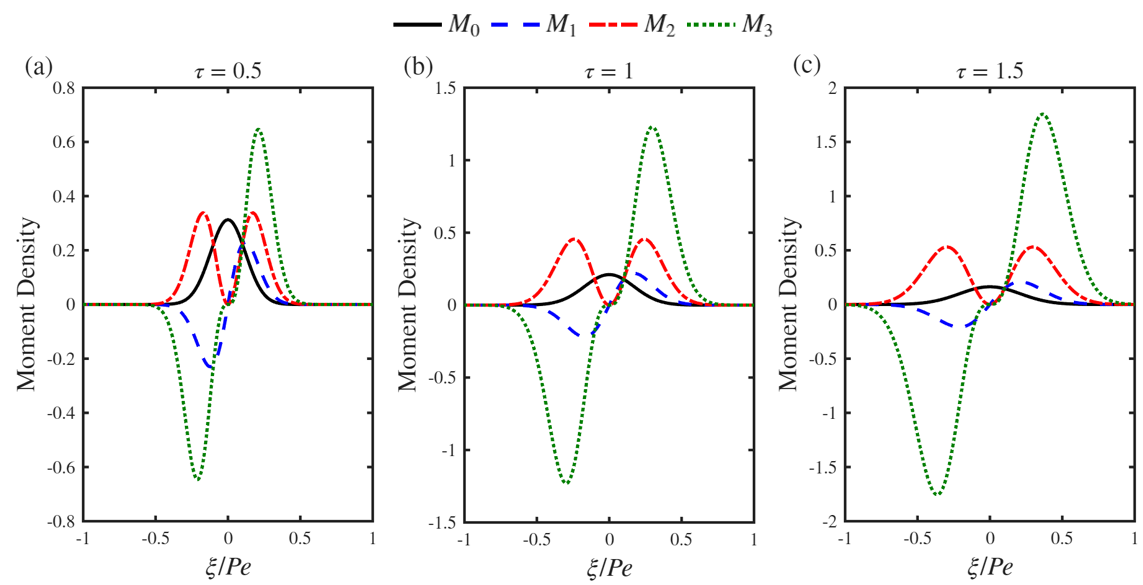}
    \vspace{0.5mm}
    \caption{Spatial distributions of mean local concentration moment densities ($M_0, M_1, M_2, M_3$) along the transformed axial coordinate $\xi/Pe$ at dimensionless times (a) $\tau = 0.5$, (b) $\tau = 1.0$, and (c) $\tau = 1.5$, evaluated for $Pe = 10$, $\hat{K}_f = 0.1$, $\hat{\beta} = 0.5$, $Gr = 2$, and $Ha = 0.5$. Specifically, the rapid growth and spreading of the antisymmetric $M_3$ extrema (reaching magnitudes near $\pm 1.8$ at $\tau = 1.5$) capture strong shear-driven advective dispersion and persistent non-Gaussian plume skewness induced by combined buoyant-magnetic forcing and boundary reaction kinetics.}
    \label{fig:momentdis}
\end{figure}

Based on the integral formulation of the statistical measures provided above,  \textcolor{blue}{figure} \ref{fig:momentdis} \DD{shows}  the spatial distribution of the local concentration moment densities, computed using (\ref{eq:momentform}). The zeroth-order moment density, $M_0$, \DD{corresponds to} the mean longitudinal concentration profile. \textcolor{blue}{Figure} \ref{fig:momentdis}\textcolor{black}{(a,b,c)} \DD{show that} the symmetric central peak \DD{decreases with time while the profile broadens, reflecting the progressive action of longitudinal Taylor dispersion. In contrast,} the first-order moment density, $M_1$, \DD{exhibits} an antisymmetric \DD{distribution arising from the} linear weighting $\xi$, \DD{thereby characterizing the longitudinal displacement of the solute cloud. Its overall shape is preserved throughout the evolution, with only a modest increase in spatial extent.
The second-order moment density, $M_2$, is governed by the quadratic weighting $\xi^2$ and therefore exhibits a symmetric bimodal structure with a minimum at the plume center. As time increases (\textcolor{blue}{figure} \ref{fig:momentdis}\textcolor{black}{(a--c)}), the two peaks move farther apart while their amplitudes increase from approximately $0.35$ at $\tau=0.5$ to about $0.5$ at $\tau=1.5$. This behaviour reflects the continuous growth of the longitudinal variance as the solute cloud spreads downstream.
Finally, the third-order moment density, $M_3$, provides a measure of the longitudinal asymmetry of the concentration field. Owing to the cubic weighting $\xi^3$, the positive and negative lobes become increasingly pronounced with time, reaching magnitudes of approximately $\pm1.8$ at $\tau=1.5$. The progressive amplification of these antisymmetric extrema indicates the persistence of plume skewness generated by the combined effects of buoyancy, magnetic forcing, and heterogeneous wall reactions.}

\begin{figure}[htbp]
    \centering
    
    \includegraphics[width=1.02\textwidth]{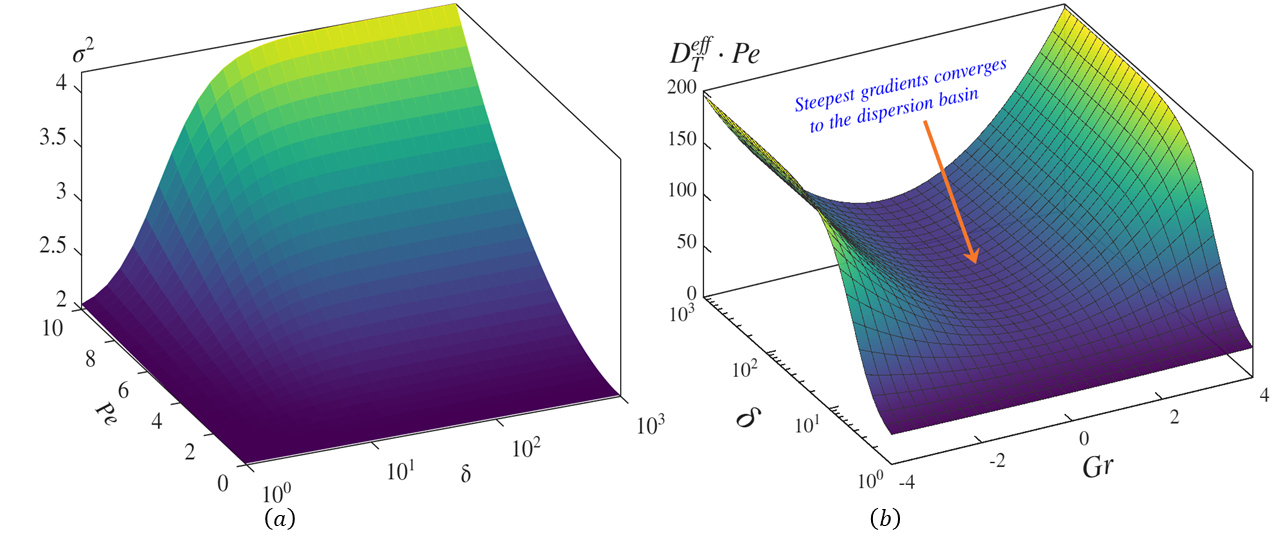}
    \caption{Parametric variation of mean spatial variance ($\sigma^2$) and effective Taylor dispersion ($D_T^{\text{eff}} \cdot Pe$) evaluated at dimensionless time $\tau = 1.0$ under magnetic forcing $Ha = 0.5$.(a) Solute variance $\sigma^2(\delta, Pe)$ at $Gr = 4$, showing monotonic growth with Péclet number ($Pe$) up to $\sigma^2 \approx 4$, alongside strong transport suppression as couple stress parameter increases ($\delta \to 0$).(b) Effective dispersion landscape $D_T^{\text{eff}} \cdot Pe(\delta, Gr)$, displaying a central "dispersion basin" towards which the steepest transport gradients converge. Maximum dispersion ($D_T^{\text{eff}} \cdot Pe \approx 200$) occurs at low couple stress ($\delta \to 0$) under intense symmetric, buoyancy effects for $\vert{}Gr\vert{} = 4$.}
    \label{fig:vardispersion}
\end{figure}

In \textcolor{blue}{figure} \ref{fig:vardispersion}, we depict the parametric surfaces representing the  spatial variance and effective Taylor dispersion landscapes, \DD{providing a global view of the competing transport mechanisms.} In \textcolor{blue}{figure} \ref{fig:vardispersion}\textcolor{black}{(a)}, the spatial variance ($\sigma^2$) \DD{increases monotonically} with the Péclet number ($Pe$). \DD{For a fixed $Pe$,} however, this spreading is profoundly constrained by the couple stress fluid rheology. \DD{In the strongly non-Newtonian regime ($\delta \rightarrow 10^0$),} the internal fluid friction drastically flattens the velocity profile. By blunting the core velocity, the fluid mathematically suppresses the transverse shear gradients ($\partial u^* / \partial z^*$) that fundamentally drives the Taylor dispersion, resulting in a severe attenuation of the total spatial variance regardless of the convective strength.  \textcolor{blue}{Figure} \ref{fig:vardispersion}\textcolor{black}{(b)} \DD{shows} the parametric surface form  by  mapping the effective Taylor dispersion ($D_T^{\text{eff}} \cdot Pe$) in the plane of both rheology ($\delta$) and mixed convection ($Gr$). It is evident from \textcolor{blue}{figure} \ref{fig:vardispersion}\textcolor{black}{(b)} that \DD{distinct minimum appears near the isothermal condition ($Gr=0$), forming a "dispersion basin".} As the buoyancy strength increases in either the assisting ($Gr = 4$) or opposing ($Gr = -4$) regimes, \DD{the dispersion increases almost symmetrically with $|Gr|$. This behaviour indicates that buoyancy enhances the transverse velocity gradients responsible for Taylor dispersion, which in turn, leads to stronger longitudinal spreading away from the isothermal state. The highest dispersion occurs for large $\delta$ (Newtonian limit) combined with strong buoyancy ($|Gr|=4$), where the shear-induced dispersion is maximized.}

\section{Summary, perspective, outlook}
\label{sec:summary}
This study develops higher order multiscale homogenization technique to investigate reactive solute dispersion over asymptotically long time scales in  a viscous fluid exhibiting couple stress rheology. The inelastic non-Newtonian fluid flows through a narrow fluidic channel formed between parallel plates under the combined effects of applied pressure gradients and induced magnetic field  \citep{poddar2021exact,dhar2021dispersion}. The present formulation incorporates an induced magnetic field, providing a more self-consistent description of the coupled magnetohydrodynamics transport problem. Additionally, the model incorporates first-order heterogeneous reactions at both channel walls together with a bulk chemical reaction within the fluid, as these two effects have also been considered in the present study. 

We first derived a closed-form analytical solution for the velocity field considering all the effects mentioned above, and undertake an effort to validate the velocity profile in the limiting behavior of Newtonian fluid against the result of \cite{saha2023solute}. Our analytical solution exhibits a singular branch at $Ha=\delta/2$ (see \textcolor{blue}{figure} \ref{fig:velsol_fail}\textcolor{black}{(a,b)}), where the discriminant vanishes ($\Delta=0$). This singularity arises from the coupled effect of the couple-stress fluid rheological behavior and induced magnetic forcing -- the effect that has not been considered in the reported work of \cite{saha2023solute}. Subsequently, we solve the coupled MHD governing equations numerically (\ref{eq:mhdfinal}-\ref{eq:momentumfinal}). To further characterize the hydrodynamic response, \textcolor{blue}{figure} \ref{fig:velcmap} provides a systematic parameter-regime map showing the effects of buoyancy, magnetic forcing and couple-stress rheology on the velocity field. Strong assisting and opposing buoyancy ($|Gr|=4$) produce pronounced velocity skewness and local flow reversal near the walls, whereas increasing $Ha$ suppresses the transverse shear through Lorentz damping. Specifically, \textcolor{blue}{figure} \ref{fig:velcmap}\textcolor{black}{(c)} presents the scaling law $Ha\approx\delta^{1/4}$, which identifies the crossover between the couple-stress-dominated and magnetically controlled regimes, with the flow progressively approaching the classical Newtonian limit as $\delta$ increases ($\delta^{-1}\to0$).

A higher-order multiscale asymptotic expansion within Mei's homogenization framework is then employed to derive the effective longitudinal dispersion coefficient and reconstruct the two-dimensional solute concentration field for the advection-diffusion-reaction equation. Our analytical framework recovers the classical Taylor-dispersion behavior in the appropriate Newtonian, non-reactive limit and agrees closely with the experimental measurements of \citep{yan2015chip} (see \textcolor{blue}{figure} \ref{fig:tayloredisp}\textcolor{black}{(a)}). To the best of our knowledge, this is the first study where a high-order analysis, with an accuracy of $\mathcal{O}(\varepsilon^3)$, has been used to simultaneously predict the effective dispersion coefficient and two-dimensional concentration structure in a couple-stress conducting fluid under simultaneous influences  of applied pressure gradients, buoyancy and induced magnetic forcing.

We further classify the saturation dynamics of $D^{\text{Taylor}}$ as a function of $\delta$ for representative values of $Pe$ to identify the onset of the Newtonian saturation regime (cf. \textcolor{blue}{figure} \ref{fig:tayloredisp}\textcolor{black}{(d)}). Notably, the saturation locus is described by the empirical relation $D_{\text{sat}} = 4.5473 (\log_{10}\delta)^2 - 11.5350 \log_{10}\delta + 8.4291$. This relation provides a comprehensive quantitative measure of the maximum attainable dispersion as the couple-stress fluid approaches the Newtonian limit. Besides, this established relation may assist in identifying suitable range of rheological parameters in future experiments. The analysis also shows that the saturation threshold increases with $Pe$, indicating that stronger solute advection delays the approach to attain the Newtonian dispersion limit. In accordance with \citep{Wu_Chen_2014,chatwin1970approach}, we quantified the transverse variation rate of the solute dispersion.

Furthermore, to gain quantitative insights into the structural evolution and removal of the solute cloud, we define the mass removal rate within a prescribed longitudinal interval, $R_{\mathrm{rate}}$, and the total removal efficiency, $\eta^{R}$). An important observation from \textcolor{blue}{figure} \ref{fig6}\textcolor{black}{(b)} is that increasing equal boundary-absorption parameters ($\hat{\beta}_1=\hat{\beta}_2$) enhances the scalar removal rate, while \textcolor{blue}{figure} \ref{fig6}\textcolor{black}{(c)} shows that the removal rate is highest near the source region, $[\xi_a,\xi_b]=[1,2]$, and decreases progressively downstream. Previous studies \citep{Wu_Chen_2014,jiang2019solute,poddar2021exact} reported symmetric transverse concentration variations under their respective configurations. In contrast, the present formulation exhibits persistent transverse asymmetry when the induced magnetic field and unequal wall absorption ($\hat{\beta}_1\neq\hat{\beta}_2$) are included, reflecting the coupled influence of the asymmetric flow field and heterogeneous wall mass transfer.

To further assess the credibility of the analytical homogenization framework, we perform Brownian dynamics simulation that provides an independent stochastic description connecting microscopic particle trajectories with the macroscopic concentration evolution (cf. \textcolor{blue}{figure} \ref{fig:brownvalid}). The simulations reproduce the principal features of the analytical solution, including longitudinal spreading, buoyancy-induced plume deformation and progressive attenuation of the concentration peak. Additionally, the near-wall adsorption phenomenon, which is associated with $\hat{\beta}_1$ and $\hat{\beta}_2$ as well as governed by the coupled effects of induced MHD forcing, buoyancy and couple-stress rheology (see \textcolor{blue}{figure} \ref{fig7p}\textcolor{black}{(a--i)}), is resolved through the stochastic ODE-based particle simulations \citep{Jiang2020}. These trajectories further demonstrate that stronger wall absorption increases the probability of irreversible particle capture at the reactive boundaries. We also employ a finite-difference formulation to resolve the higher-order concentration field and to examine the influence of the aspect ratio ($\varepsilon=H/L$), since the multiscale formulation may lose quantitative fidelity at larger aspect ratios \citep{ling2021macroscale,teng2023diffusioosmotic}. The numerical results show that, although the principal transverse concentration structure is retained, the higher-order longitudinal perturbations become distributed over a substantially wider streamwise region as $\varepsilon$ increases, highlighting the complementary role of full numerical solutions in regimes where the asymptotic approximation becomes less reliable. 

Inspired by the seminal work \citep{latini2001transient}, we characterize the transient spreading of the solute cloud using statistical measures, including the global moments, mean displacement and spatial variance. It is evident from \textcolor{blue}{figure} \ref{fig:vardispersion}\textcolor{black}{(a)} that the variance ($\sigma^2$) increases monotonically with $Pe$ and is strongly suppressed by couple-stress rheology, reflecting the reduction of transverse velocity gradients responsible for shear-induced Taylor dispersion. Remarkably, \textcolor{blue}{figure} \ref{fig:vardispersion}\textcolor{black}{(b)} reveals a distinct dispersion basin centered at the isothermal condition ($Gr=0$), with dispersion increasing approximately symmetrically as $|Gr|$ increases. This behavior demonstrates that the magnitude of buoyancy-induced shear, rather than its direction, primarily controls the enhancement of longitudinal dispersion. 

We believe that the inferences of this endeavor will provide insights into thermal buoyancy induced cooling in micro-reactors, effective strategies to mitigate hydrometallurgical gangue contamination \citep{davidson_meta}, MHD driven chromatography \citep{eijkel2003circular}, magnetic field stimulated PCR devices \citep{west2002application} and chemical/food manufacturing processes to minimize undesired by-products \citep{srivastava2025magnetic}. Although the present study is focused on the development of a theoretical framework for understanding the solute dispersion phenomenon under multifaceted effects, experimental investigation mimicking the present study is worth pursuing in future endeavors. Potential improvements of the developed theoretical framework considering multiphase flow configurations with heterogeneous reactive components, which finds relevance to biomedical/biochemical applications, remain a challenge for future studies in this research paradigm.

{\bf Acknowledgments.}
A.R. gratefully acknowledges Professor 
\href{https://sites.google.com/view/avishekranjan/home}{Avishek Ranjan},
Indian Institute of Technology Bombay, India, for introducing him to the theoretical aspects of magnetohydrodynamics (MHD) and the numerical methods employed for its analysis. The author also thanks him for his valuable guidance and insightful discussions on topics related to different fields of MHD, which provided a strong motivation for the present work.  The authors gratefully acknowledge Mr. Kaushal
Agrawal, a senior PhD scholar at the School of Agro and Rural Technology,
Indian Institute of Technology Guwahati, for his help in preparing the supplementary movie.

\vspace{0.5mm}
{\bf Declaration of interests.} The authors declare that there is no conflict of interest.
\label{appenA}
\section*{Appendix A: Details of first order perturbation solution}
\label{appenA}
The leading-order perturbation term $A_1$ is determined by the differential equation
\begin{equation}
    \frac{d^2 A_1}{dz^{*2}} = u^* - \langle u^* \rangle, \quad -1 \le z^* \le 1
    \tag{A1}
    \label{eq:A1_gov}
\end{equation}
which is subject to Neumann boundary conditions at the walls and a zero-mean integral constraint, given by
\begin{equation}
    \left.\frac{dA_1}{dz^*}\right|_{z^*=\pm 1} = 0, \qquad \langle A_1 \rangle = 0
    \tag{A2}
    \label{eq:A1_bc}
\end{equation}
Similarly, the subsequent term $A_2(z^*)$ is obtained by solving
\begin{equation}
    \frac{d^2 A_2}{dz^{*2}} = -1, \quad -1 \le z^* \le 1
    \tag{A3}
    \label{eq:A2_gov}
\end{equation}
with its corresponding boundary conditions and integral constraint expressed as
\begin{equation}
    \left.\frac{dA_2}{dz^*}\right|_{z^*=-1} = \frac{2\beta_2^*}{\beta_1^* + \beta_2^*}, \qquad
    \left.\frac{dA_2}{dz^*}\right|_{z^{*}=1} = -\frac{2\beta_1^*}{\beta_1^* + \beta_2^*}, \qquad
    \langle A_2 \rangle = 0
    \tag{A4}
    \label{eq:A2_bc}
\end{equation}
\section*{Appendix B: Details of second order perturbation solution}
The governing differential equation for $A_{3}$ is given by
\begin{equation}
    \frac{d^{2}A_{3}}{d{z^{*}}^{2}} = u^{*}A_{1} -\langle u^{*}\rangle A_{1} -\langle u^{*}A_{1}\rangle
    \tag{B1}
    \label{eq:A3_gov}
\end{equation}
which is subject to Neumann boundary conditions at the channel walls and a zero-mean integral constraint, expressed as
\begin{equation}
    \left.\frac{dA_{3}}{dz^{*}}\right|_{z^{*}=\pm 1} = 0, \qquad \langle A_{3}\rangle = 0
    \tag{B2}
    \label{eq:A3_bc}
\end{equation}
Proceeding to the next order, the differential equation governing $A_{4}$ takes the form
\begin{equation}
    \frac{d^{2}A_{4}}{d{z^{*}}^{2}} = -A_{1} -\langle u^{*}\rangle A_{2} + u^{*}A_{2} -\langle u^{*}A_{2}\rangle -\frac{\beta_{1}^{*}A_{1}(1)+\beta_{2}^{*}A_{1}(-1)}{\beta_{1}^{*}+\beta_{2}^{*}}
    \tag{B3}
    \label{eq:A4_gov}
\end{equation}
This equation must satisfy the corresponding boundary conditions at the walls, along with the zero-mean integral constraint, given by
\begin{equation}
    \left.\frac{dA_{4}}{dz^{*}}\right|_{z^{*}=-1} = \frac{2\beta_{2}^{*}A_{1}(-1)}{\beta_{1}^{*}+\beta_{2}^{*}}, \qquad
    \left.\frac{dA_{4}}{dz^{*}}\right|_{z^{*}=1} = -\frac{2\beta_{1}^{*}A_{1}(1)}{\beta_{1}^{*}+\beta_{2}^{*}}, \qquad
    \langle A_{4}\rangle = 0
    \tag{B4}
    \label{eq:A4_bc}
\end{equation}
Similarly, advancing to the next order, the differential equation governing the perturbation term $A_{5}$ is
\begin{equation}
    \frac{d^{2}A_{5}}{d{z^{*}}^{2}} = -A_{2} -\frac{\beta_{1}^{*}A_{2}(1)+\beta_{2}^{*}A_{2}(-1)}{\beta_{1}^{*}+\beta_{2}^{*}}
    \tag{B5}
    \label{eq:A5_gov}
\end{equation}
The associated boundary conditions and integral constraint for $A_{5}$ follow an analogous form:
\begin{equation}
    \left.\frac{dA_{5}}{dz^{*}}\right|_{z^{*}=-1} = \frac{2\beta_{2}^{*}A_{2}(-1)}{\beta_{1}^{*}+\beta_{2}^{*}}, \qquad
    \left.\frac{dA_{5}}{dz^{*}}\right|_{z^{*}=1} = -\frac{2\beta_{1}^{*}A_{2}(1)}{\beta_{1}^{*}+\beta_{2}^{*}}, \qquad
    \langle A_{5}\rangle = 0
    \tag{B6}
    \label{eq:A5_bc}
\end{equation}
\section*{Appendix C: Details of third order perturbation solution}
The boundary-value problem governing the transverse function $A_6(z^*)$ is given by the differential equation
\begin{equation}
    \frac{d^2 A_6}{d{z^*}^2} = u^* A_3 - \langle u^* \rangle A_3 - \langle u^* A_3 \rangle - \langle u^* A_1 \rangle A_1
    \tag{C1}
    \label{eq:A6_gov}
\end{equation}
which is subject to homogeneous Neumann boundary conditions at the channel walls along with the zero-mean integral constraint:
\begin{equation}
    \left.\frac{dA_6}{dz^*}\right|_{z^*=\pm 1} = 0, \qquad \langle A_6 \rangle = 0
    \tag{C2}
    \label{eq:A6_bc}
\end{equation}
Similarly, the differential equation for the next higher-order transverse function, $A_7(z^*)$, is expressed as
\begin{equation}
    \begin{aligned}
        \frac{d^2 A_7}{d{z^*}^2} &= u^* A_4 - \langle u^* \rangle A_4 - \langle u^* A_4 \rangle - A_3 - A_2 \langle u^* A_1 \rangle - A_1 \langle u^* A_2 \rangle \\
        &\quad - A_1 \frac{\beta_1^* A_1(1) + \beta_2^* A_1(-1)}{\beta_1^* + \beta_2^*} - \frac{\beta_1^* A_3(1) + \beta_2^* A_3(-1)}{\beta_1^* + \beta_2^*}
    \end{aligned}
    \tag{C3}
    \label{eq:A7_gov}
\end{equation}
The corresponding boundary conditions and the integral normalization constraint for $A_7(z^*)$ are given by
\begin{equation}
    \left.\frac{dA_7}{dz^*}\right|_{z^*=-1} = \frac{2\beta_2^* A_3(-1)}{\beta_1^* + \beta_2^*}, \qquad
    \left.\frac{dA_7}{dz^*}\right|_{z^*=1} = -\frac{2\beta_1^* A_3(1)}{\beta_1^* + \beta_2^*}, \qquad
    \langle A_7 \rangle = 0
    \tag{C4}
    \label{eq:A7_bc}
\end{equation}
Continuing the expansion, the governing differential equation for the transverse function $A_8(z^*)$ is given by
\begin{equation}
    \begin{aligned}
        \frac{d^2 A_8}{d{z^*}^2} &= u^* A_5 - \langle u^* \rangle A_5 - \langle u^* A_5 \rangle - A_4 - \langle u^* A_2 \rangle A_2 \\
        &\quad - A_1 \frac{\beta_1^* A_2(1) + \beta_2^* A_2(-1)}{\beta_1^* + \beta_2^*} - A_2 \frac{\beta_1^* A_1(1) + \beta_2^* A_1(-1)}{\beta_1^* + \beta_2^*} - \frac{\beta_1^* A_4(1) + \beta_2^* A_4(-1)}{\beta_1^* + \beta_2^*}
    \end{aligned}
    \tag{C5}
    \label{eq:A8_gov}
\end{equation}
This equation is subject to the following non-homogeneous Neumann boundary conditions at the channel walls and the integral constraint:
\begin{equation}
    \left.\frac{dA_8}{dz^*}\right|_{z^*=-1} = \frac{2\beta_2^* A_4(-1)}{\beta_1^* + \beta_2^*}, \qquad
    \left.\frac{dA_8}{dz^*}\right|_{z^*=1} = -\frac{2\beta_1^* A_4(1)}{\beta_1^* + \beta_2^*}, \qquad
    \langle A_8 \rangle = 0
    \tag{C6}
    \label{eq:A8_bc}
\end{equation}
Finally, the boundary-value problem for the remaining transverse function, $A_9(z^*)$, is expressed as
\begin{equation}
    \frac{d^2 A_9}{d{z^*}^2} = -A_5 - A_2 \frac{\beta_1^* A_2(1) + \beta_2^* A_2(-1)}{\beta_1^* + \beta_2^*} - \frac{\beta_1^* A_5(1) + \beta_2^* A_5(-1)}{\beta_1^* + \beta_2^*}
    \tag{C7}
    \label{eq:A9_gov}
\end{equation}
with the associated wall boundary conditions and zero-mean condition specified as
\begin{equation}
    \left.\frac{dA_9}{dz^*}\right|_{z^*=-1} = \frac{2\beta_2^* A_5(-1)}{\beta_1^* + \beta_2^*}, \qquad
    \left.\frac{dA_9}{dz^*}\right|_{z^*=1} = -\frac{2\beta_1^* A_5(1)}{\beta_1^* + \beta_2^*}, \qquad
    \langle A_9 \rangle = 0
    \tag{C8}
    \label{eq:A9_bc}
\end{equation}
Since the velocity field $u^*$ does not admit a tractable closed-form solution, obtaining exact analytical expressions for the transverse functions $A_i \; \left(i = 1,2,3,...,7,8,9 \right)$ is unfeasible. Consequently, we employ a semi-analytical approach, solving the system of boundary-value problems via numerical computation.
\bibliographystyle{jfm2}
\begingroup
\def\bibfont{\small}
\setlength{\bibsep}{1.1ex}
\bibliography{mhdcite}
\endgroup	
\end{document}